\documentclass[a4paper,oneside,11pt]{article}
\usepackage{cprform}
\usepackage{amsmath,amstext,amsfonts,amsbsy,amssymb,amscd,bbm,epsfig,lscape,slashed,color}

\usepackage{epsfig,cite}
\usepackage{amssymb}
\usepackage{subfigure}
\usepackage{hyperref}
\usepackage{lscape}
\usepackage[font={small}]{caption}
\usepackage[margin=1.1in]{geometry}
\usepackage{graphicx}
\usepackage{multirow,booktabs}
\usepackage{placeins}

\usepackage{tikz}
\usetikzlibrary{arrows,shapes}
\usetikzlibrary{trees}
\usetikzlibrary{matrix,arrows}                          
\usetikzlibrary{positioning}                            
\usetikzlibrary{calc,through}                           
\usetikzlibrary{decorations.pathreplacing}  
\usetikzlibrary{decorations.pathmorphing}       
\usetikzlibrary{decorations.markings}
\tikzset{
    vector/.style={decorate, decoration={snake}, draw},
        provector/.style={decorate, decoration={snake,amplitude=2.5pt}, draw},
        antivector/.style={decorate, decoration={snake,amplitude=-2.5pt}, draw},
    fermion/.style={draw=black, postaction={decorate},
        decoration={markings,mark=at position .55 with {\arrow[draw=black]{>}}}},
    fermionbar/.style={draw=black, postaction={decorate},
        decoration={markings,mark=at position .55 with {\arrow[draw=black]{<}}}},
    fermionnoarrow/.style={draw=black},
    gluon/.style={decorate, draw=black,
        decoration={coil,amplitude=4pt, segment length=5pt}},                           
    scalar/.style={dashed,draw=black, postaction={decorate},
        decoration={markings,mark=at position .55 with {\arrow[draw=black]{>}}}},
    scalarbar/.style={dashed,draw=black, postaction={decorate},
        decoration={markings,mark=at position .55 with {\arrow[draw=black]{<}}}},
    scalarnoarrow/.style={dashed,draw=black},
    electron/.style={draw=black, postaction={decorate},
        decoration={markings,mark=at position .55 with {\arrow[draw=black]{>}}}},
        bigvector/.style={decorate, decoration={snake,amplitude=4pt}, draw},
}

\newcommand{\texp}{{\rm T}\kern-2pt\exp}

\newcommand{\ba}{\begin{array}}
\newcommand{\ea}{\end{array}}

\newcommand{\req}[1]{Eq.~(\ref{#1})}
\newcommand{\res}[1]{Section~\ref{#1}}
\newcommand{\reapp}[1]{Appendix~\ref{#1}}

\newcommand{\refig}[1]{Fig.~\ref{#1}}

\newcommand{\dif}{{\rm d}}

\newcommand{\Dslash}{\relax{\kern+.25em / \kern-.70em D}}

\newcommand{\GeV}{{\rm GeV}}

\newcommand{\Real}{\relax{\mathsf{\Gamma\kern-.35em R}}}
\newcommand{\Int}{\relax{\mathsf{Z\kern-.40em Z}}}

\newcommand{\CF}{C_{\rm F}}

\newcommand{\half}{{\scriptstyle{{1\over 2}}}}

\newcommand{\ihalf}{{\scriptstyle{{i\over 2}}}}

\newcommand{\NC}{N}
\newcommand{\NF}{N_\mathrm{\scriptstyle f}}

\newcommand{\MSbar}{{\overline{\rm MS}}}
\newcommand{\SF}{{\rm SF}}

\newcommand{\gbar}{\kern1pt\overline{\kern-1pt g\kern-0pt}\kern1pt}
\newcommand{\mbar}{\kern2pt\overline{\kern-1pt m\kern-1pt}\kern1pt}
\newcommand{\obar}[1]{\kern3pt\overline{\kern-2pt #1\kern-0pt}\kern1pt}
\newcommand{\gren}{g_{\rm R}}
\newcommand{\mren}[1]{m_{{\rm R} #1}}

\newcommand{\oren}[1]{#1_{\rm R}}

\newcommand{\orgi}[1]{\hat #1}

\newcommand{\mcrit}{m_{\rm cr}}

\newcommand{\alphas}{\alpha_{\rm\scriptscriptstyle s}}

\newcommand{\Oa}{\mbox{O}(a)}

\newcommand{\icsw}{c_{\rm sw}}

\newcommand{\abar}{\kern1pt\overline{\kern-1pt a\kern-0.5pt}\kern1pt}

\newcommand{\cA}{{\cal A}}

\newcommand{\cO}{{\cal O}}

\newcommand{\cQ}{{\cal Q}}

\newcommand{\cS}{{\cal S}}

\newcommand{\cW}{{\cal W}}
\newcommand{\cX}{{\cal X}}

\newcommand{\cZ}{{\cal Z}}

\newcommand{\vx}{\mathbf{x}}
\newcommand{\vy}{\mathbf{y}}
\newcommand{\vz}{\mathbf{z}}

\begin{document}


\begin{titlepage}


\vspace*{-30truemm}
\begin{flushright}
IFT-UAM/CSIC-16-140\\
FTUAM-16-48\\[10pt]
{\large January 2018}
\end{flushright}
\vspace{15truemm}


\centerline{\Bigrm On the perturbative renormalisation of}
\vskip 3 true mm
\centerline{\Bigrm four-quark operators for new physics}
\vskip 7 true mm
\begin{center}
\includegraphics[width=25mm]{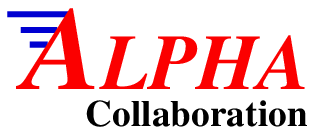}
\end{center}
\centerline{\bigrm  M.~Papinutto,$^a$  C.~Pena,$^{b,c}$ D.~Preti$^c$}
\vskip 4 true mm
\centerline{\it $^a$ Dipartimento di Fisica, ``Sapienza'' Universit\`a di Roma, and INFN, Sezione di Roma}
\centerline{\it Piazzale A. Moro 2, I-00185 Roma, Italy}
\vskip 3 true mm
\centerline{\it $^b$ Departamento de F\'{\i}sica Te\'orica, Universidad Aut\'onoma de Madrid}
\centerline{\it Cantoblanco E-28049 Madrid, Spain}
\vskip 3 true mm
\centerline{\it $^c$ Instituto de F\'{\i}sica Te\'orica UAM-CSIC}
\centerline{\it c/Nicol\'as Cabrera 13-15, Universidad Aut\'onoma de Madrid}
\centerline{\it Cantoblanco E-28049 Madrid, Spain}
\vskip 30 true mm


\noindent{\tenbf Abstract:}
{\tenrm
We discuss the renormalisation properties of the full set of $\Delta F=2$ operators involved in BSM processes,
including the definition of RGI versions of operators that exhibit mixing under RG transformations.
As a first step for a fully non-perturbative determination of the scale-dependent renormalization factors and their runnings,
we introduce a family of appropriate Schr\"odinger Functional schemes, and study them in perturbation theory.
This allows, in particular, to determine the NLO anomalous dimensions of all $\Delta F=1,2$ operators
in these schemes. Finally, we discuss the systematic uncertainties related to the use of NLO perturbation
theory for the RG running of four-quark operators to scales in the GeV range, in both our SF schemes
and standard $\MSbar$ and RI-MOM schemes. Large truncation effects are found for some of the operators considered.
}
\vspace{10truemm}

\eject
\end{titlepage}

\section{Introduction}
\label{sec:intro}

Hadronic matrix elements of four-quark operators play an important r\^ole in the study of flavour physics within the Standard Model (SM),
as well as in searches for new physics. In particular, they are essential to the study of CP-violation in the hadron sector
in both the SM and beyond-the-SM
(BSM) models, where they parametrise the effect of new interactions.
A key ingredient of these studies is the renormalisation of the operators, including their renormalisation group (RG) running
from low-energy hadronic scales $\mathcal{O}(\Lambda_{\rm\scriptscriptstyle QCD})$ to the high-energy electroweak or new physics scales,
where contact with the fundamental underlying theory is made.

In this paper we prepare the ground for a full non-perturbative computation of the low-energy renormalisation and RG running of all possible four-quark operators with net flavour change, by introducing appropriate Schr\"odinger Functional (SF) renormalisation schemes. In order to connect them with standard $\MSbar$ schemes at high energies,
as well as with renormalisation group invariant (RGI) operators, it is however still necessary to compute
the relevant scheme matching factors perturbatively. We compute the latter at one loop, which, in particular,
allows us to determine the complete set of next-to-leading (NLO) anomalous dimensions in our SF schemes.

An interesting byproduct of our computation is the possibility to study the systematic uncertainties related to the use of NLO
perturbation theory in the computation of the RG running of four-quark operators to hadronic scales. This is a common feature
of the phenomenological literature, and the question can be posed whether perturbative truncation effects can have an impact
in physics analyses. The latter are studied in detail in our SF schemes, as well as in the $\MSbar$ and RI-MOM schemes
that have been studied in the literature.
One of our main conclusions is that perturbative truncation effects in RG running can be argued to be significantly large.
This makes a very strong case for a fully non-perturbative RG running programme for these operators.

The structure of the paper is as follows. In \res{sec:renorm} we provide a short review of the renormalisation properties of the
full basis of $\Delta F=2$ four-quark operators, stressing how considering it also allows to obtain the anomalous
dimensions of $\Delta F=1$ operators. We focus on the operators that appear in BSM physics, which exhibit scale-dependent
mixing under renormalisation, and discuss the definition of RGI operators in that case.
In \res{sec:sf} we introduce our SF schemes, and explain the strategy to obtain NLO anomalous dimensions in the latter
through a one-loop computation of the relevant four- and two-point correlation functions. Finally, in \res{sec:rgpt}
we carry out a systematic study of the perturbative RG running in several schemes, and provide estimates of the
resulting truncation uncertainty at scales in the few GeV range. In order to improve readability, several tables
and figures are collected after the main text, and a many technical details are discussed in appendices.

\section{Renormalisation of four-quark operators}
\label{sec:renorm}

\subsection{Mixing of four-quark operators under renormalisation}

The mixing under renormalisation of four-quark operators that do not
require subtraction of lower-dimensional operators has been determined
in full generality in~\cite{Donini:1999sf}. The absence of subtractions
is elegantly implemented by using a formalism in which the operators
are made of four different quark flavours; a complete set of Lorentz-invariant
operators is
\begin{gather}
\label{eq:rel_ops}
\ba{l@{}l@{}l@{\hspace{20mm}}l@{}l@{}l}
Q_1^\pm &\,\,=\,\,& \cO^\pm_{\rm VV+AA}\,, \quad &\cQ_1^\pm &\,\,=\,\,& \cO^\pm_{\rm VA+AV}\,,\\[1.0ex]
Q_2^\pm &\,\,=\,\,& \cO^\pm_{\rm VV-AA}\,, \quad &\cQ_2^\pm &\,\,=\,\,& \cO^\pm_{\rm VA-AV}\,,\\[1.0ex]
Q_3^\pm &\,\,=\,\,& \cO^\pm_{\rm SS-PP}\,, \quad &\cQ_3^\pm &\,\,=\,\,& \cO^\pm_{\rm PS-SP}\,,\\[1.0ex] 
Q_4^\pm &\,\,=\,\,& \cO^\pm_{\rm SS+PP}\,, \quad &\cQ_4^\pm &\,\,=\,\,& \cO^\pm_{\rm PS+SP}\,,\\[1.0ex] 
Q_5^\pm &\,\,=\,\,& -2\cO^\pm_{\rm TT}\,,    \quad &\cQ_5^\pm &\,\,=\,\,& -2\cO^\pm_{\rm T\tilde{T}}\,,
\ea
\end{gather}
where
\begin{gather}
\label{eq:gen_4f}
\cO^\pm_{\Gamma_1\Gamma_2} = \frac{1}{2}\left[
(\bar\psi_1\Gamma_1\psi_2)(\bar\psi_3\Gamma_2\psi_4)\,\pm\,
(\bar\psi_1\Gamma_1\psi_4)(\bar\psi_3\Gamma_2\psi_2)
\right]\,,
\end{gather}
$\cO^\pm_{\Gamma_1\Gamma_2\pm\Gamma_2\Gamma_1}\equiv\cO^\pm_{\Gamma_1\Gamma_2}\pm\cO^\pm_{\Gamma_2\Gamma_1}$,
and the labeling is adopted ${\rm V}\to\gamma_\mu$, ${\rm A}\to\gamma_\mu\gamma_5$, ${\rm S}\to\mathbf{1}$,
${\rm P}\to\gamma_5$, ${\rm T}\to\sigma_{\mu\nu}$, ${\rm \tilde{T}}\to\half\varepsilon_{\mu\nu\rho\tau}\sigma_{\rho\tau}$,
with $\sigma_{\mu\nu}=\ihalf [\gamma_\mu,\gamma_\nu]$.
In the above expression round parentheses indicate spin and colour scalars, and subscripts are flavour labels.
Note that operators $Q_k^\pm$ are parity-even, and $\cQ_k^\pm$ are parity-odd.

It is important to stress that this framework is fairly general. For instance,
with the assignments
\begin{gather}
\psi_1=\psi_3=s\,,~~~~~\psi_2=\psi_4=d
\end{gather}
the operators $Q_k^-$ vanish, while $Q_1^+$ enters the SM amplitude
for $K^0$--$\bar K^0$ mixing, and $Q_{2,\ldots,5}^+$ the contributions to the same amplitude
from arbitrary extensions of the SM. Idem for $B_{(s)}^0$--$\bar B_{(s)}^0$ mixing
with
\begin{gather}
\psi_1=\psi_3=b\,,~~~~~\psi_2=\psi_4=d/s\,.
\end{gather}
If one instead chooses the assignments
\begin{gather}
\begin{split}
&\psi_1=s\,,~~~\psi_2=d\,,~~~\psi_3=\psi_4=u,c\,,\\
&\psi_1=s\,,~~~\psi_4=d\,,~~~\psi_2=\psi_3=u,c\,,
\end{split}
\end{gather}
the resulting $Q_1^\pm$ will be the operators in the SM $\Delta S=1$ effective weak Hamiltonian
with an active charm quark, which, in the chiral limit, do not mix with lower-dimensional
operators. By proceeding in this way, essentially all possible four-quark effective interactions
with net flavour change can be easily seen to be comprised within our scope.

In the following we will assume a mass-independent renormalisation scheme.
Renormalised operators can be written as
\begin{gather}
\label{eq:ren_relativistic}
\begin{split}
\bar Q_k^\pm &= Z_{kl}^{\pm}(\delta_{lm}+\Delta_{lm}^{\pm}) Q_m^\pm\,,\\
\bar\cQ_k^\pm &= \cZ_{kl}^{\pm}(\delta_{lm}+\mbox{\textcyr{D}}_{lm}^{\pm})\cQ_m^\pm
\end{split}
\end{gather}
(summations over $l,m$ are implied\footnote{to simplify the notation from now on we suppress the superscript "$\pm$"}), where the matrices $Z,\cZ$ are scale-dependent and reabsorb
logarithmic divergences, while $\Delta,\mbox{\textcyr{D}}$ are (possible) finite subtraction coefficients
that only depend on the bare coupling. They have the generic structure
\begin{gather}
\label{eq:ren_pattern}
Z=\left(\ba{ccccc}
Z_{11} & 0 & 0 & 0 & 0 \\
0 & Z_{22} & Z_{23} & 0 & 0 \\
0 & Z_{32} & Z_{33} & 0 & 0 \\
0 & 0 & 0 & Z_{44} & Z_{45} \\
0 & 0 & 0 & Z_{54} & Z_{55}
\ea\right)\,,\qquad
\Delta=\left(\ba{ccccc}
0 & \Delta_{12} & \Delta_{13} & \Delta_{14} & \Delta_{15} \\
\Delta_{21} & 0 & 0 & \Delta_{24} & \Delta_{25} \\
\Delta_{31} & 0 & 0 & \Delta_{34} & \Delta_{35} \\
\Delta_{41} & \Delta_{42} & \Delta_{43} & 0 & 0 \\
\Delta_{51} & \Delta_{52} & \Delta_{53} & 0 & 0
\ea\right)\,,
\end{gather}
and similarly in the parity-odd sector. If chiral symmetry is preserved by the regularization,
both $\Delta$ and $\mbox{\textcyr{D}}$ vanish. The main result of~\cite{Donini:1999sf} is
that $\mbox{\textcyr{D}}=0$ even when a lattice regularisation that breaks chiral symmetry
explicitly through the Wilson term is employed, due to the presence of
residual discrete flavour symmetries.
In particular, the left-left operators $\cO_{\rm VA+AV}^\pm$
that mediate Standard Model-allowed transitions renormalise multiplicatively,
while operators that appear as effective interactions in extensions of the
Standard Model do always mix.\footnote{The use of twisted mass Wilson regularisations
leads to a different chiral symmetry breaking pattern, which changes the mixing
properties. This can be exploited in specific cases to achieve favorable mixing
patterns, see e.g.~\cite{Frezzotti:2000nk,Pena:2004gb,Frezzotti:2004wz}.}

Interestingly, in~\cite{Donini:1999sf} some identities are derived that relate the
renormalisation matrices for $(\cQ_2^+,\cQ_3^+)$ and $(\cQ_2^-,\cQ_3^-)$
in RI-MOM schemes. In~\reapp{app:symm} we discuss the underlying symmetry
structure in some more detail, and show how it can be used to derive
constraints between matrices of anomalous dimensions in generic schemes.

\subsection{Callan-Symanzik equations}

Theory parameters and operators are renormalised at the renormalisation scale $\mu$.
The scale dependence of renormalised quantities is then governed by renormalisation
group evolution. We will consider QCD with $\NF$ quark flavours and $\NC$ colours.
The Callan-Symanzik equations satisfied by the gauge coupling and quark masses
are of the form
\begin{align}
q\frac{\rm d}{{\rm d}q}\gbar(q) &= \beta(\gbar(q))\,,\\
q\frac{\rm d}{{\rm d}q}\mbar_{\rm f}(q) &= \tau(\gbar(q))\mbar_{\rm f}(q)\,,
\end{align}
respectively, and satisfy the initial conditions
\begin{align}
\gbar(\mu) &= \gren\,,\\
\mbar_{\rm f}(\mu) &= \mren{,\rm f}\,,
\end{align}
where ${\rm f}$ is a flavour label. Mass-independence of the scheme is reflected
in the fact that the beta function and mass anomalous dimension $\tau$ depend
on the coupling and the number of flavours, but not on quark masses. Asymptotic perturbative expansions read
\begin{align}
\beta(g) &\underset{g \sim 0}{\approx} -g^3(b_0+b_1g^2+\ldots)\,,\\
\tau(g)  &\underset{g \sim 0}{\approx} -g^2(d_0+d_1g^2+\ldots)\,.
\end{align}
The universal coefficients of the perturbative beta function and mass anomalous
dimension are
\begin{gather}
\begin{split}
b_0 &= \frac{1}{(4\pi)^2}\left[\frac{11}{3}\NC-\frac{2}{3}\NF\right]\,,\\
b_1 &= \frac{1}{(4\pi)^4}\left[\frac{34}{3}\NC^2-\left(\frac{13}{3}\NC-\frac{1}{\NC}\right)\NF\right]\,,\\
d_0 &= \frac{1}{(4\pi)^2}\,\frac{3(\NC^2-1)}{\NC}\,.
\end{split}
\end{gather}

We will deal with Euclidean correlation functions of gauge-invariant composite operators.
Without loss of generality, let us consider correlation functions of the form
\begin{gather}
G_k(x;y_1,\ldots,y_n) = \langle O_k(x)\cO_1(y_1)\cdots\cO_n(y_n))\rangle\,,
\end{gather}
with $x \neq y_j~\forall j,~y_j \neq y_k~\forall j \neq k$, where $\{O_k\}$
is a set of operators that mix under renormalisation, and where $\cO_k$ are
multiplicatively renormalisable operators.\footnote{To avoid burdening the notation,
we have omitted the dependence of $G_k$ on coupling and masses, as well as on the renormalisation scale.}
Renormalised correlation functions satisfy a system of Callan-Symanzik equations
obtained by imposing that $G_k$ is independent of the renormalisation scale $\mu$, viz.
\begin{gather}
\label{eq:rge_compact}
\mu\frac{\dif}{\dif\mu} \bar G_j =
\sum_k\left[\gamma_{jk}(\gren)+\sum_{l=1}^n\tilde\gamma_l(\gren)\delta_{jk}\right]\bar G_k\,,
\end{gather}
which, expanding the total derivative, leads to
\begin{gather}
\label{eq:rge_detailed}
\left\{
\mu\frac{\partial}{\partial\mu} +
\beta(\gren)\frac{\partial}{\partial\gren} +
\beta_\lambda(\gren)\lambda\,\frac{\partial}{\partial\lambda} +
\sum_{{\rm f}=1}^{\NF}\tau(\gren)\mren{,\rm f}\frac{\partial}{\partial\mren{,\rm f}} -
\sum_{l=1}^n\tilde\gamma_l(\gren)
\right\}\bar G_j =
\sum_k\gamma_{jk}(\gren)\,\bar G_k \,,
\end{gather}
where $\gamma$ is a matrix of anomalous dimensions describing the mixing of $\{O_k\}$,
and $\tilde\gamma_l$ is the anomalous dimension of $\cO_l$.
For completeness, we have included a term which takes into account the dependence
on the gauge parameter $\lambda$ in covariant gauges; this term is absent in schemes
like $\MSbar$ (irrespective of the regularisation prescription) or the SF schemes
we will introduce, but is present in the RI schemes we will also be concerned with later.
The RG function $\beta_\lambda$ is given by
\begin{gather}
q\frac{\rm{d}}{{\rm d} q}\lambda(q) = \beta_\lambda(\gbar(q))\lambda(q)\,,
\end{gather}
and its perturbative expansion has the form
\begin{gather}
\beta_\lambda(g) = -g^2(b_0^\lambda + b_1^\lambda g^2 + \ldots)\,,
\end{gather}
where the universal coefficient is given by
\begin{gather}
b_0^\lambda = \frac{1}{(4\pi)^2}\left[\left(\lambda-\frac{13}{3}\right)\NC+\frac{4}{3}\NF\right]\,.
\end{gather}
In the Landau gauge ($\lambda=0$) the term with $\beta_\lambda$ always vanishes.
From now on, in order to avoid unnecessary complications, we will assume that whenever
RI anomalous dimensions are employed they will be in Landau gauge, and consequently drop
terms with $\beta_\lambda$ in all equations.

From now on, in order to simplify the notation we will use the shorthand notation
\begin{gather}
\label{eq:rge_oper}
q\frac{\rm{d}}{{\rm d} q}\obar{O}_j(q) = \sum_k\gamma_{jk}(\gbar(q))\obar{O}_k(q)
\end{gather}
for the Callan-Symanzik equation satisfied by the insertion of a composite operator
in a renormalised, on-shell correlation function (i.e. \req{eq:rge_oper} is
to be interpreted in the sense provided by \req{eq:rge_detailed}).
The corresponding initial condition can be written as
\begin{gather}
\obar{O}_k(\mu) = {\oren{O}}_{,k}\,,
\end{gather}
and the perturbative expansion of the anomalous dimension matrix $\gamma$ as
\begin{gather}
\gamma(g) \underset{g \sim 0}{\approx} -g^2(\gamma_0+\gamma_1g^2+\ldots)\,.
\end{gather}
The universal, one-loop coefficients of the anomalous dimension matrix for four-fermion operators
were first computed in~\cite{Ciuchini:1997bw,Bagger:1997gg} and~\cite{Narison:1983kn}. With our notational conventions the non-zero entries read
\begin{gather}
\label{eq:load}
\begin{split}
\gamma^{+,(0)}_{11} &= \left ( 6-\frac{6}{\NC} \right )(4\pi)^{-2} \,,\\
\gamma^{+,(0)}_{22} &= \left ( \frac{6}{\NC}\right )(4\pi)^{-2} \,,\\
\gamma^{+,(0)}_{23} &= 12 (4\pi)^{-2}\,,\\
\gamma^{+,(0)}_{33} &=\left ( -6\NC+\frac{6}{\NC}\right )(4\pi)^{-2} \,,\\
\gamma^{+,(0)}_{44} &=\left ( 6-6\NC+\frac{6}{\NC}\right )(4\pi)^{-2} \,,\\
\gamma^{+,(0)}_{45} &= \left ( \frac{1}{2}-\frac{1}{\NC}\right )(4\pi)^{-2}\,,\\
\gamma^{+,(0)}_{54} &=\left ( -24-\frac{48}{\NC}\right )(4\pi)^{-2} \,,\\
\gamma^{+,(0)}_{55} &=\left ( 6+2\NC-\frac{2}{\NC}\right )(4\pi)^{-2} \,,
\end{split}
\quad \quad \quad \quad
\begin{split}
\gamma^{-,(0)}_{11} &=\left ( -6-\frac{6}{\NC}\right )(4\pi)^{-2} \,,\\
\gamma^{-,(0)}_{22} &=\left ( \frac{6}{\NC}\right )(4\pi)^{-2} \,,\\
\gamma^{-,(0)}_{23} &= -12(4\pi)^{-2}\,,\\
\gamma^{-,(0)}_{33} &=\left ( -6\NC+\frac{6}{\NC}\right )(4\pi)^{-2} \,,\\
\gamma^{-,(0)}_{44} &=\left ( -6-6\NC+\frac{6}{\NC}\right )(4\pi)^{-2} \,,\\
\gamma^{-,(0)}_{45} &=\left (  -\frac{1}{2}-\frac{1}{\NC}\right )(4\pi)^{-2}\,,\\
\gamma^{-,(0)}_{54} &=\left ( 24-\frac{48}{\NC}\right )(4\pi)^{-2} \,,\\
\gamma^{-,(0)}_{55} &=\left ( -6+2\NC-\frac{2}{\NC}\right )(4\pi)^{-2} \,.
\end{split}
\end{gather}

\subsection{Formal solution of the RG equation}

Let us now consider the solution to~\req{eq:rge_oper}. To that purpose we start by
introducing the (matricial) renormalisation group evolution operator $U(\mu_2,\mu_1)$
that evolves renormalised operators between the scales\footnote{Restricting the evolution operator
to run towards the IR avoids unessential algebraic technicalities below. The running towards
the UV can be trivially obtained by taking $\left[U(\mu_2,\mu_1)\right]^{-1}$.}
$\mu_1$ and $\mu_2<\mu_1$,
\begin{gather}
\label{eq:evol}
\obar{O}_i(\mu_2) = U_{ij}(\mu_2,\mu_1) \obar{O}_j(\mu_1)\,.
\end{gather}
By substituting into~\req{eq:rge_oper} one has the equation for $U(\mu_2,\mu_1)$
\begin{gather}
\label{eq:rg_evol}
\mu_2\,\frac{\rm {d}}{{\rm d}\mu_2}\,U(\mu_2,\mu_1) = \gamma[\gbar(\mu_2)]U(\mu_2,\mu_1)\,,
\end{gather}
(n.b. the matrix product on the r.h.s.) with initial condition $U(\mu_1,\mu_1)=\mathbf{1}$.
Following a standard procedure, this differential equation for $U$ can be converted into a Volterra-type integral
equation and solved iteratively, viz.
\begin{gather}
U(\mu_2,\mu_1) = \texp\left\{
\int_{\gbar(\mu_1)}^{\gbar(\mu_2)}\kern-8pt\dif g\,\frac{1}{\beta(g)}\,\gamma(g)
\right\}\,,
\end{gather}
where as usual the notation $\texp$ refers to a definition in terms of the Taylor expansion
of the exponential function with ``powers'' of the integral involving argument-ordered integrands ---
explicitly, for a generic matrix function $M$, one has
\begin{gather}
\label{eq:texp}
\begin{split}
\texp\left\{\int_{x_-}^{x_+}\kern-8pt\dif x\,M(x)\right\} \equiv \mathbf{1}
&+ \int_{x_-}^{x_+}\kern-8pt\dif x\,M(x) \\
&+ \int_{x_-}^{x_+}\kern-8pt\dif x_1\,M(x_1)\int_{x_-}^{x_1}\kern-8pt\dif x_2\,M(x_2)\\
&+ \int_{x_-}^{x_+}\kern-8pt\dif x_1\,M(x_1)\int_{x_-}^{x_1}\kern-8pt\dif x_2\,M(x_2)\int_{x_-}^{x_2}\kern-8pt\dif x_3\,M(x_3)\\
&+ \ldots\\
= \mathbf{1}
&+ \int_{x_-}^{x_+}\kern-8pt\dif x\,M(x) \\
&+ \frac{1}{2!}\int_{x_-}^{x_+}\kern-8pt\dif x_1\int_{x_-}^{x_+}\kern-8pt\dif x_2\,\Big\{\theta(x_1-x_2)M(x_1)M(x_2)+\\
&~~~~~~~~~~~~~~~~~~~~~~~~~~~~~\theta(x_2-x_1)M(x_2)M(x_1)\Big\}\\
&+ \ldots
\end{split}
\end{gather}

\subsection{RGI in the absence of mixing}

Let us now consider an operator $O$ that renormalises multiplicatively. In that case,
both $\gamma$ and $U$ are scalar objects, and~\req{eq:evol} can be manipulated as
\begin{gather}
\begin{split}
\obar{O}(\mu_2) &=
\exp\left\{
\int_{\gbar(\mu_1)}^{\gbar(\mu_2)}\kern-8pt\dif g\,\frac{\gamma(g)}{\beta(g)}
\right\}\obar{O}(\mu_1)\\
&=\exp\left\{
\int_{\gbar(\mu_1)}^{\gbar(\mu_2)}\kern-8pt\dif g\,\frac{\gamma_0}{b_0g}
\right\}
\exp\left\{
\int_{\gbar(\mu_1)}^{\gbar(\mu_2)}\kern-8pt\dif g\,\left[\frac{\gamma(g)}{\beta(g)}-\frac{\gamma_0}{b_0g}\right]
\right\}\obar{O}(\mu_1)\\
&=\left[\frac{\gbar^2(\mu_2)}{\gbar^2(\mu_1)}\right]^{\frac{\gamma_0}{2b_0}}
\exp\left\{
\int_{\gbar(\mu_1)}^{\gbar(\mu_2)}\kern-8pt\dif g\,\left[\frac{\gamma(g)}{\beta(g)}-\frac{\gamma_0}{b_0g}\right]
\right\}\obar{O}(\mu_1)\,,
\end{split}
\end{gather}
yielding the identity\footnote{We introduce an --- otherwise arbitrary ---
constant overall normalisation factor to match standard conventions in the literature.}
\begin{gather}
\left[\frac{\gbar^2(\mu_2)}{4\pi}\right]^{-\frac{\gamma_0}{2b_0}}\obar{O}(\mu_2) =
\left[\frac{\gbar^2(\mu_1)}{4\pi}\right]^{-\frac{\gamma_0}{2b_0}}\exp\left\{
\int_{\gbar(\mu_1)}^{\gbar(\mu_2)}\kern-8pt\dif g\,\left[\frac{\gamma(g)}{\beta(g)}-\frac{\gamma_0}{b_0g}\right]
\right\}\obar{O}(\mu_1)\,.
\end{gather}
The advantage of having rewritten~\req{eq:evol} in this way is that now the integral in the
exponential is finite as either integration limit is taken to zero; in particular,
the r.h.s. is well-defined when $\mu_2\to\infty~\Leftrightarrow~\gbar(\mu_2)\to 0$, and
therefore so is the l.h.s. Thus, we define the RGI operator insertion as
\begin{gather}
\orgi{O} \equiv \lim_{\mu\to\infty}\left[\frac{\gbar^2(\mu)}{4\pi}\right]^{-\frac{\gamma_0}{2b_0}}\obar{O}(\mu)\,,
\end{gather}
upon which we have an explicit expression to retrieve the RGI operator from the renormalised
one at any value of the renormalisation scale $\mu$, provided the anomalous dimension and the
beta function are known for scales $\geq\mu$,
\begin{gather}
\label{eq:rgi_nomix}
\orgi{O} = \left[\frac{\gbar^2(\mu)}{4\pi}\right]^{-\frac{\gamma_0}{2b_0}}\exp\left\{
-\int_{0}^{\gbar(\mu)}\kern-8pt\dif g\,\left[\frac{\gamma(g)}{\beta(g)}-\frac{\gamma_0}{b_0g}\right]
\right\}\obar{O}(\mu)\,.
\end{gather}
Starting from the latter equation, it is easy to check explicitly that $\orgi{O}$
is invariant under a change of renormalisation scheme.

Note that the crucial step in the manipulation has been to add and subtract the term $\frac{\gamma_0}{b_0g}$
in the integral that defines the RG evolution operator, which allows to obtain a quantity
that is UV-finite by removing the logarithmic divergence induced at small $g$ by the perturbative behaviour
$\gamma(g)/\beta(g) \sim 1/g$. When $\gamma$ is a matrix of anomalous dimensions this step
becomes non-trivial, since in general $[\gamma(g),\gamma_0]\neq 0$; the derivation has thus
to be performed somewhat more carefully.

\subsection{RGI in the presence of mixing}

Let us start by studying the UV behaviour of the matricial RG evolution operator
in~\req{eq:evol}, using its actual definition in~\req{eq:texp}. To that purpose,
we first observe that by taking the leading-order approximation for $\gamma(g)/\beta(g)$
the T-exponential becomes a standard exponential, since $[\gamma_0 g_1^2,\gamma_0 g_2^2]=0~\forall g_1,g_2$.
One can then perform the integral trivially and write
\begin{gather}
\label{eq:U_LO}
U(\mu_2,\mu_1) \underset{\rm LO}{=} \left[\frac{\gbar^2(\mu_2)}{\gbar^2(\mu_1)}\right]^{\frac{\gamma_0}{2b_0}}
\equiv U_{\rm\scriptscriptstyle LO}(\mu_2,\mu_1)\,.
\end{gather}
When next-to-leading order corrections are included the T-exponential becomes non-trivial.
In order to make contact with the literature (see e.g.~\cite{Buras:2000if,Ciuchini:1997bw}),
we write\footnote{The property underlying this equation is that the evolution
operator can actually be factorised, in full generality, as
\begin{gather}
\label{eq:Utilde}
U(\mu_2,\mu_1) = \left[\tilde U(\mu_2)\right]^{-1} \tilde U(\mu_1)\,,~~~~~~
\tilde U(\mu) = \left[\frac{\gbar^2(\mu)}{4\pi}\right]^{-\frac{\gamma_0}{2b_0}}W(\mu)
\end{gather}
with a $W(\mu)$ that satisfies~\req{eq:rg_W} below.}
\begin{gather}
\label{eq:def_W}
U(\mu_2,\mu_1) \equiv \left[W(\mu_2)\right]^{-1}\, U_{\rm\scriptscriptstyle LO}(\mu_2,\mu_1) W(\mu_1)\,.
\end{gather}
Upon inserting~\req{eq:def_W} in~\req{eq:rg_evol} we obtain for $W$ the RG equation
\begin{gather}
\label{eq:rg_W}
\begin{split}
\mu\frac{\rm{d}}{{\rm d}\mu}W(\mu) &= -W(\mu)\gamma(\gbar(\mu))+\beta(\gbar(\mu))\frac{\gamma_0}{b_0\gbar(\mu)}W(\mu) \\
&= [\gamma(\gbar(\mu)),W(\mu)] - \beta(\gbar(\mu))\left(
\frac{\gamma(\gbar(\mu))}{\beta(\gbar(\mu))}-\frac{\gamma_0}{b_0\gbar(\mu)}
\right)W(\mu)\,.
\end{split}
\end{gather}
The matrix $W$ can be interpreted as the piece of the evolution operator
containing contributions beyond the leading perturbative order.
It is easy to check by expanding perturbatively (see below)
that $W$ is regular in the UV, and that all the logarithmic divergences
in the evolution operator are contained in $U_{\rm\scriptscriptstyle LO}$; in particular,
\begin{gather}
\label{eq:W_ic}
W(\mu)\underset{\mu\to \infty}{=}\mathbf{1}\,.
\end{gather}
Note also that in the absence of mixing~\req{eq:rg_W} can be solved
explicitly to get (using $W(\infty)=1$)
\begin{gather}
\label{eq:W_nomix}
W(\mu)\underset{\rm no~mixing}{=}\exp\left\{
-\int_{0}^{\gbar(\mu)}\kern-8pt\dif g\,\left[
\frac{\gamma(g)}{\beta(g)}-\frac{\gamma_0}{b_0g}
\right]
\right\}\,.
\end{gather}

Now it is easy, by analogy with the non-mixing case, to define RGI operators.
We rewrite~\req{eq:evol} as
\begin{gather}
\left[\frac{\gbar^2(\mu_2)}{4\pi}\right]^{-\frac{\gamma_0}{2b_0}}W(\mu_2)\obar{O}(\mu_2)
= \left[\frac{\gbar^2(\mu_1)}{4\pi}\right]^{-\frac{\gamma_0}{2b_0}}W(\mu_1)\obar{O}(\mu_1)\,,
\end{gather}
where $\obar{O}$ is a vector of renormalised operators on which the RG evolution matrix acts,
cf. \req{eq:evol}.
The l.h.s. (resp. r.h.s.) is obviously finite for $\mu_1\to\infty$ (resp. $\mu_2\to\infty$),
which implies that the vector of RGI operators can be obtained as
\begin{gather}
\label{eq:rgi_mix}
\orgi{O} = \left[\frac{\gbar^2(\mu)}{4\pi}\right]^{-\frac{\gamma_0}{2b_0}}W(\mu)\obar{O}(\mu)
\equiv \tilde U(\mu)\obar{O}(\mu)\,.
\end{gather}
When there is no mixing, the use of~\req{eq:W_nomix} immediately brings back~\req{eq:rgi_nomix}.

\subsection{Perturbative expansion of RG evolution functions}


By expanding~\req{eq:rg_W} perturbatively, with\footnote{It is easy to check that this is indeed
the correct form of the expansion for $W$. If terms with odd powers $\gbar^{2k+1},~k=0,1,\ldots$ are allowed,
the consistency between the left- and right-hand sides of~\req{eq:rg_W} requires them to vanish.
The same applies if a dependence of $J_n$ on $\mu$ is allowed --- consistency then requires
$\frac{\partial J_n}{\partial\mu}=0$.}
\begin{gather}
\label{eq:Wpert}
W(\mu) \approx \mathbf{1} + \gbar^2(\mu)J_1 + \gbar^4(\mu)J_2 + \gbar^6(\mu)J_3 + \gbar^8(\mu)J_4 + \ldots
\end{gather}
we find for the first four orders in the expansion the conditions
\begin{align}
\label{eq:J1}
2b_0J_1-\left[\gamma_0,J_1\right] &= \frac{b_1}{b_0}\gamma_0-\gamma_1\,,\\
\label{eq:J2}
4b_0J_2-\left[\gamma_0,J_2\right]
&= J_1\left(\frac{b_1}{b_0}\gamma_0-\gamma_1\right)
+\left(\frac{b_2}{b_0}-\frac{b_1^2}{b_0^2}\right)\gamma_0+\frac{b_1}{b_0}\gamma_1-\gamma_2\,,\\
\label{eq:J3}
6b_0J_3-\left[\gamma_0,J_3\right]
&= J_2\left(\frac{b_1}{b_0}\gamma_0-\gamma_1\right)+J_1\left\{\left(\frac{b_2}{b_0}-\frac{b_1^2}{b_0^2}\right)\gamma_0+\frac{b_1}{b_0}\gamma_1-\gamma_2\right\}+\nonumber\\
&\ \ +\left(\frac{b_3}{b_0}-\frac{2b_2b_1}{b_0^2}+\frac{b_1^3}{b_0^3}\right)\gamma_0+\left(\frac{b_2}{b_0}-\frac{b_1^2}{b_0^2}\right)\gamma_1+\frac{b_1}{b_0}\gamma_2-\gamma_3\,,\\
8b_0J_4-\left[\gamma_0,J_4\right]
&= J_3\left(\frac{b_1}{b_0}\gamma_0-\gamma_1\right)+J_2\left\{\left(\frac{b_2}{b_0}-\frac{b_1^2}{b_0^2}\right)\gamma_0+\frac{b_1}{b_0}\gamma_1-\gamma_2\right\}+\nonumber\\
&\ \ +J_1\left\{\left(\frac{b_3}{b_0}-\frac{2b_2b_1}{b_0^2}+\frac{b_1^3}{b_0^3}\right)\gamma_0+\left(\frac{b_2}{b_0}-\frac{b_1^2}{b_0^2}\right)\gamma_1+\frac{b_1}{b_0}\gamma_2-\gamma_3\right\}+\nonumber\\
&\ \ +\left(-\frac{b_1^4}{b_0^4} + 3\frac{b_1^2b_2}{b_0^3} -\frac{b_2^2}{b_0^2} - 2\frac{b_1b_3}{b_0^2} + \frac{b_4}{b_0}\right)\gamma_0+\left(\frac{b_3}{b_0}-\frac{2b_2b_1}{b_0^2}+\frac{b_1^3}{b_0^3}\right)\gamma_1+\nonumber\\ 
&\ \ +\left(\frac{b_2}{b_0}-\frac{b_1^2}{b_0^2}\right)\gamma_2+\frac{b_1}{b_0}\gamma_3-\gamma_4\,.
\label{eq:J4}
\end{align}
Modulo sign and normalisation conventions (involving powers of $4\pi$ related
to expanding in $\gbar^2$ rather than $\alpha/(4\pi)$), and the dependence on gauge fixing (which does
not apply to our context), \req{eq:J1} coincides with Eq.~(24) of~\cite{Ciuchini:1997bw}.
All four equations, as well as those for higher orders, can be easily solved to obtain 
$J_n$ for given values the coefficients in the perturbative expansion of $\gamma$.
The LO, NLO, and NNLO and NNNLO matching for the RGI operators is thus obtained from \req{eq:rgi_mix}
by using the expansion in powers of $\gbar^2$ in \req{eq:Wpert} up to zeroth, first, second, and third order,
respectively.

\section{Anomalous dimensions in SF schemes}
\label{sec:sf}

\subsection{Changes of renormalisation scheme}

Let us now consider a change to a different mass-independent renormalisation scheme,
indicated by primes. The relation between renormalised quantities in either scheme
amounts to finite renormalisations of the form
\begin{align}
\label{eq:schemechange_g}
\gren' &= \sqrt{\cX_{\rm g}(\gren)}\,\gren\,,\\
\label{eq:schemechange_m}
\mren{,\rm f}' &= \cX_{\rm m}(\gren)\,\mren{,\rm f}\,,\\
\label{eq:schemechange_O}
{\oren{O}}_{,j}' &= (\cX_{O})_{jk}\,{\oren{O}}_{,k}\,.
\end{align} 
The scheme-change factors $\cX$ can be expanded perturbatively as
\begin{gather}
\cX(g) \underset{g \sim 0}{\approx} 1 + \cX^{(1)} g^2 + \ldots\,.
\end{gather}
By substituting Eqs.~(\ref{eq:schemechange_g}-\ref{eq:schemechange_O}) into the
corresponding Callan-Symanzik equations, the relation between the RG evolution
functions in different schemes is found
\begin{align}
\beta'(g')  &= \left[\beta(g)\frac{\partial g'}{\partial g}\right]_{g=g(g')}\,,\\
\tau'(g')   &= \left[\tau(g)+\beta(g)\frac{\partial}{\partial g}\ln\cX_{\rm m}(g)\right]_{g=g(g')}\,,\\
\gamma'(g') &= \left[\gamma(g)+\beta(g)\frac{\partial}{\partial g}\ln\cX_O(g)\right]_{g=g(g')}\,.
\end{align}
One can then plug in perturbative expansions and obtain explicit formulae relating
coefficients in different schemes. In particular, it is found that $b_0,b_1$ are
scheme-independent, and the same applies to $d_0$ and $\gamma^{(0)}$. The relation
between next-to-leading order coefficients for quark masses and operator anomalous
dimensions are given by
\begin{align}
\label{eq:scheme_change_mass}
d_1' &= d_1 + 2b_0\cX_{\rm m}^{(1)}-d_0\cX_{\rm g}^{(1)}\,,\\
\label{eq:scheme_change_op}
{\gamma'}^{(1)} &=
\gamma^{(1)} +
[\cX_O^{(1)},\gamma^{(0)}] +
2b_0\cX_O^{(1)} +
b_0^\lambda\frac{\partial}{\partial\lambda}\cX_O^{(1)} -
\gamma^{(0)}\cX_{\rm g}^{(1)}\,.
\end{align}
Therefore, if the anomalous dimension is known at two loops in some scheme, in order
to obtain the same result in a different scheme it is sufficient to compute the
one-loop relation between them.

\subsection{Strategy for the computation of NLO anomalous dimensions in SF schemes}

\req{eq:scheme_change_op} will be the key ingredient for our computation
of anomalous dimensions to two loops in SF schemes, using as starting point
known two-loop results in $\MSbar$ or RI schemes. Indeed, our strategy will
be to compute the one-loop matching coefficient between the SF schemes that we will
introduce presently, and the continuum schemes where $\gamma^{(1)}$ is known.
$\gamma^{(1);\MSbar}$ can be found in~\cite{Buras:1991jm,Buras:1992tc,Buras:2000if}, while $\gamma^{(1);\rm RI}$ can be computed from both~\cite{Ciuchini:1997bw} and~\cite{Buras:2000if}; we gather them in~\reapp{app:gammacont}.

One practical problem arises due to the dependence of the scheme definition in
the continuum on the regulator employed (usually some form of dimensional regularisation).
This implies that one-loop computation in SF schemes needed to obtain the matching coefficient
should be carried out using the same regulator as in the continuum scheme.
However, the lattice is the only regularisation currently available for the SF.
As a consequence, it is necessary to employ a third, intermediate reference scheme,
which we will dub ``lat'', where the $\MSbar$ or RI prescription is applied to the lattice-regulated theory.
One can then proceed in two steps:
\begin{enumerate}
\item Compute the matching coefficient $[\cX_O^{(1)}]_{\rm SF;lat}$ between SF and lat schemes.
As we will see later, the latter is retrieved by computing SF renormalisation constants at
one loop.
\item Retrieve the one-loop matching coefficients between the lattice- and dimensionally-regulated
versions of the continuum scheme ``cont'' (i.e. $\MSbar$ or RI), $[\cX_O^{(1)}]_{\rm cont;lat}$,
and obtain the matching coefficient that enters~\req{eq:scheme_change_op} as
\begin{gather}
\label{eq:match_practical}
[\cX_O^{(1)}]_{\rm SF;cont} = [\cX_O^{(1)}]_{\rm SF;lat}-[\cX_O^{(1)}]_{\rm cont;lat}\,.
\end{gather}
\end{enumerate}
The one-loop matching coefficients $[\cX_O^{(1)}]_{\rm cont;lat}$ that we will need can
be extracted from the literature. For the RI-MOM scheme they can be found in~\cite{Constantinou:2010zs}, while for the $\MSbar$ scheme they can be extracted from~\cite{Gupta:1996yt,Kim:2014tda,Capitani:2000xi}~\footnote{We are grateful to S.~Sharpe for having converted for us, in the case of Fierz~+ operators, the $\MSbar$ scheme used in~\cite{Gupta:1996yt} to the one defined in~\cite{Buras:2000if}.}). We gather the RI-MOM results in Landau gauge in~\reapp{app:finite_RI}. $\chi^{(1)}_{\rm g}$ can be found in~\cite{Sint:1995ch}. 

\subsection{SF renormalisation conditions}
\label{sec:sfren}

We now consider the problem of specifying suitable renormalisation
conditions on four-quark operators, using the Schr\"odinger Functional
formalism. The latter~\cite{Luscher:1992an}, initially developed to
produce a precise determination of the running coupling~\cite{Luscher:1992zx,Luscher:1993gh,DellaMorte:2004bc,Tekin:2010mm,Brida:2016flw,DallaBrida:2016kgh},
has been extended along the years to various other phenomenological
contexts, like e.g. quark masses~\cite{Capitani:1998mq,DellaMorte:2005kg,Campos:2015fka} or heavy-light currents
relevant for B-physics, among others~\cite{Blossier:2010jk,Bernardoni:2014fva} . In the context of four-quark operators,
the first applications involved the multiplicatively renormalisable
operators $\cQ_1^\pm$ of~\req{eq:rel_ops} (which, as explained above,
enter Standard Model effective Hamiltonians for $\Delta F=1$ and $\Delta F=2$
processes)~\cite{Guagnelli:2005zc,Palombi:2005zd,Dimopoulos:2006ma,Dimopoulos:2007ht}, as well as generic $\Delta B=2$ operators in the
static limit~\cite{Palombi:2007dr,Dimopoulos:2007ht}. The latter studies are extended in this paper
to cover the full set of relativistic operators.
It is important to stress that, while these schemes will be ultimately employed
in the context of a non-perturbative lattice computation of renormalisation constants
and anomalous dimensions, the definition of the schemes is fully independent of any
choice of regulator.

We use the standard SF setup as described in~\cite{Luscher:1996sc}, where the reader
is referred for full details including unexplained notation.
We will work on boxes with spacial extent $L$ and time extent $T$; in practice,
$T=L$ will always be set.
Source fields are made up of boundary quarks and antiquarks,
\begin{align}
\cO_{\alpha\beta}[\Gamma]  &= a^6\sum_{\vy,\vz}\bar\zeta_\alpha(\vy)\Gamma\zeta_\beta(\vz)\,,\\
\cO'_{\alpha\beta}[\Gamma] &= a^6\sum_{\vy,\vz}\bar\zeta'_\alpha(\vy)\Gamma\zeta'_\beta(\vz)\,,
\end{align}
where $\alpha,\beta$ are flavour indices, unprimed (primed) fields live at the
$x_0=0$ ($x_0=T$) boundary, and $\Gamma$ is a spin matrix that must
anticommute with $\gamma_0$, so that the boundary fermion field does not vanish.
This is a consequence of the structure of the conditions imposed
on boundary fields,
\begin{gather}
\zeta(\vx)=\half(\mathbf{1}-\gamma_0)\zeta(\vx)\,,~~~~~~~~~~
\bar\zeta(\vx)=\bar\zeta(\vx)\half(\mathbf{1}+\gamma_0)\,,
\end{gather}
and similarly for primed fields. The resulting limitations on the possible
Dirac structures for these operators imply e.g. that it is not possible to
use scalar bilinear operators, unless non-vanishing angular momenta are introduced.
This can however be expected to lead to poor numerical behaviour; thus,
following our earlier studies~\cite{Guagnelli:2005zc,Palombi:2005zd,Palombi:2007dr,Dimopoulos:2007ht}, we will work with zero-momentum
bilinears and combine them suitable to produce the desired quantum numbers.

\begin{figure}[t!]
\begin{center}
\includegraphics[width=100mm]{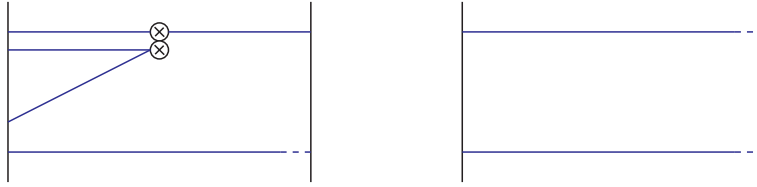}
\end{center}
\vspace{-5mm}
\caption{Feynman diagrams for the four-quark correlation functions $F_{k;s}$
and the boundary-to-boundary correlators $f_1,k_1$ at tree level.
Euclidean time goes from left to right.
The double blob indicates the four-quark operator insertion, and dashed lines indicate
the explicit time-like link variable involved in boundary-to-boundary quark propagators.}
\label{fig:diagtl}
\end{figure}

Renormalisation conditions will be imposed in the massless theory, in order
to obtain a mass-independent scheme by construction. They will furthermore
be imposed on parity-odd four-quark operators, since working in the parity-even
sector would entail dealing with the extra mixing due to explicit chiral
symmetry breaking with Wilson fermions, cf.~\req{eq:ren_pattern}.
In order to obtain non-vanishing SF correlation functions, we will then
need a product of source operators with overall negative parity;
taking into account the above observation about boundary fields, and
the need to saturate flavour indices, the minimal structure involves
three boundary bilinear operators and the introduction of an extra,
``spectator'' flavour (labeled as number 5, keeping with the notation in~\req{eq:gen_4f}).
We thus end up with correlation functions of the generic form
\begin{align}
F_{k;s}^\pm(x_0) &= \langle \cQ_k^\pm(x) \cS_s \rangle\,,\\
G_{k;s}^\pm(T-x_0) &= \eta_k\langle \cQ_k^\pm(x) \cS'_s \rangle\,,
\end{align}
where $\cS_s$ is one of the five source operators
\begin{align}
\cS_1 &= \cW[\gamma_5,\gamma_5,\gamma_5]\,,\\
\cS_2 &= \frac{1}{6}\sum_{k,l,m=1}^3\epsilon_{klm}\cW[\gamma_k,\gamma_l,\gamma_m]\,,\\
\cS_3 &= \frac{1}{3}\sum_{k=1}^3\cW[\gamma_5,\gamma_k,\gamma_k]\,,\\
\cS_4 &= \frac{1}{3}\sum_{k=1}^3\cW[\gamma_k,\gamma_5,\gamma_k]\,,\\
\cS_5 &= \frac{1}{3}\sum_{k=1}^3\cW[\gamma_k,\gamma_k,\gamma_5]
\end{align}
with
\begin{gather}
\cW[\Gamma_1,\Gamma_2,\Gamma_3] = L^{-3}\cO'_{21}[\Gamma_1]\cO'_{45}[\Gamma_2]\cO_{53}[\Gamma_3]\,,
\end{gather}
and similarly for $\cS'_s$. The constant $\eta_k$ is a sign that ensures
$F_{k;s}^\pm(x_0)=G_{k;s}^\pm(x_0)$ for all possible indices; it is easy
to check that $\eta_2=-1,~\eta_{s \neq 2}=+1$.\footnote{Since the correlation functions
$F$ and $G$ are related by invariance under time reversal, and are thus identical
only after integration over the whole ensemble of gauge
fields, computing both of them in a numerical simulation and averaging
the results will allow to reduce statistical noise at negligible computational cost.
From now on, we will proceed by using only $F$, and leave possible usage of $G$ at the
numerical level, or for cross-checks of results, implicit.}
We will also use the two-point functions of boundary sources
\begin{align}
f_1 &= -\,\frac{1}{2L^6}\langle\cO'_{21}[\gamma_5]\cO_{12}[\gamma_5]\rangle\,,\\
k_1 &= -\,\frac{1}{6L^6}\sum_{k=1}^3\langle\cO'_{21}[\gamma_k]\cO_{12}[\gamma_k]\rangle\,.
\end{align}
Finally, we define the ratios
\begin{gather}
\label{eq:corr_ratio}
\cA_{k;s,\alpha} = \frac{F_{k;s}(T/2)}{f_1^{\scriptscriptstyle{\frac{3}{2}}-\alpha}k_1^\alpha}\,,
\end{gather}
where $\alpha$ is an arbitrary real parameter.
The structure of $F_{k;s}$ and $f_1,k_1$ is illustrated in~\refig{fig:diagtl}.

We then proceed to impose renormalisation conditions at bare coupling $g_0$
and scale $\mu=1/L$ by generalising the condition introduced in~\cite{Guagnelli:2005zc,Palombi:2005zd}
for the renormalisable multiplicative operators $\cQ_1^\pm$: the latter reads
\begin{gather}
\label{eq:mult_ren}
\mathcal{Z}_{11;s,\alpha}\,\cA_{1;s,\alpha} =
\left.\cA_{1;s,\alpha}\right|_{g_0^2=0}\,,
\end{gather}
while for operators that mix in doublets, we impose\footnote{These renormalisation conditions were first introduced by S. Sint~\cite{ssprivate}.}
\begin{gather}
\label{eq:mat_ren}
\left(\ba{cc}
\mathcal{Z}_{22;s_1,s_2,\alpha} & \mathcal{Z}_{23;s_1,s_2,\alpha} \\
\mathcal{Z}_{32;s_1,s_2,\alpha} & \mathcal{Z}_{33;s_1,s_2,\alpha} \\
\ea\right)
\left(\ba{cc}
\cA_{2;s_1,\alpha} & \cA_{2;s_2,\alpha} \\
\cA_{3;s_1,\alpha} & \cA_{3;s_2,\alpha} \\
\ea\right)
=
\left(\ba{cc}
\cA_{2;s_1,\alpha} & \cA_{2;s_2,\alpha} \\
\cA_{3;s_1,\alpha} & \cA_{3;s_2,\alpha} \\
\ea\right)_{g_0^2=0}
\,,
\end{gather}
and similarly for $\cQ_{4,5}$.
The products of boundary-to-boundary correlators in the denominator of~\req{eq:corr_ratio}
cancels the renormalisation of the boundary operators in $F_{k;s}$, and therefore
$\mathcal{Z}_{k;s,\alpha}$ only contains anomalous dimensions of four-fermion operators.
Following~\cite{Capitani:1998mq,Guagnelli:2005zc,Dimopoulos:2007ht}, conditions are imposed on renormalisation functions
evaluated at $x_0=T/2$, and the phase that parameterises spatial boundary
conditions on fermion fields is fixed to $\theta=0.5$.
Together with the $L=T$ geometry of our finite box, this fixes the
renormalisation scheme completely, up to the choice of boundary source,
indicated by the index $s$, and the parameter $\alpha$. The latter can
in principle take any value, but we will restrict to the choices $\alpha=0,1,3/2$.

One still has to check that renormalisation conditions are well-defined
at tree-level. While this is straightforward for~\req{eq:mult_ren}, it is not so
for~\req{eq:mat_ren}: it is still possible that
the matrix of ratios $\cA$ has zero determinant at tree-level, rendering
the system of equations for the matrix of renormalisation constants
ill-conditioned. This is indeed the obvious case for $s_1 = s_2$,
but the determinant turns out to be zero also for other non-trivial
choices $s_1 \neq s_2$. In practice, out of the ten possible schemes
one is only left with six, viz.\footnote{Note that schemes obtained by
exchanging $s_1 \leftrightarrow s_2$ are trivially related to each other.}
\begin{gather}
(s_1,s_2) \in \{(1,2),(1,4),(1,5),(2,3),(3,4),(3,5)\}\,.
\end{gather}
It has to be stressed that this property is independent of the choice
of $\theta$ and $\alpha$. Thus, we are left with a total of 15 schemes
for $\cQ_1^\pm$, and 18 for each of the pairs $(\cQ_2^\pm,\cQ_3^\pm)$
and $(\cQ_4^\pm,\cQ_5^\pm)$.

\subsection{One-loop results in the SF}

Let us now carry out a perturbative computation of the SF renormalisation matrices
introduced above, using a lattice regulator.
For any of the correlation functions discussed in~\res{sec:sf}, the perturbative expansion reads
\begin{gather}
X=X^{(0)} + g_0^2\left[
X^{(1)}+\mcrit^{(1)}\frac{\partial X^{(0)}}{\partial m_0}
\right]
+\cO(g_0^4)\,,
\end{gather}
where $X$ is one of $F_{k;s}^\pm(x_0),~f_1,~k_1$, or some combination thereof;
where $m_0$ is the bare quark mass; and $\mcrit^{(1)}$ the one-loop coefficient in the
perturbative expansion of the critical mass. The derivative term in the square bracket is 
needed to set the correlation function $X$ to zero renormalised quark mass, when
every term in the r.h.s. of the equation is computed at vanishing bare mass.
We use the values for the critical mass provided in~\cite{Sint:1995rb},
\begin{gather}
\begin{split}
\mcrit^{(1)} &= -0.20255651209\,\CF~~~(\icsw=1)\,,\\
\mcrit^{(1)} &= -0.32571411742\,\CF~~~(\icsw=0)\,,
\end{split}
\end{gather}
with $\CF=(\NC^2-1)/(2\NC)$, and
the (tree-level) value of the Sheikholeslami-Wohlert (SW) coefficient $\icsw$ indicating whether
the computation is performed with or without an $\Oa$-improved action.

The entries of the renormalisation matrix admit a similar expansion,
\begin{gather}
\mathcal{Z}(g_0,L/a) = 1 + g_0^2 \mathcal{Z}^{(1)}(L/a) + \cO(g_0^4)\,,
\end{gather}
where we have indicated explicitly the dependence of the quantities
on the bare coupling and the lattice spacing-rescaled renormalisation scale $a\mu=a/L$.
The explicit expression of the one-loop order coefficient $\mathcal{Z}^{(1)}$ for
the multiplicatively renormalisable operators $\cQ_1^\pm$ is
\begin{gather}
\begin{split}
\mathcal{Z}^{(1)} & =-\left \{ \frac{F^{(1)}}{F^{(0)}} + \frac{F^{(1)}_b}{F^{(0)}} + \mcrit^{(1)} \frac{\partial}{\partial m_0} \log F^{(0)} \right \}\\
& + \left ( \frac{3}{2}-\alpha \right )\left [ \frac{f_1^{(1)}}{f_1^{(0)}} + \frac{f_{1;b}^{(1)}}{f_1^{(0)}} + \mcrit^{(1)}\frac{\partial}{\partial m_0}\log f_1^{(0)} \right ] \\
& + \alpha \left [ \frac{k_1^{(1)}}{k_1^{(0)}} + \frac{k_{1;b}^{(1)}}{k_1^{(0)}} + \mcrit^{(1)}\frac{\partial}{\partial m_0} \log k_1^{(0)} \right ] \, ,
\label{eq:zpert_mul}
\end{split}
\end{gather}
while for the entries of each $2 \times 2$ submatrix that renormalises operator pairs one has
\begin{gather}
\label{eq:zpert_mat}
\mathcal{Z}_{ij}^{(1)}=-\cA_{ik}^{(1)}\left [ \left ( \cA^{(0)} \right )^{-1} \right ]_{kj}\, ,
\end{gather}
with
\begin{gather}
\begin{split}
\cA_{ij}^{(0)}= &\frac{F_{ij}^{(0)}}{\left [f_1^{(0)}\right ]^{3/2-\alpha}\left [k_1^{(0)} \right ]^{\alpha}}\, , \\
\cA_{ij}^{(1)}  =&  \left \{ \left [F_{ij}^{(1)} + F_{ij;b}^{(1)} + \mcrit^{(1)}\frac{\partial}{\partial m_0}F_{ij}^{(0)} \right ] \right . \\
&  -\left (\frac{3}{2}-\alpha \right ) \left [ \frac{f_1^{(1)}}{f_1^{(0)}} +  \frac{f_{1;b}^{(1)}}{f_1^{(0)}} +  \mcrit^{(1)}\frac{\partial}{\partial m_0} \log f_1^{(0)} \right ] F_{ij}^{(0)}\\
& -\alpha  \left . \left  [ \frac{k_1^{(1)}}{k_1^{(0)}} + \frac{k_{1;b}^{(1)}}{k_1^{(0)}} + \mcrit^{(1)}\frac{\partial}{\partial m_0} \log k_1^{(0)} \right ] F_{ij}^{(0)} \right \} \left [f_1^{(0)}\right ]^{\alpha-3/2}\left [k_1^{(0)}\right ]^{-\alpha} \, .
\end{split}
\label{eq:Z1loop_mix}
\end{gather}
Contributions with the label ``b'' arise from the boundary terms that are needed
in addition to the SW term in order to achieve full $\Oa$ improvement of the action
in the SF~\cite{Luscher:1996sc}. They obviously vanish in the unimproved case. We will set them
to zero in the improved case as well, since they vanish in the continuum limit and
thus will not contribute to our results for NLO anomalous dimensions.\footnote{These
terms do enter perturbative cutoff effects. Note however that we will not include
in our computation the required subtractions of dimension $7$ operators to render
correlation functions $\Oa$ improved, and therefore the scaling to the continuum limit
will be dominated by terms linear in $a$ up to logarithms --- cf. \req{eq:asymp_z} below.
The missing boundary contributions
are actually expected to be subdominant with respect to the missing counterterms to four-fermion operators.}

\begin{figure}[t!]
\begin{center}
\includegraphics[width=100mm]{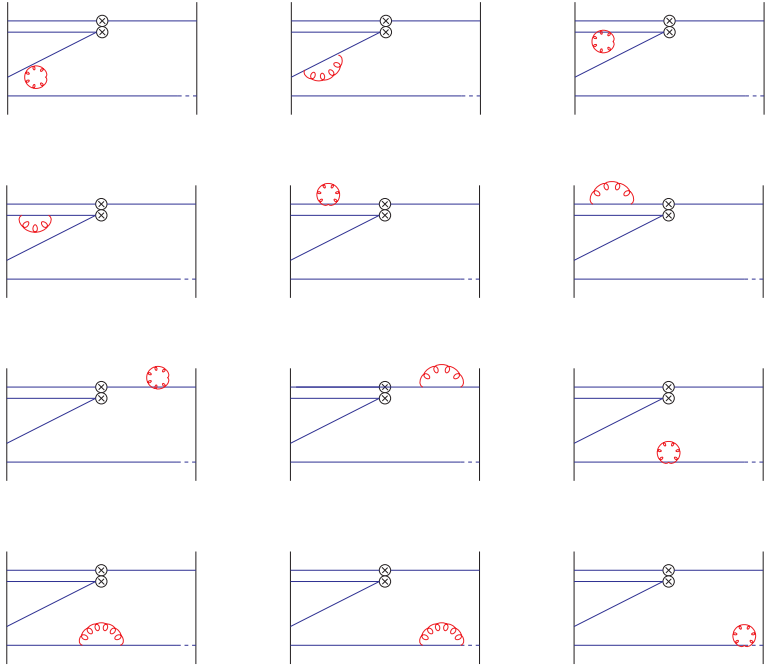}
\end{center}
\vspace{-5mm}
\caption{Feynman diagrams of the self-energy type entering the one-loop
computation of $F_{k;s}^\pm$.}
\label{fig:diag1}
\end{figure}

\begin{figure}[t!]
\begin{center}
\includegraphics[width=100mm]{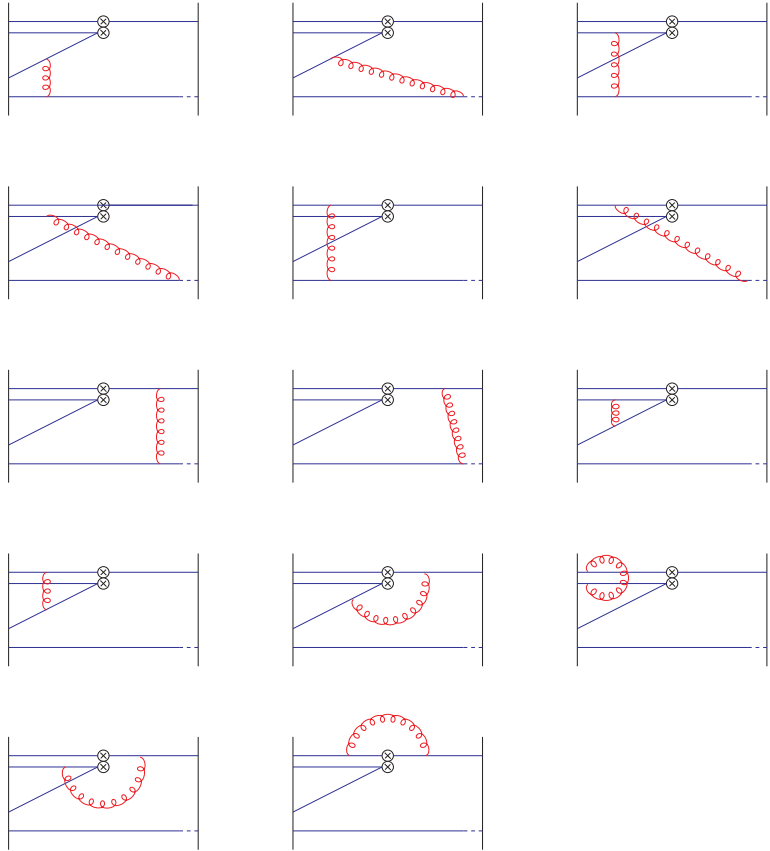}
\end{center}
\vspace{-5mm}
\caption{Feynman diagrams with gluon exchanges between quark lines entering the one-loop
computation of $F_{k;s}^\pm$.}
\label{fig:diag2}
\end{figure}

The computation of the r.h.s. of the four-quark operator correlators $F_{k;s}^\pm$
requires the evaluation of the Feynman diagrams in Figure~\ref{fig:diagtl} at tree level,
and of those in Figures~\ref{fig:diag1} and~\ref{fig:diag2} at one loop.
The one-loop expansion of the boundary-to-boundary correlators $f_1$ and $k_1$
is meanwhile known from~\cite{Luscher:1996vw}. Each diagram can be written as a loop sum
of a Dirac trace in time-momentum representation, where the Fourier transform
is taken over space only. The sums have been performed numerically in double
precision arithmetics using a Fortran 90 code, for all even lattice sizes ranging
from $L/a=4$ to $L/a=48$. The results have been cross-checked against those of an independent
C++ code, also employing double precision arithmetics.

The expected asymptotic expansion for the one-loop coefficient of renormalisation
constants is (operator and scheme indices not explicit)
\begin{gather}
\label{eq:asymp_z}
\mathcal{Z}^{(1)} (L/a) = \sum_{n=0}^\infty \left(\frac{a}{L}\right)^n\left\{
r_n + s_n\,\ln(L/a)
\right\}\,.
\end{gather}
In particular, the coefficient $s_0$ of the log that survives the continuum
limit will be the corresponding entry of the anomalous dimension matrix,
while the finite part $r_0$ will contribute to the one-loop matching coefficients
we are interested in. In particular, one has
\begin{gather}
[\cX_O^{(1)}]_{\rm SF;lat} = r_0\,,
\end{gather}
which is the required input for the matching condition in~\req{eq:match_practical}.
We thus proceed as follows:
\begin{enumerate}
\item Compute tree-level and one-loop diagrams for all correlation functions.
\item Construct one-loop renormalisation constants using~\req{eq:zpert_mul}
and~\req{eq:zpert_mat}.
\item Fit the results to the ansatz in~\req{eq:asymp_z} as a function of $(a/L)$, using the known value
of the entries of the leading-order anomalous dimension matrix $\gamma^{(0)}$
as fixed parameters, and extract $r_0$.
\end{enumerate}
The description of the procedure employed to extract the finite parts as well as our results are provided in~\reapp{app:finite}.

\subsection{NLO SF anomalous dimensions}

Having collected $[\cX_O^{(1)}]_{\rm SF;lat}$, $[\cX_O^{(1)}]_{\rm cont;lat}$, $\gamma^{(1);\rm cont}$ and $\cX^{(1)}_{\rm g}$ we have finally been able to compute the matrix $\gamma^{(1);\rm SF}$ for both the ``+'' and the ``-'' operator basis and for all the 18 schemes presented in Section~\ref{sec:sfren}.
The results are collected in \reapp{app:nload}.

We have performed two strong consistency checks of our calculation:
\begin{itemize}
\item In our one-loop perturbative computation, we have obtained $[\cX_O^{(1)}]_{\rm SF;lat}$ for both $\icsw=0$ and $\icsw=1$ values. The results for $[\cX_O^{(1)}]_{\rm cont;lat}$ are known for generic values of $\icsw$. We have thus been able to compute $[\cX_O^{(1)}]_{\rm SF;cont}$ for both $\icsw=0$ and $\icsw=1$ in such a way to check its independence from $\icsw$. 
\item For the ``+'' operators, we have checked the independence of $\gamma^{(1);\rm SF}$ from the reference scheme used (either the RI-MOM or the $\MSbar$). This is a strong check of the calculations from the literature of the NLO anomalous dimensions $\gamma^{(1);\rm cont}$ and one-loop matching coefficients $[\cX_O^{(1)}]_{\rm cont;lat}$ in both the RI-MOM and $\MSbar$ scheme.
\end{itemize}

The resulting values of $\gamma^{(1)}$ exhibit a strong scheme dependence.
In order to define a reference scheme for each operator, we have devised a criterion
that singles out those schemes with the smallest NLO corrections: given the matrix
\begin{gather}
16 \pi^2\, \gamma^{(1);\rm SF}\, (\gamma^{(0)})^{-1}\,,
\end{gather}
we compute the the trace and the determinant of each non-trivial submatrix, and
look for the smallest absolute value of both quantities. Remarkably, in all cases
(2-3 and 4-5 operator doublets, both in the Fierz $+$ and $-$ sectors) this
is satisfied by the scheme given by $s=6,~\alpha=3/2$.

\section{Renormalisation group running in perturbation theory}
\label{sec:rgpt}

In this section we will discuss the perturbative computation of the
RG running factor $\tilde U(\mu)$ in \req{eq:rgi_mix}. The main purpose of this exercise
is to understand the systematic of perturbative truncation, both in view
of our own non-perturbative computation of the RG running factor~\cite{Papinutto:2014xna}
(which involves a matching to NLO perturbation theory around the electroweak
scale), and in order to assess the extensive use of NLO RG running down to hadronic
scales in the phenomenological literature.
In view of our upcoming publication of a non-perturbative determination of the anomalous
dimensions for QCD with $\NF=2$, the analysis below will be performed for that case; the qualitative
conclusions are independent of the precise value of $\NF$. The scale will be fixed using
the value $\Lambda_{\rm\scriptscriptstyle QCD}^{\MSbar;\NF=2}=310(20)~{\rm MeV}$ quoted in~\cite{Fritzsch:2012wq}.

\begin{figure}[h!]
\vspace{-6mm}
\begin{center}
\includegraphics[width=130mm]{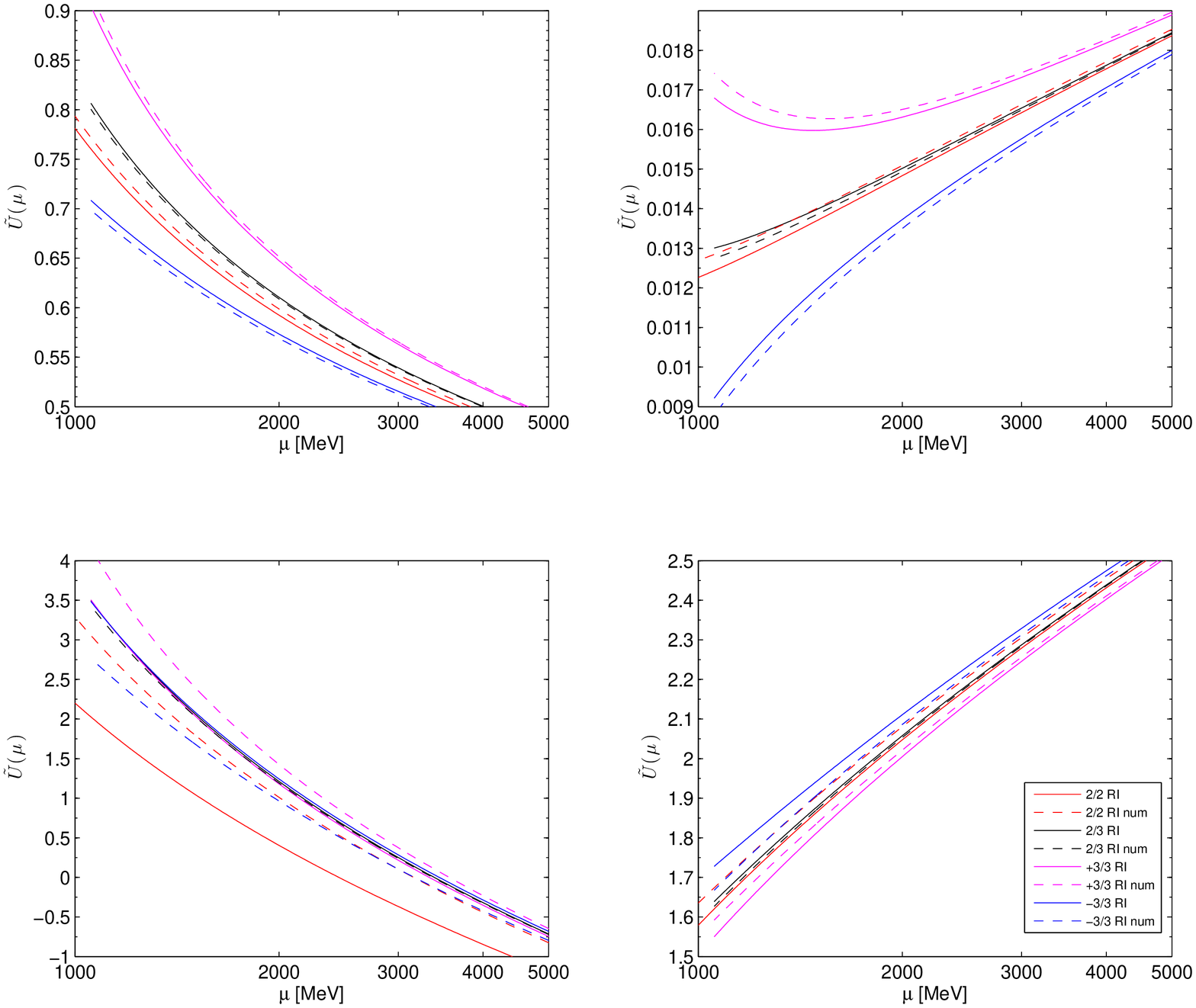}
\end{center}
\vspace{-5mm}
\begin{center}
\includegraphics[width=130mm]{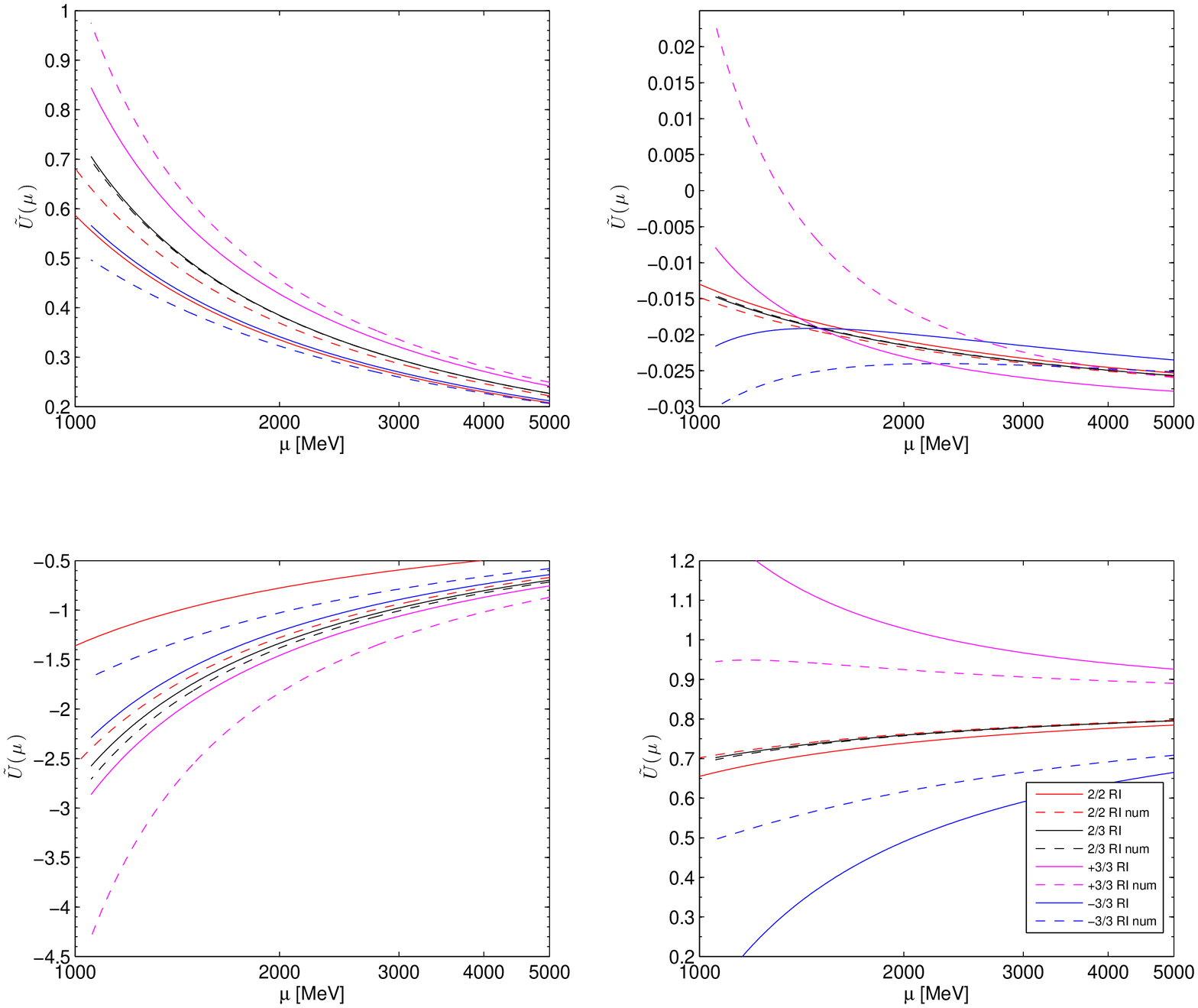}
\end{center}
\vspace{-3mm}
\caption{RG running matrix for the Op $4,5$ in the RI scheme. Top half (a): Fierz $+$. Bottom half (b): Fierz $-$. The four cases $n_\gamma/n_\beta = \{2/2, 2/3, +3/3, -3/3\}$ are plotted respectively in red, black, magenta and blue. Dashed lines correspond to the 
numerical integration of $W(\mu)$. Solid lines correspond to the perturbative expansion up to ${\mbox O}(g^2)$ (i.e. $J_1$) for the 2/2 case and up to ${\mbox O}(g^4)$ (i.e. $J_2$) for the 2/3, $+3/3$ and $-3/3$ cases.}  
\label{Fig:runJ1J2}
\end{figure}

At leading order in perturbation theory the running factor is given
by $U_{\rm\scriptscriptstyle LO}$ in~\req{eq:U_LO}. 
Beyond LO, the running factor is given by~\req{eq:Utilde}, where $W(\mu)$ satisfies~\req{eq:rg_W}.
In the computation of $W$, the $\beta$ and the $\gamma$ functions are known only up to 3 loops and 2 loops, respectively. 
In order to asses the systematic, we will compute the running factor for several 
approximations that will be labeled through a pair of numbers ``$n_\gamma/n_\beta$'' 
where $n_\gamma$ is the order used for the $\gamma$ function while $n_\beta$ is the order 
used for the $\beta$ function. We will consider the following cases:   
\begin{enumerate}
\item ``1/1'', i.e. the LO approximation in which $W\equiv 1$;
\item ``2/2'', in which both $\gamma$ and $\beta$ are taken at NLO; 
\item ``2/3'', in which $\beta$ is taken at NNLO and $\gamma$ at NLO;
\item ``+3/3'', in which $\beta$ is taken at NNLO and for the NNLO coefficient $\gamma_2$ we use a guesstimate given by $\gamma_2\gamma_1^{-1}=\gamma_1\gamma_0^{-1}$;
\item ``-3/3'', in which $\beta$ is taken at the NNLO and for the NNLO coefficient $\gamma_2$ we use a guesstimate given by $\gamma_2\gamma_1^{-1}=-\gamma_1\gamma_0^{-1}$;
\end{enumerate} 
Beyond LO we have first computed the perturbative expansion of the running factor,~\req{eq:Utilde} and \req{eq:Wpert}, by 
including all the $J_n$'s corresponding to the highest order used in the combinations 
of $\beta$/$\gamma$ functions chosen above. 
The $J_n$ have been computed from~\req{eq:J1} and~\req{eq:J2} setting 
the unknown coefficients to zero. Namely: $J_1$ in the 2/2 case, $J_1$ 
and $J_2$ (with $\gamma_2=0$) in the 2/3 case, $J_1$ and $J_2$ with $\gamma_2$ set to the 
guesstimates above in the +3/3 and -3/3 cases.
We have compared these results with the numerical solution of~\req{eq:rg_W} in which the perturbative expansions for $\gamma$ and $\beta$ at the chosen orders are plugged in.
We have chosen two cases in which perturbation theory seems particularly ill-behaved, namely the matrix for operators 4 and 5 with both Fierz + and - in the RI-MOM scheme, and we show the 
comparison in Fig.~\ref{Fig:runJ1J2}. As one can see, the two methods are not in very good 
agreement in the region of few GeV scales. This is obvious, because by expanding $W$ in powers of 
$g^2$ and including only the first/second coefficients $J_1$, $J_2$,
substantial information is lost.

\begin{figure}[h!]
\vspace{-6mm}
\begin{center}
\includegraphics[width=130mm]{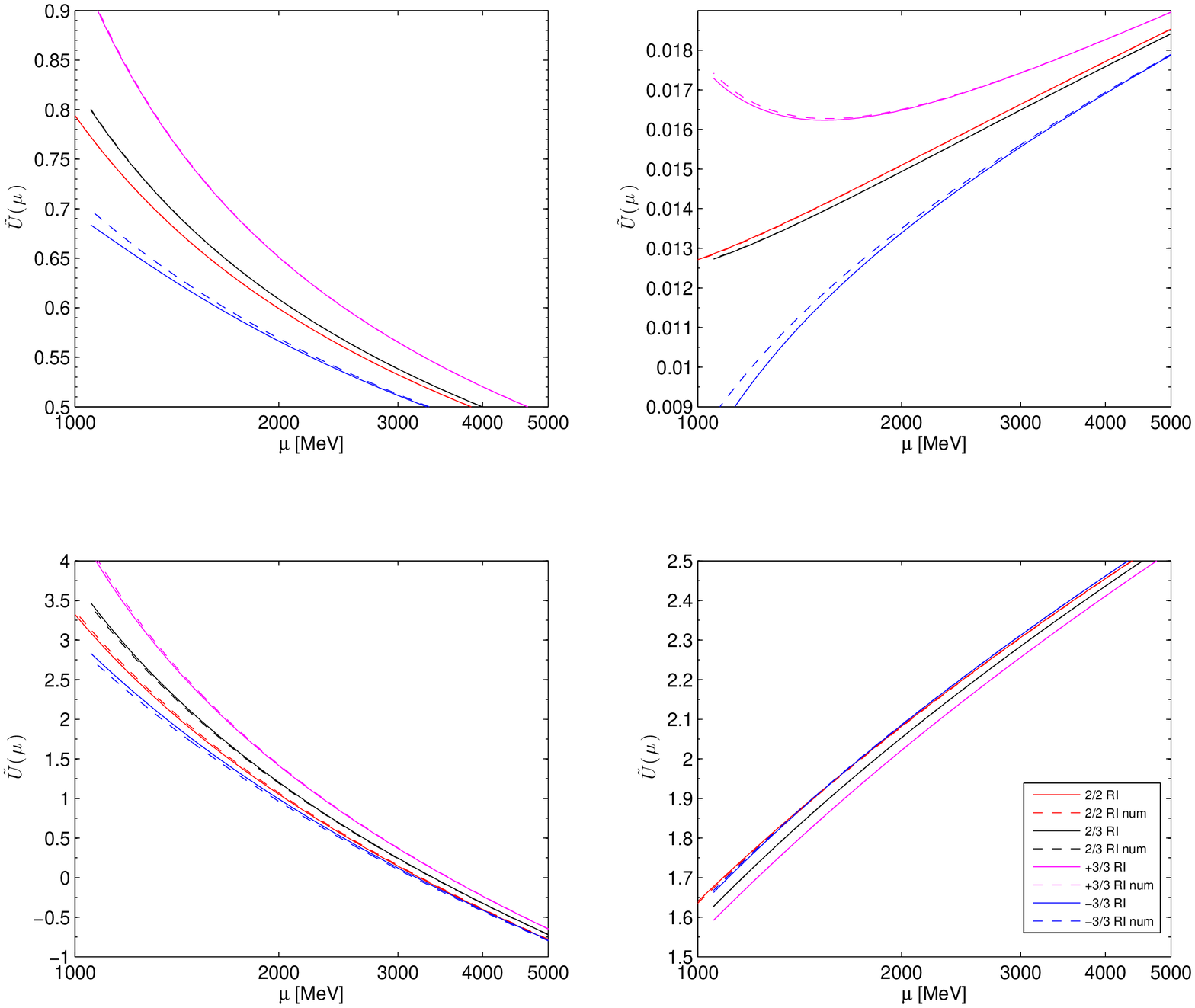}
\end{center}
\vspace{-5mm}
\begin{center}
\includegraphics[width=130mm]{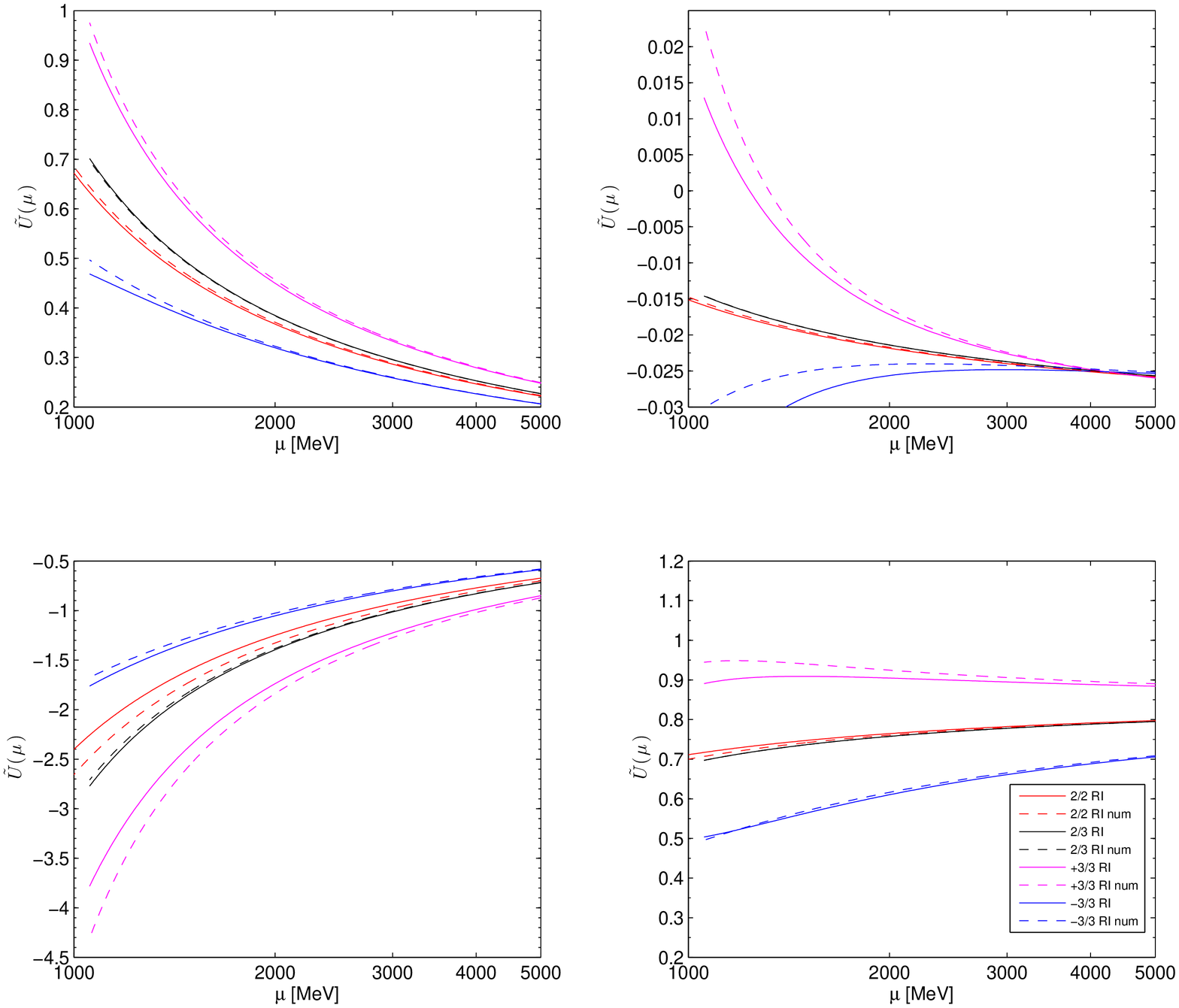}
\end{center}
\vspace{-3mm}
\caption{RG running matrix for the Op $4,5$ in the RI scheme. Top half (a): Fierz $+$. Bottom half (b): Fierz $-$. The four cases $n_\gamma/n_\beta = \{2/2, 2/3, +3/3, -3/3\}$ are plotted respectively in red, black, magenta and blue. Dashed lines correspond to the 
numerical integration of $W(\mu)$. Solid lines correspond to the perturbative expansion up to ${\mbox O}(g^4)$ (i.e. $J_2$) for the 2/2 case and up to ${\mbox O}(g^6)$ (i.e. $J_3$) for the 2/3, $+3/3$ and $-3/3$ cases.}
\label{Fig:runJ1J2J3}
\end{figure}

We have then included in the perturbative expansion the next order, computed from \req{eq:J2} and~\req{eq:J3}, setting again the unknown coefficients to zero. Namely: $J_2$ (with $b_2=\gamma_2=0$) in the 2/2 case, $J_3$ (with $b_3=\gamma_3=\gamma_2=0$) in the 2/3 case, $J_3$ (with $b_3=\gamma_3=0$ and $\gamma_2$ set to the guesstimates above) in the +3/3 and -3/3 cases.
The comparison, again with the corresponding numerical solution of~\req{eq:rg_W} (which remains unchanged), is shown in Fig.~\ref{Fig:runJ1J2J3} and shows a reasonable agreement for the Fierz + matrix while still noticeable desagreement for some of the Fierz - matrix elements.  

\begin{figure}[h!]
\vspace{-8mm}
\begin{center}
\includegraphics[width=130mm]{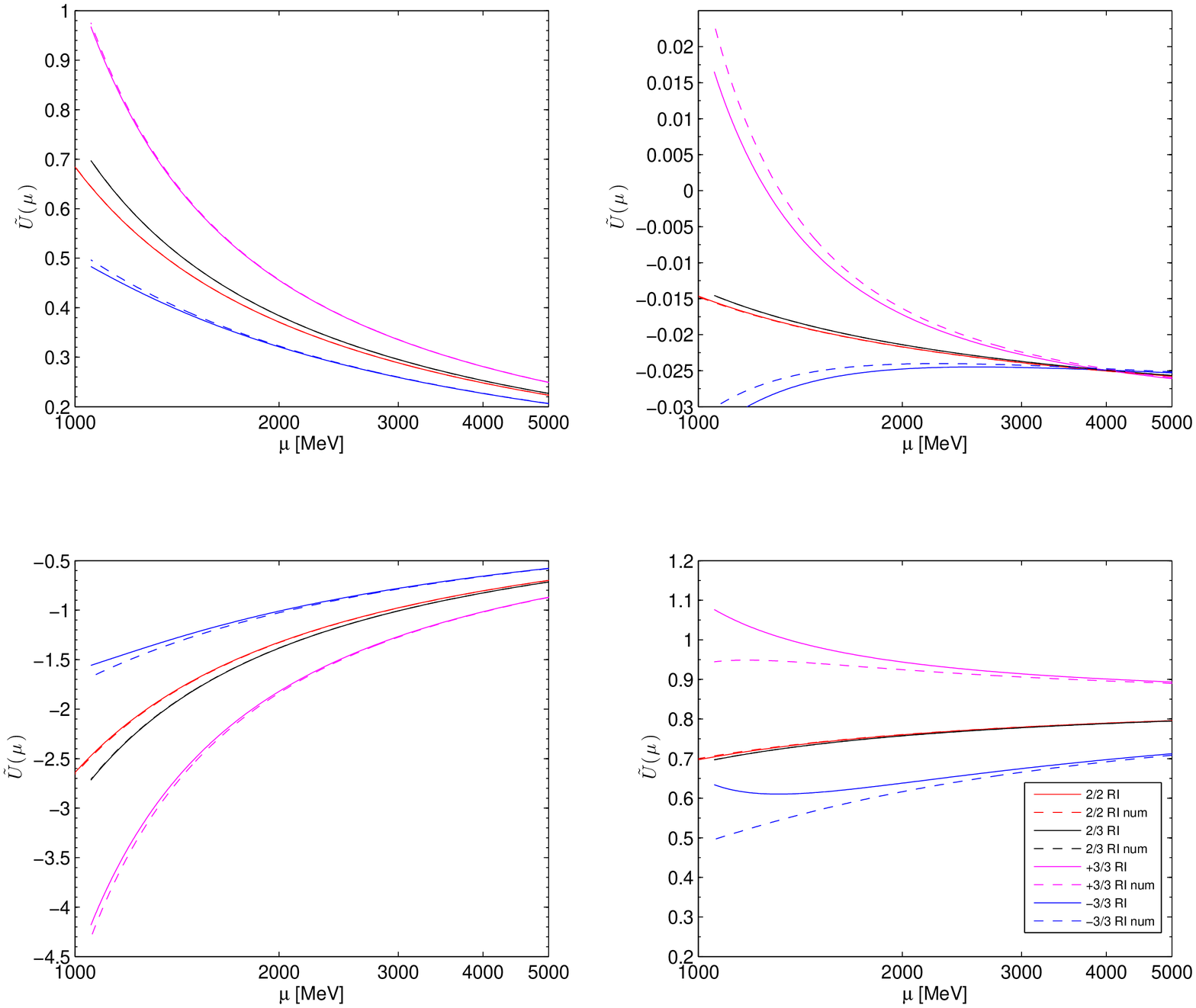}
\end{center}
\vspace{-5mm}
\begin{center}
\includegraphics[width=130mm]{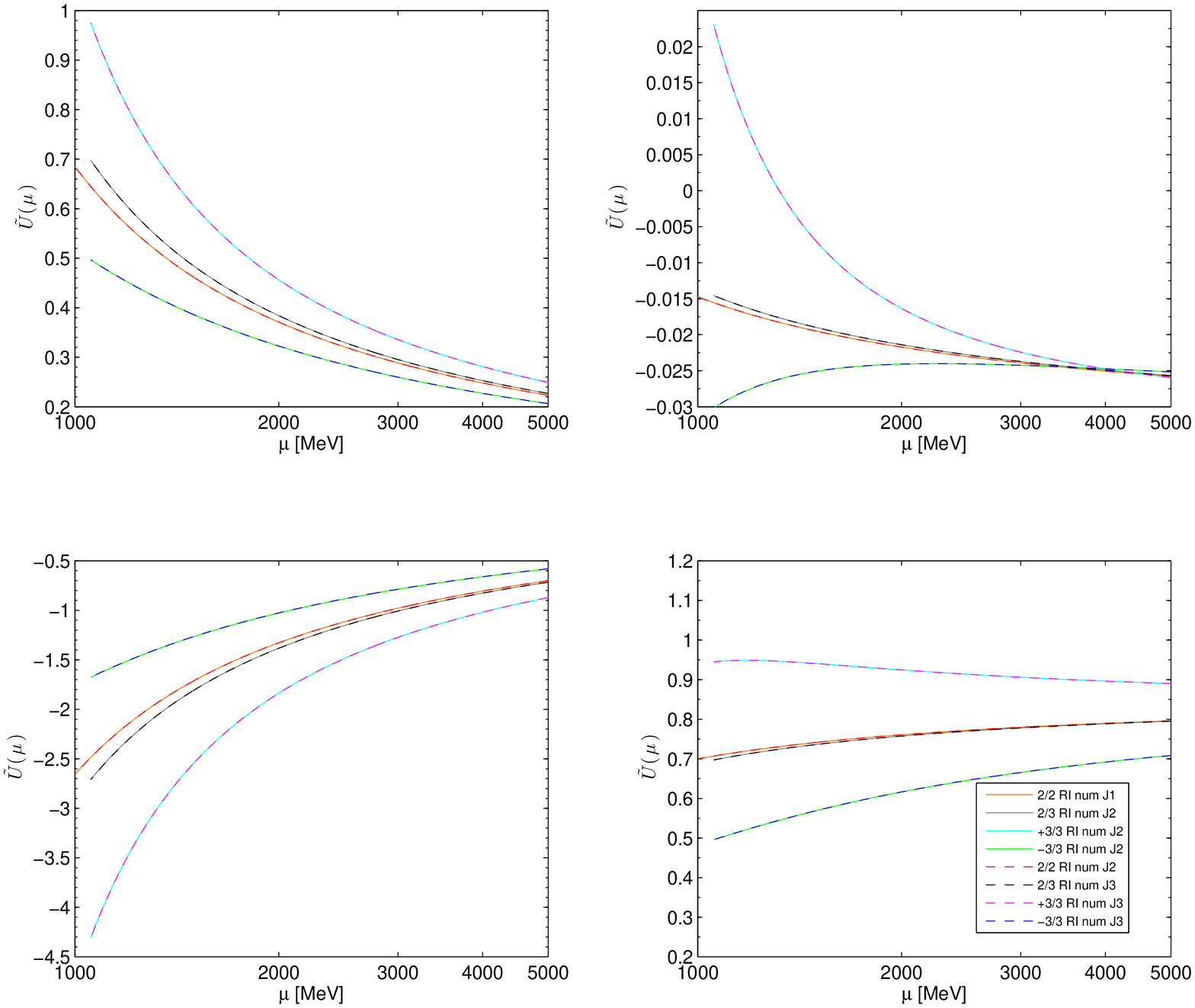}
\end{center}
\vspace{-4mm}
\caption{RG running matrix for the Op $4,5$ Fierz $-$ in the RI scheme. Top half (a): the four cases $n_\gamma/n_\beta = \{2/2, 2/3, +3/3, -3/3\}$ are plotted respectively in red, black, magenta and blue. Dashed lines correspond to the numerical integration of $W(\mu)$. Solid lines correspond to the perturbative expansion up to ${\mbox O}(g^6)$ (i.e. $J_3$) for the 2/2 case and up to ${\mbox O}(g^8)$ (i.e. $J_4$) for the 2/3, $+3/3$ and $-3/3$ cases. Bottom half (b): comparison of the results for the numerical integration of $W(\mu)$ when matched at $\gbar^2_{\MSbar}(M_P)$ with the perturbative expansion at the order used in Fig.~\ref{Fig:runJ1J2} (solid lines) and Fig.~\ref{Fig:runJ1J2J3} (dashed lines).}
\label{Fig:runJ1J2J3J4}
\end{figure}

In the the Fierz - case we have thus proceeded by introducing the next order, namely: $J_3$ (with 
$b_2=\gamma_2=b_3=\gamma_3=0$) in the 2/2 case, $J_4$ (with 
$b_4=b_3=\gamma_4=\gamma_3=\gamma_2=0$) in the 2/3 case, $J_4$ (with 
$b_4=b_3=\gamma_4=\gamma_3=0$ and $\gamma_2$ set to the guesstimates 
above) in the +3/3 and -3/3 cases. The comparison, again with the corresponding numerical 
solution of~\req{eq:rg_W}, is shown in Fig.~\ref{Fig:runJ1J2J3J4}a. The agreement between the numerical solution and the perturbative expansion further improves in all cases except for the 55 matrix element in the $\pm3/3$ cases where the perturbative expansion further moves away from the numerical solution. From both examples of Fierz $\pm$ 4-5 matrix, we understand that by including more and more orders in the perturbative expansion of $W(\mu)$~\req{eq:Wpert}, we approximate better and better the numerical solution of~\req{eq:rg_W}~\footnote{Except for the 55 matrix element where, in presence of a non-zero $\gamma_2$ the expansion looks like an alternating series.}, which can thus be considered the best approximation of the running factor given a fixed order computation of the $\beta$ and $\gamma$ functions. 
 
There is still a subtle technical issue concerning the numerical integration 
of~\req{eq:rg_W} which needs to be discussed, because it becomes relevant in practice.
Since $\gamma$ and $\beta$ have simple expressions in terms of $\gbar(\mu)$
rather than in terms of $\mu$, \req{eq:rg_W} is most easily
solved by rewriting it in terms of the derivative with respect
to the coupling, viz.
\begin{gather}
\label{eq:rg_W_g}
\widetilde W'(g) = -\widetilde W(g)\,\frac{\gamma(g)}{\beta(g)}
+\frac{\gamma_0}{b_0g}\,\widetilde W(g)\,,
\end{gather}
where $\widetilde W(\gbar(\mu)) \equiv W(\mu)$.
While both terms on the right hand side diverge as $g \to 0$, the divergence
cancels in the sum due to \req{eq:W_ic}. However, it is not straightforward
to implement this latter initial condition at the level of the numerical
solution to \req{eq:rg_W_g}:
a stable numerical solution requires fixing the initial condition \req{eq:W_ic}
at an extremely small value of the coupling, and consequently the use of a sophisticated and computationally expensive integrator.
A simpler solution is to substitute \req{eq:W_ic}
by an initial condition of the form
\begin{gather}
\label{eq:in_cond}
\widetilde W(g_{\rm i}) = \mathbf{1} + g_{\rm i}^2 J_1 + g_{\rm i}^4 J_2 + g_{\rm i}^6 J_3 + g_{\rm i}^8 J_4 + \ldots \,,
\end{gather}
at some very perturbative coupling $g_{\rm i}$ (but still a significantly larger value than required by \req{eq:W_ic}),
where we include exactly the same coefficients $J_n$, $n=1,\ldots$ that we use in the 
perturbative expansion of the running factor, and which are computed by using the same amount of 
perturbative information employed in the ratio $\gamma/\beta$ used for 
the numerical integration.\footnote{For $\NF=3$ \req{eq:in_cond} is not practical, and \req{eq:W_ic} becomes
mandatory, cf. \reapp{app:nf3}.}
Note that indeed the numerical value of $g_{\rm i}$ needs not be
extremely small for this to make physical sense, e.g.
for $\NF=2$ (which will be of particular interest to us) and at the Planck scale one has
$\gbar^2_{\MSbar}(M_P)\approx 0.221 \leftrightarrow \alpha_{\rm s}^{\MSbar}(M_P) \approx 0.0176$ and $\gbar^2_{\rm SF}(M_P)$ differs with respect to $\gbar^2_{\MSbar}(M_P)$ only on the third decimal digit.

In Fig.~\ref{Fig:runJ1J2J3J4}b we compare the results for the numerical integration of $W(\mu)$ 
when matched at $g_{\rm i}$ with the perturbative expansion at the order used in Fig.~\ref{Fig:runJ1J2} and Fig.~\ref{Fig:runJ1J2J3} respectively and the results turn out to be indistiguishable. We have also changed the value of the coupling chosen for the matching in a broad range of $\gbar^2$ without observing any noticeable difference in the solution. 
These checks prove the stability of the numerical procedure and give us confidence in the 
corresponding results, which will be used below to assess the systematic uncertainties. 
In the following we won't consider anymore the perturbative expansion of the running factor 
except for the 2/2 case where only $J_1$ is included (we will call this $2/2$ at 
$\mbox{O}(g^2)$), which is the case usually considered in literature, both for 
phenomenological application and in lattice computations.   

According to the previous discussion, we have chosen to quote as our best estimate of the running factors the 2/3 results (which encode the maximum of information at our disposal for the 
$\beta$ and $\gamma$ functions) obtained through numerical integration. They are presented in Tab.~\ref{tab:runnum} at the scale $\mu=3\,\GeV$. In alternative we quote also the results for the 2/2 case perturbatively expanded at $\mbox{O}(g^2)$ (i.e. including $J_1$ only), which are the results usually considered in literature. We present them in Tab.~\ref{tab:run} again at the scale 
$\mu=3\,\GeV$. 

The systematic uncertainties in Tab.~\ref{tab:runnum} (respectively Tab.~\ref{tab:run}) are estimated by considering the maximal deviation of the 2/3 case (respectively the 2/2, $\mbox{O}(g^2)$ case) from the other 3 (respectively 4) numerical cases. 

\begin{figure}[h!]
\vspace{-4mm}
\begin{center}
\includegraphics[width=130mm]{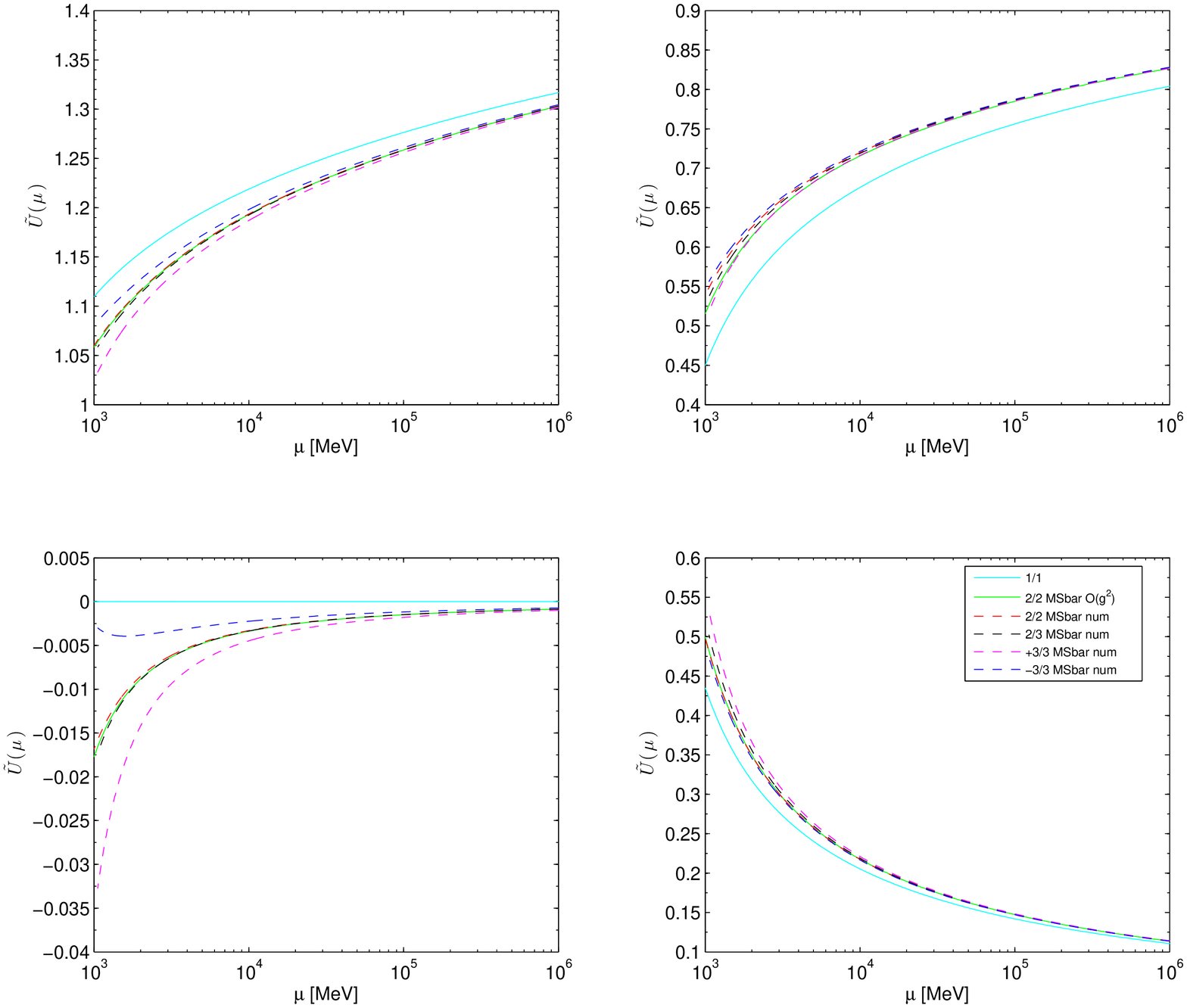}
\end{center}
\vspace{-5mm}
\begin{center}
\includegraphics[width=130mm]{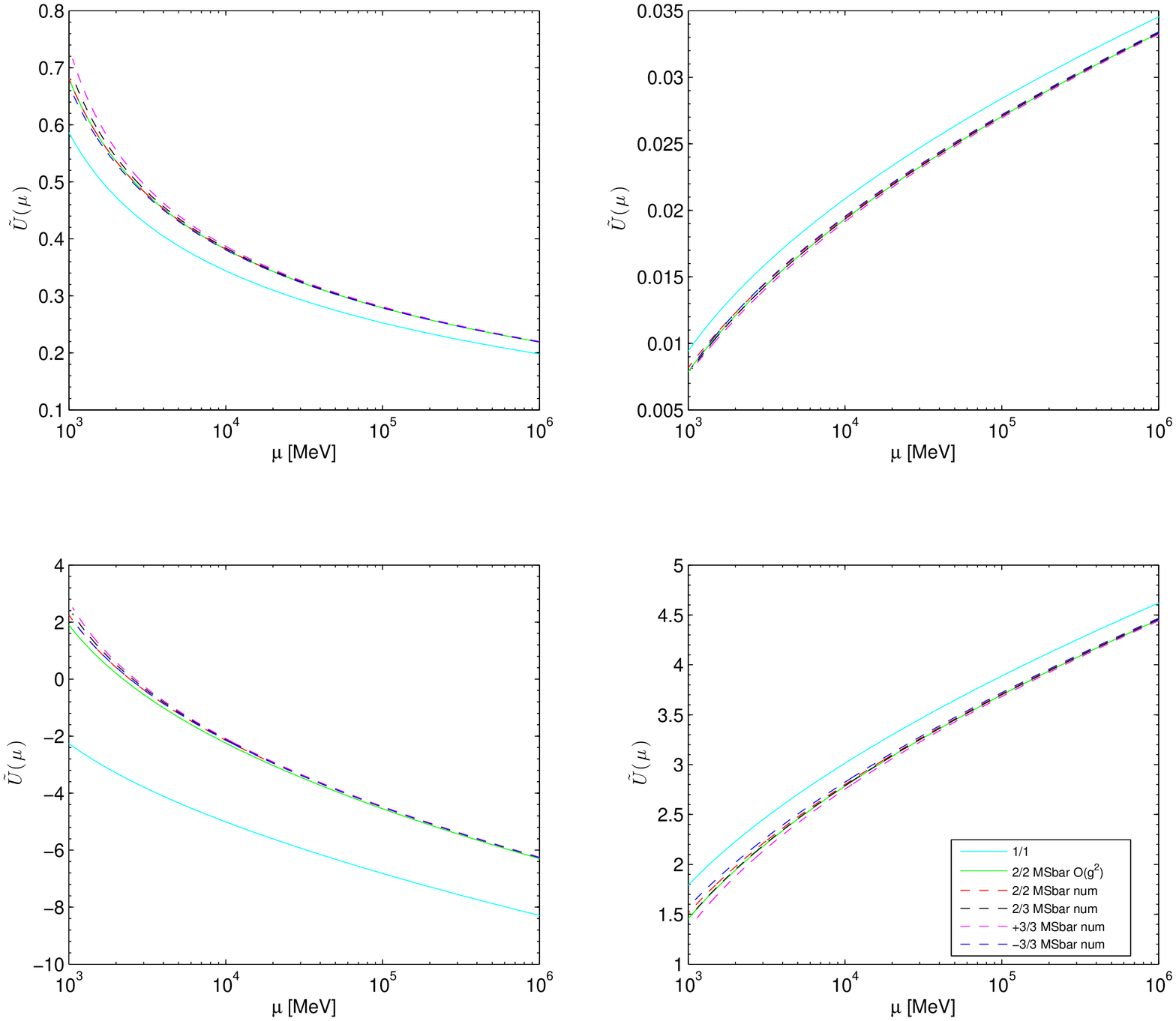}
\end{center}
\vspace{-4mm}
\caption{RG running matrices for the Fierz $+$ Op. $2,3$ (top half) and Op. $4,5$ (bottom half) in the $\MSbar$ scheme. Solid lines correspond to the LO plotted (cyan) and the perturbative expansion for the NLO 2/2 case up to ${\mbox O}(g^2)$ - i.e. including $J_1$ (green). Dashed lines correspond to the numerical solution for $W(\mu)$ in the cases $n_\gamma/n_\beta = \{2/2, 2/3, +3/3, -3/3\}$ respectively in red, black, magenta and blue.}
\label{Fig:runMSbar+}
\end{figure}

\begin{figure}[h!]
\vspace{-4mm}
\begin{center}
\includegraphics[width=130mm]{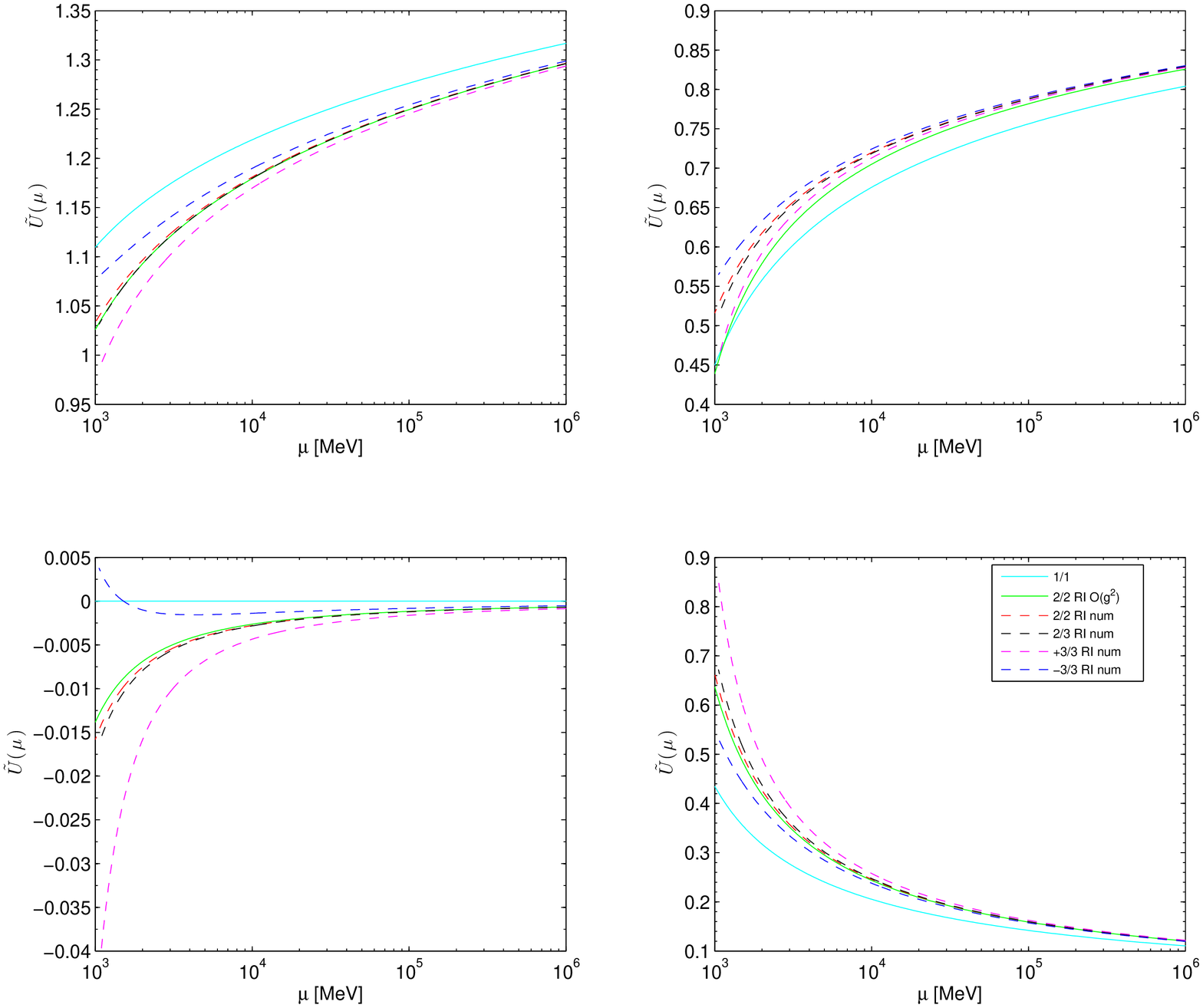}
\end{center}
\vspace{-5mm}
\begin{center}
\includegraphics[width=130mm]{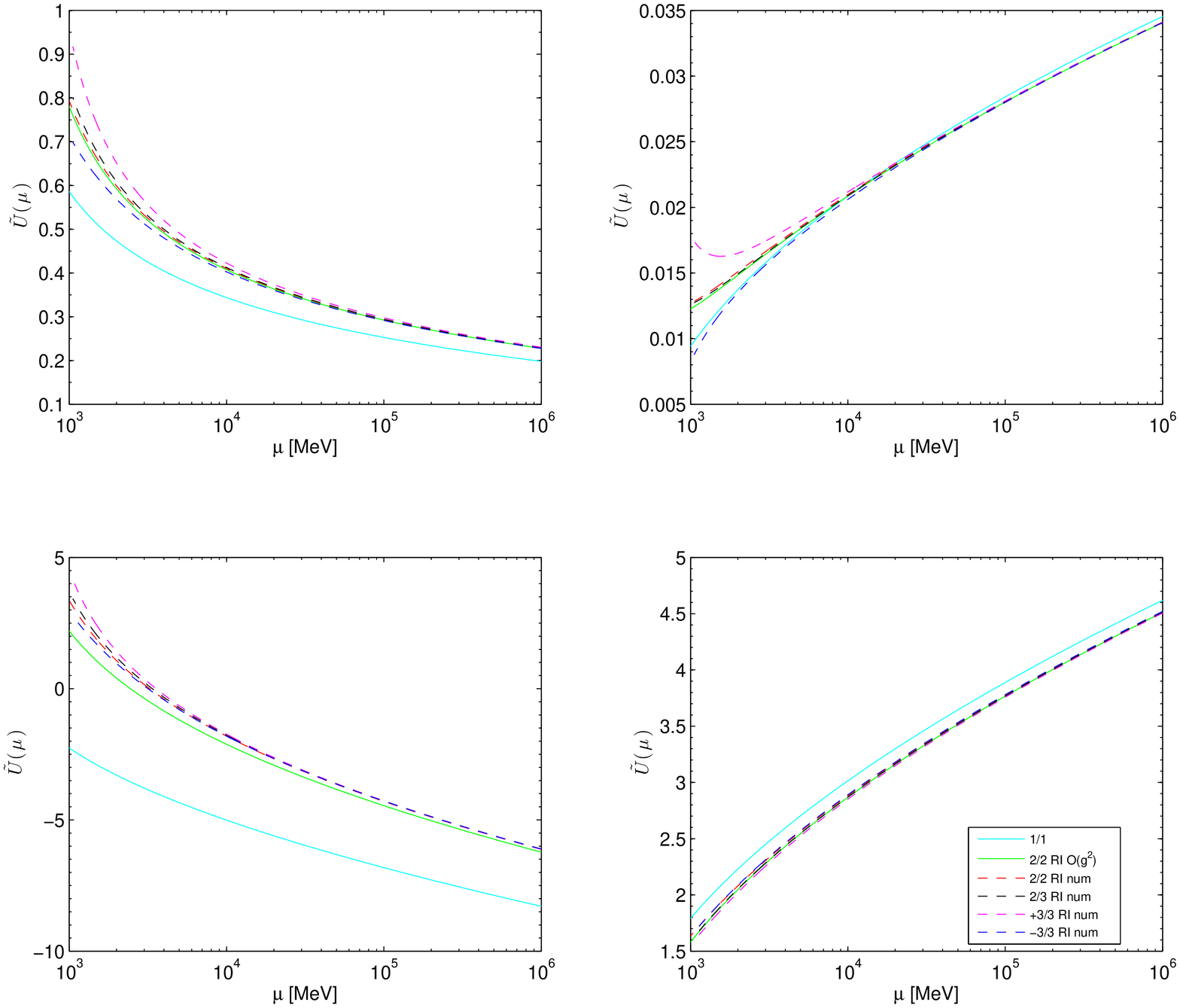}
\end{center}
\vspace{-4mm}
\caption{RG running matrices for the Fierz $+$ Op. $2,3$ (top half) and Op. $4,5$ (bottom half) in the RI scheme. Solid lines correspond to the LO (cyan) and the perturbative expansion for the NLO 2/2 case up to ${\mbox O}(g^2)$ - i.e. including $J_1$ (green). Dashed lines correspond to the numerical solution for $W(\mu)$ in the cases $n_\gamma/n_\beta = \{2/2, 2/3, +3/3, -3/3\}$ respectively in red, black, magenta and blue.}
\label{Fig:runRI+}
\end{figure}

\begin{figure}[h!]
\vspace{-4mm}
\begin{center}
\includegraphics[width=130mm]{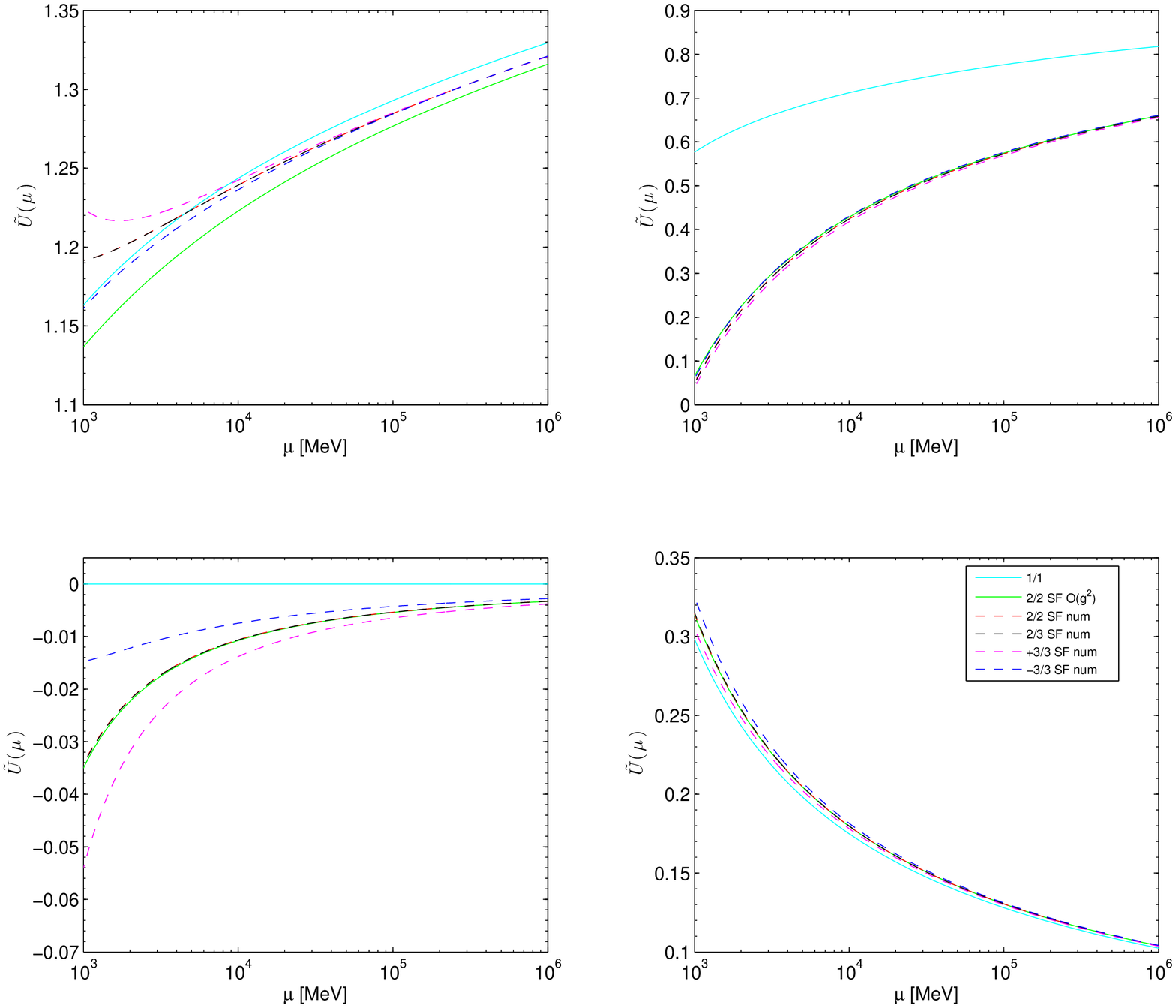}
\end{center}
\vspace{-5mm}
\begin{center}
\includegraphics[width=130mm]{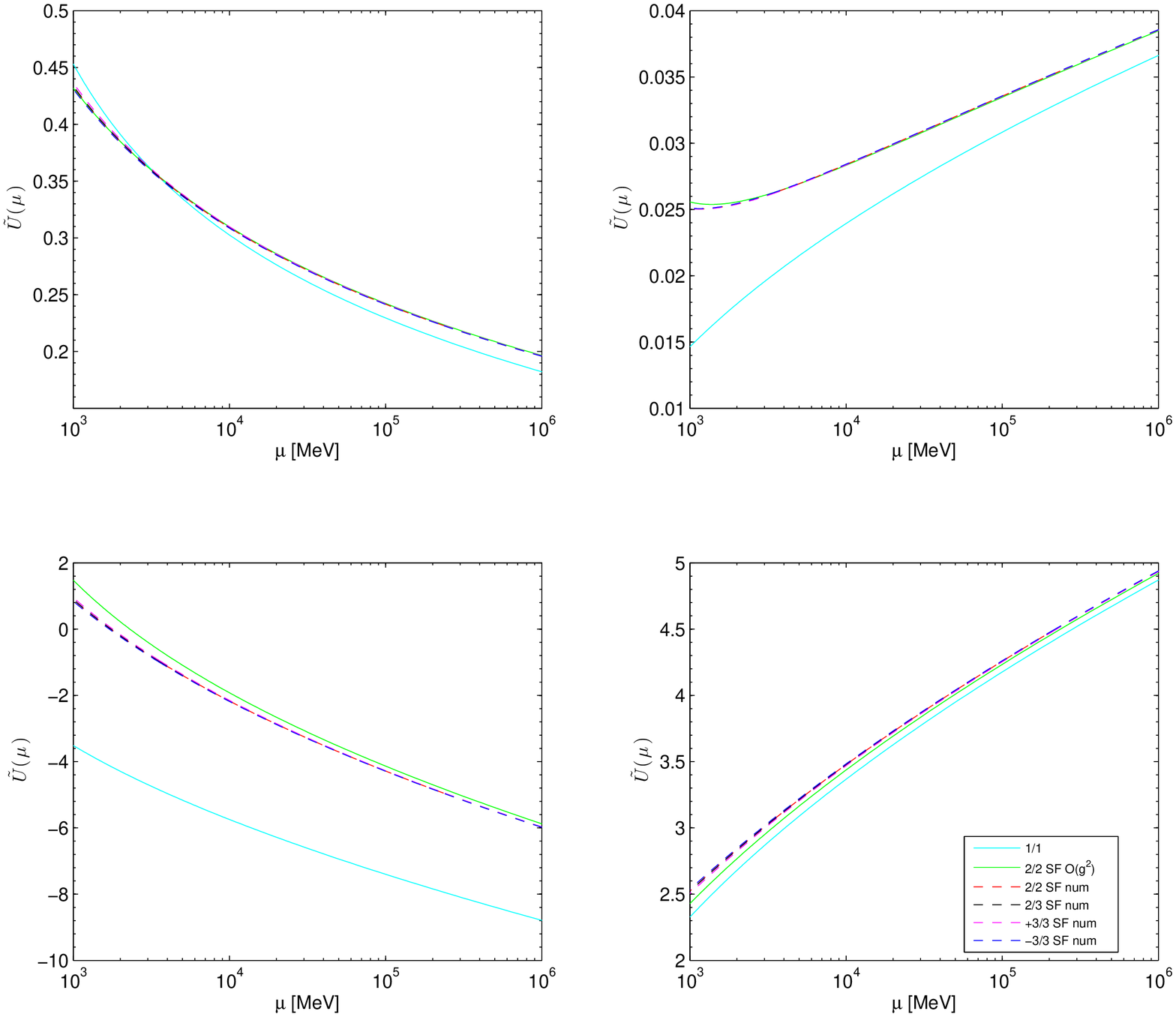}
\end{center}
\vspace{-4mm}
\caption{RG running matrices for the Fierz $+$ Op. $2,3$ (top half) and Op. $4,5$ (bottom half) in the SF scheme. Solid lines correspond to the LO (cyan) and the perturbative expansion for the NLO 2/2 case up to ${\mbox O}(g^2)$ - i.e. including $J_1$ (green). Dashed lines correspond to the numerical solution for $W(\mu)$ in the cases $n_\gamma/n_\beta = \{2/2, 2/3, +3/3, -3/3\}$ respectively in red, black, magenta and blue.}
\label{Fig:runSF+}
\end{figure}

\begin{figure}[t!]
\vspace{-4mm}
\begin{center}
\includegraphics[width=130mm]{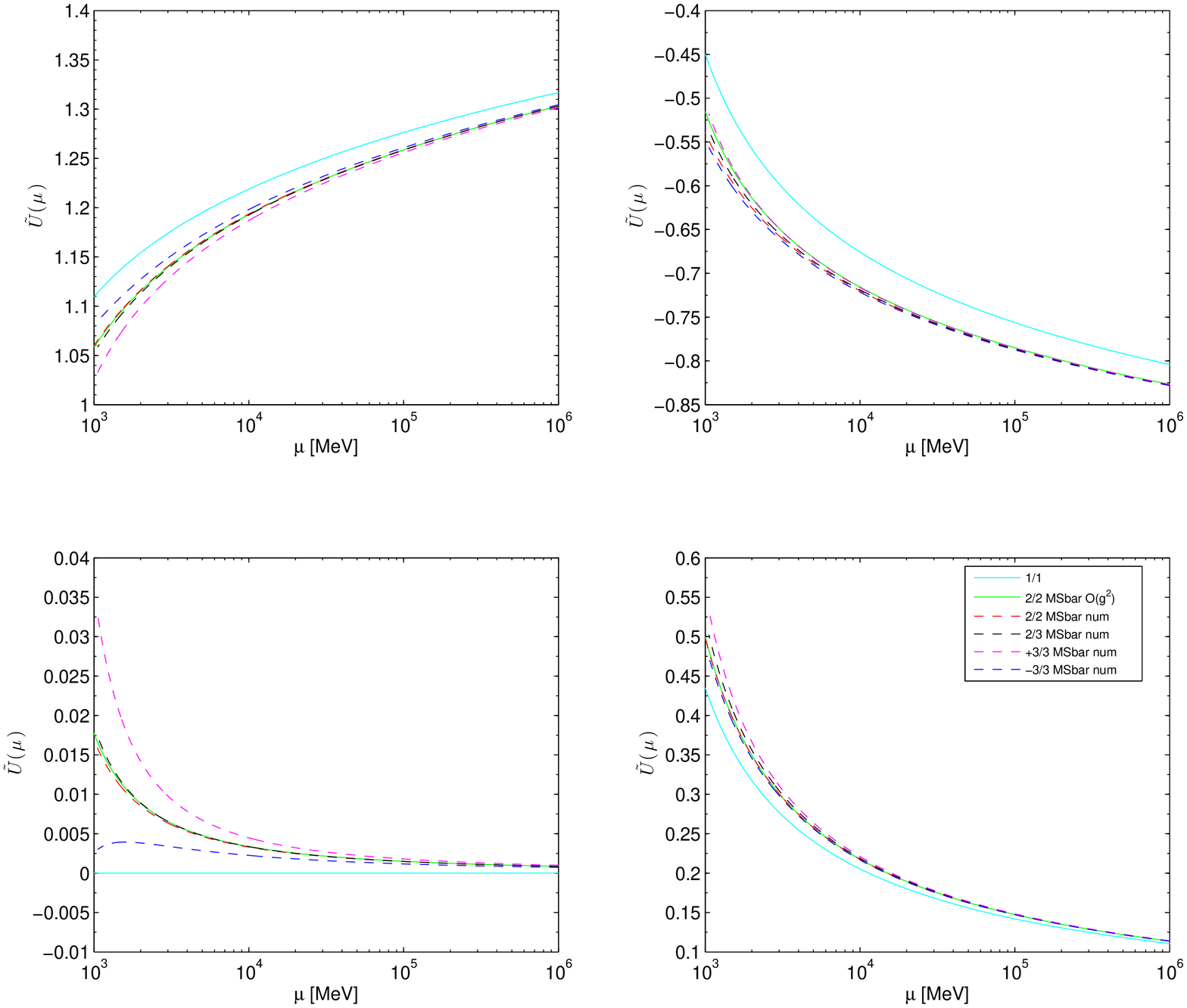}
\end{center}
\vspace{-5mm}
\begin{center}
\includegraphics[width=130mm]{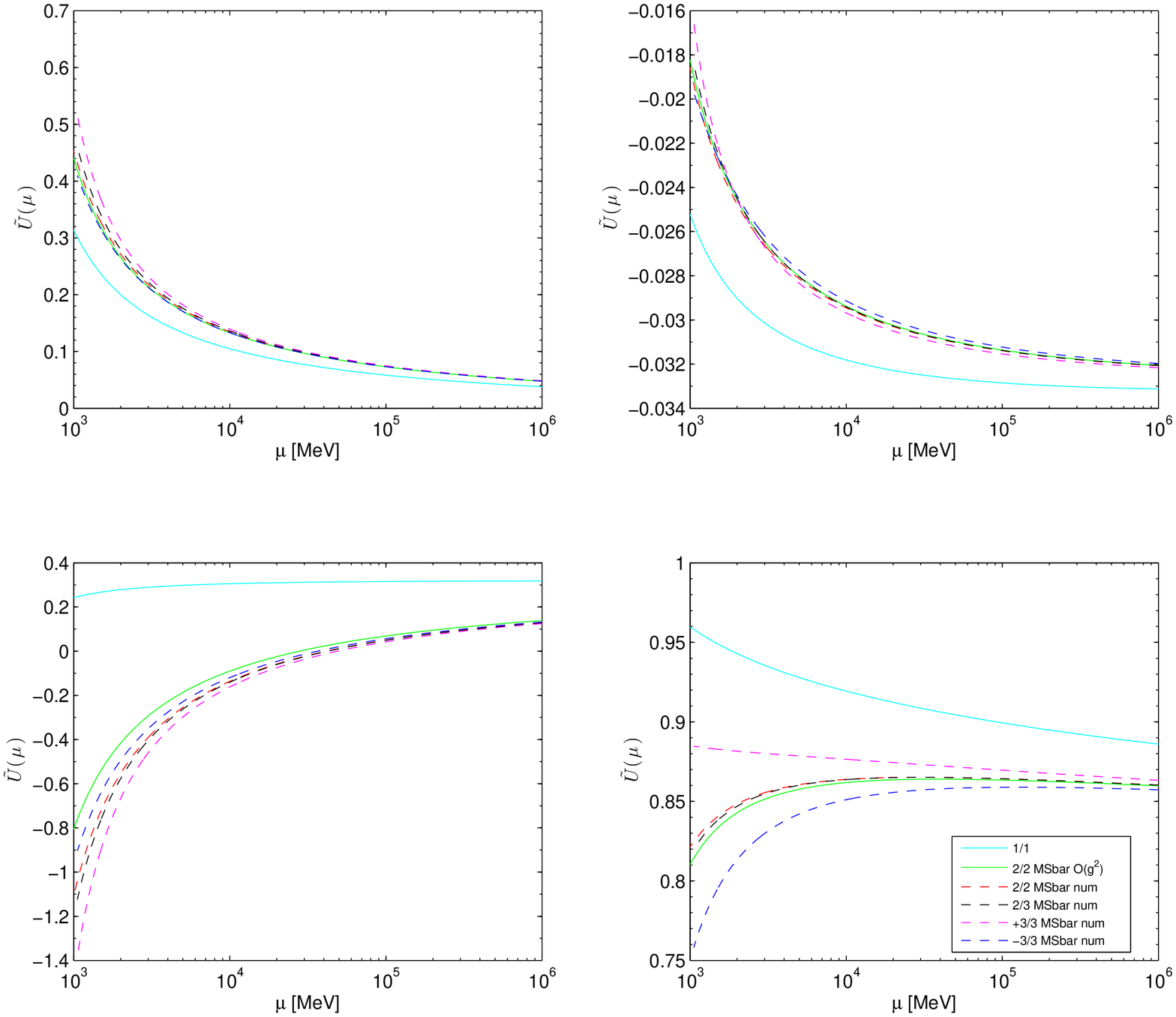}
\end{center}
\vspace{-4mm}
\caption{RG running matrices for the Fierz $-$ Op. $2,3$ (top half) and Op. $4,5$ (bottom half) in the $\MSbar$ scheme. Solid lines correspond to the LO (cyan) and the perturbative expansion for the NLO 2/2 case up to ${\mbox O}(g^2)$ - i.e. including $J_1$ (green). Dashed lines correspond to the numerical solution for $W(\mu)$ in the cases $n_\gamma/n_\beta = \{2/2, 2/3, +3/3, -3/3\}$ respectively in red, black, magenta and blue.}
\label{Fig:runMSbar-}
\end{figure}

\begin{figure}[t!]
\vspace{-4mm}
\begin{center}
\includegraphics[width=130mm]{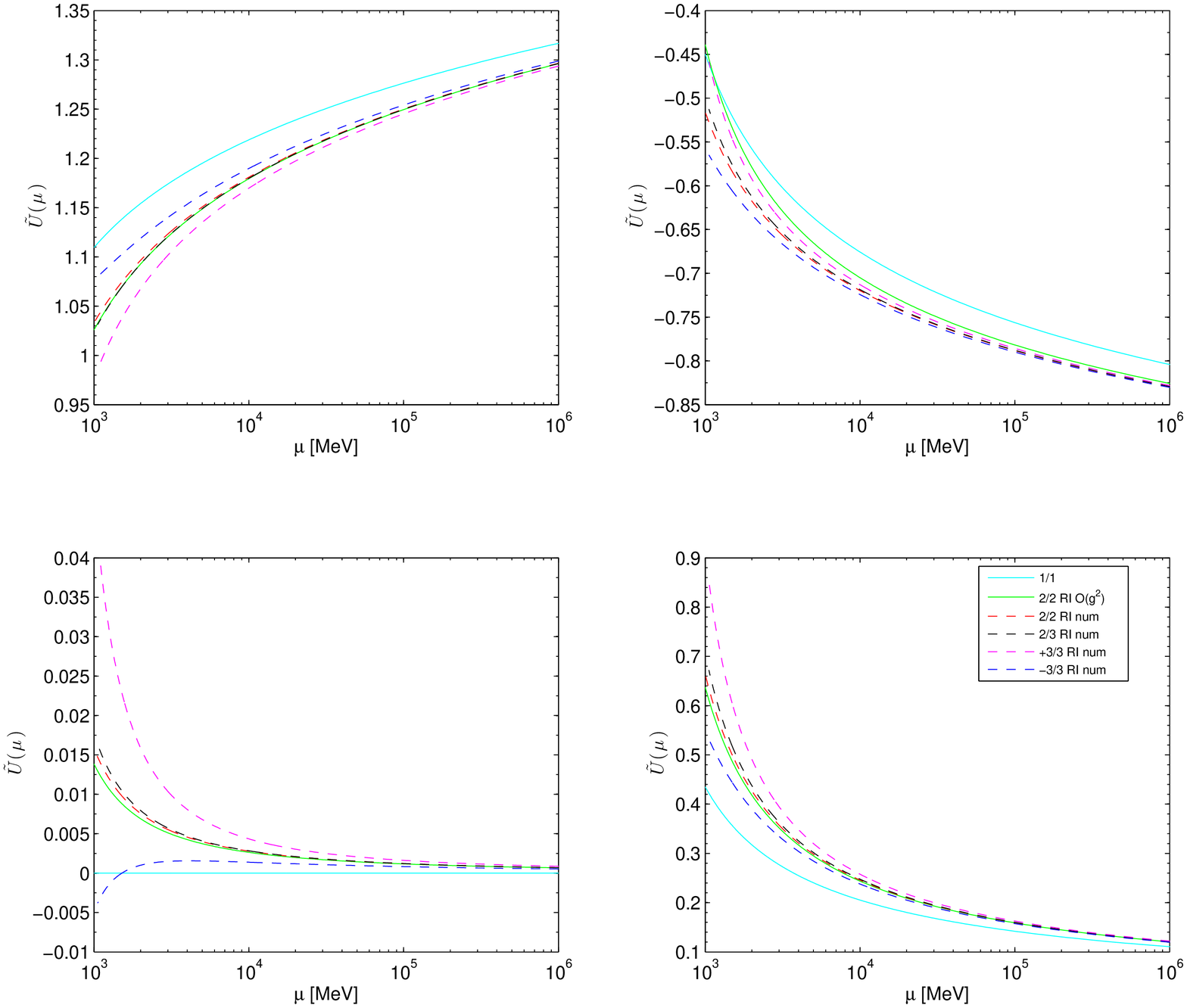}
\end{center}
\vspace{-5mm}
\begin{center}
\includegraphics[width=130mm]{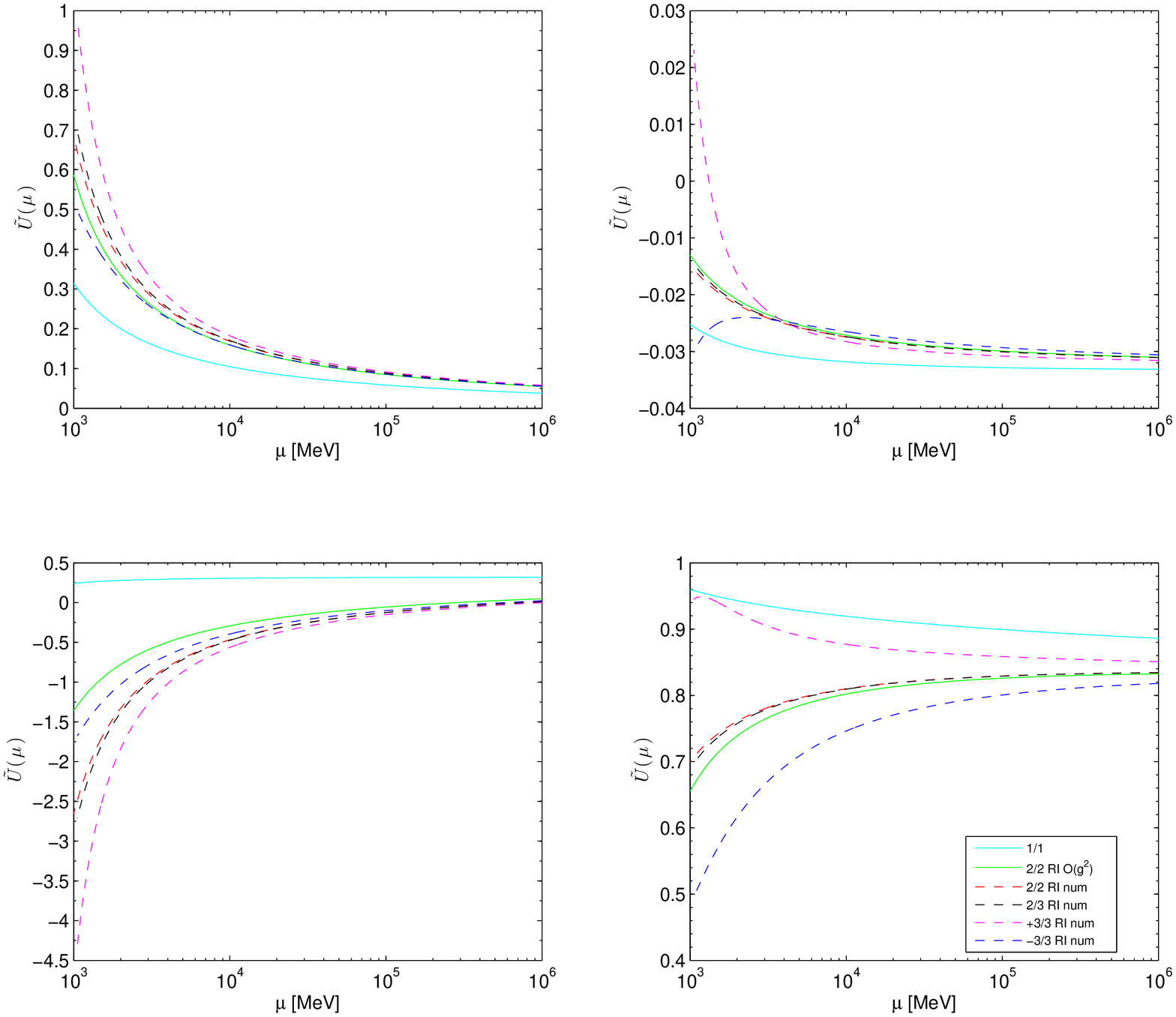}
\end{center}
\vspace{-4mm}
\caption{RG running matrices for the Fierz $-$ Op. $2,3$ (top half) and Op. $4,5$ (bottom half) in the RI scheme. Solid lines correspond to the LO (cyan) and the perturbative expansion for the NLO 2/2 case up to ${\mbox O}(g^2)$ - i.e. including $J_1$ (green). Dashed lines correspond to the numerical solution for $W(\mu)$ in the cases $n_\gamma/n_\beta = \{2/2, 2/3, +3/3, -3/3\}$ respectively in red, black, magenta and blue.}
\label{Fig:runRI-}
\end{figure}

\begin{figure}[h!]
\vspace{-4mm}
\begin{center}
\includegraphics[width=130mm]{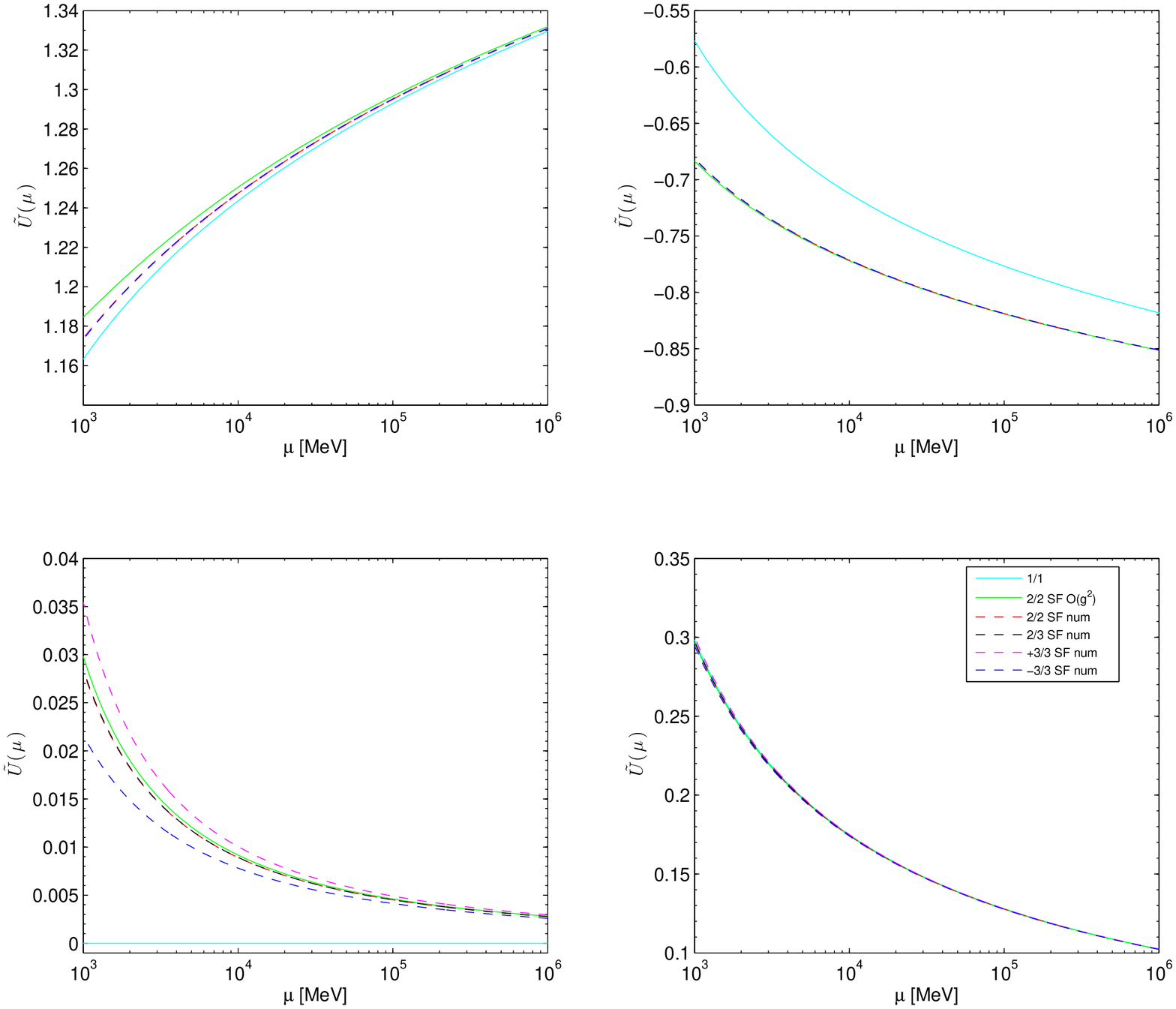}
\end{center}
\vspace{-5mm}
\begin{center}
\includegraphics[width=130mm]{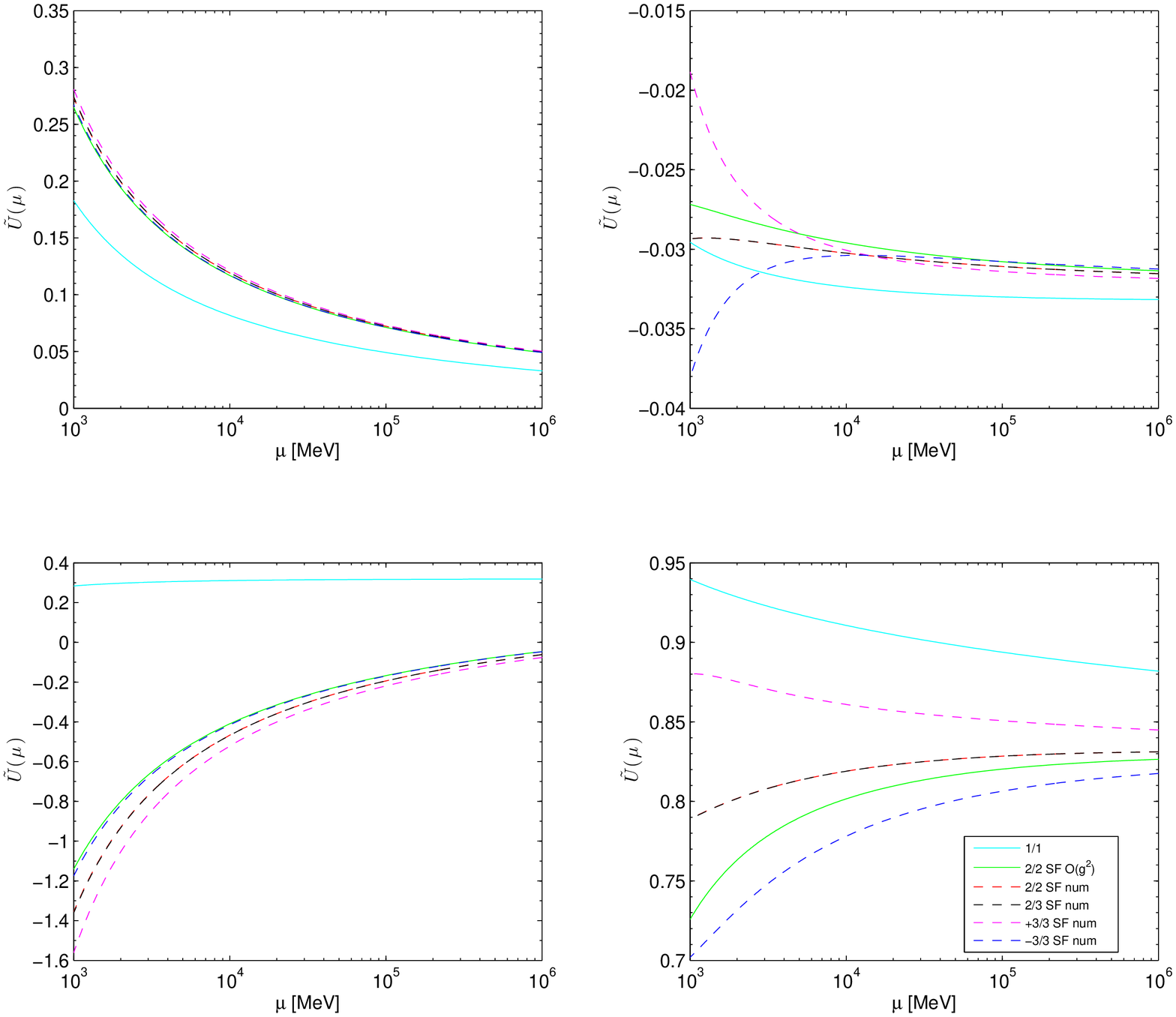}
\end{center}
\vspace{-4mm}
\caption{RG running matrices for the Fierz $-$ Op. $2,3$ (top half) and Op. $4,5$ (bottom half) in the SF scheme. Solid lines correspond to the LO (cyan) and the perturbative expansion for the NLO 2/2 case up to ${\mbox O}(g^2)$ - i.e. including $J_1$ (green). Dashed lines correspond to the numerical solution for $W(\mu)$ in the cases $n_\gamma/n_\beta = \{2/2, 2/3, +3/3, -3/3\}$ respectively in red, black, magenta and blue.}
\label{Fig:runSF-}
\end{figure}

The results for the LO running factor $U_{\rm\scriptscriptstyle LO}(\mu)$~\req{eq:U_LO} and the numerically integrated $\tilde U(\mu)$ running factors beyond LO (ii)-(v) described above are illustrated in Figs.~(\ref{Fig:runMSbar+}), (\ref{Fig:runRI+}), (\ref{Fig:runSF+}), (\ref{Fig:runMSbar-}), (\ref{Fig:runRI-}), (\ref{Fig:runSF-}) together with the 2/2 $\mbox{O}(g^2)$ perturbative 
expansion, for the four doublets of operators and three different schemes ($\MSbar$, RI 
and a chosen SF scheme). 

Some important observations are:
\begin{itemize}
\item The convergence of LO respect to NLO and NNLO seems to be slow in all the schemes under investigation for almost all the operators. In particular, for the matrix elements involving tensor current (4-5 sub-matrices) the convergence is very poor. Note that the LO anomalous dimensions for these submatrices are already very large compared with the others.
\item the 2/3 numerical running factors have always symmetric systematic errors, because most of the systematics is due to the inclusion of the guesstimate for $\gamma_2$ with + and - sign, and these effects turn out to be always symmetric with respect to the 2/3 (and also 2/2) cases.
\item the 2/2 $\mbox{O}(g^2)$ running factors are, for several matrix elements, quite far from the 2/3 (and also the 2/2) numerical ones. Possibly even further away than the $\pm3/3$ and have thus very large, asymmetric errors.
\item For both 4-5 sub-matrices (Fierz + and -) the ratio $\gamma_1\gamma_0^{-1}$ turns out to have large matrix elements. As a consequence, our plausibility argument for the guesstimates $\gamma_2\gamma_1^{-1}=\pm\gamma_1\gamma_0^{-1}$ leads to large systematic uncertainties. In particular, for the 54 matrix element the error is huge in the RI scheme and large also in the $\MSbar$ and SF schemes, already for the 2/3 numerical solution (for the 2/2 $\mbox{O}(g^2)$ perturbative
expansion the situation is much worse). This obviously poses serious doubts on all the 
computations of $\Delta F=2$ matrix elements beyond the Standard Model which uses 
perturbative running (in all cases through the 2/2 $\mbox{O}(g^2)$ expansion) 
down to scales of 3 $\GeV$ or less.     
\end{itemize}

\section{Conclusions}
\label{sec:concl}

In this paper we have reviewed the renormalisation and RG running properties of the four-quark operators
relevant for BSM analyses, and introduced a family of SF schemes that allow to compute them in a fully non-perturbative way.
Our non-perturbative results for $\NF=2$ QCD will be presented in a separate publication~\cite{nppaper}.\footnote{A comparison of perturbative and non-perturbative results for the running of these operators
in RI-MOM schemes for a small region in the few-GeV ballpark can be found in~\cite{Arthur:2011cn}.}
Here we have focused on the perturbative matching of our schemes to commonly used perturbative schemes and to RGI operators. One of our main results in this context is the full set of NLO operator anomalous dimensions in our SF schemes.

We have also conducted a detailed analysis of perturbative truncation effects in operator RG running
in both the SF schemes introduced here, and in commonly used $\MSbar$ and RI-MOM schemes.
We conclude that when NLO perturbation theory is used to run the operators from high-energy scales
down to the few GeV range, large truncation effects appear. One striking example is the mixing
of tensor-tensor and scalar-scalar operators, where all the available indications point to extremely
large anomalous dimensions and very poor perturbative convergence. One important point
worth stressing is that, in the computation of the running factor $W(\mu)$,
the use of the truncated perturbative expansion in \req{eq:Wpert} leads to
a significantly worse behaviour than the numerical integration of \req{eq:rg_W}
with the highest available orders for $\gamma$ and $\beta$.

A context where these findings might have an important impact is e.g. the computation of BSM
contributions to neutral kaon mixing. At present, few computations of the relevant $\Delta S=2$
operators exist with dynamical fermions~\cite{Boyle:2012qb,Bertone:2012cu,Carrasco:2015pra,Jang:2015sla,Garron:2016mva}, all of which use perturbative RG running
(and, in the case of~\cite{Jang:2015sla}, perturbative operator renormalisation as well). There are substantial
discrepancies between the various results in~\cite{Boyle:2012qb,Bertone:2012cu,Carrasco:2015pra,Jang:2015sla,Garron:2016mva}, which may be speculated to stem, at least
in part, from perturbative truncation effects.
Another possible contribution to the discrepancy is the delicate pole subtraction required in the RI-MOM scheme
--- indeed, results involving perturbative renormalisation and non-perturbative renormalisation
constants in RI-SMOM schemes are consistent.
At any rate, future efforts to settle this issue, as well as similar
studies for $\Delta B=2$ amplitudes, should put a strong focus on non-perturbative renormalisation.

\section*{Acknowledgments}

We are indebted to Stefan Sint and Tassos Vladikas for their role in the origin of this project,
and in the development of the formalism, originally applied to~\cite{Guagnelli:2005zc,Palombi:2005zd}.
We thank Steve Sharpe for having kindly converted for us the $\MSbar$ one-loop matching 
coefficients of Fierz $+$ operators from the scheme defined in~\cite{Gupta:1996yt} to the one used 
in~\cite{Buras:2000if}. We also thank Rainer Sommer for discussions.
M.P. acknowledges partial support by the MIUR-PRIN grant 2010YJ2NYW and by the INFN SUMA
project.
C.P. and D.P. acknowledge support by Spanish MINECO grants FPA2012-31686 and FPA2015-68541-P (MINECO/FEDER),
and MINECO's ``Centro de Excelencia Severo Ochoa'' Programme under grant SEV-2012-0249.


\begin{landscape}
\begin{table}[h!]
   \centering
\begin{scriptsize}
   \begin{tabular}{c|cccc} 
      \toprule
      $\widetilde U^{2/3}_{\mathcal{Q}_{1}\mathcal{Q}_2}(3~\rm GeV)$ & Fierz & RI & $\MSbar$ & SF \\  
      \midrule
      \multirow{2}{*}{$23$} & + & 
      $\begin{pmatrix}  1.121_{\ -0.019}^{\ +0.019} & 0.650_{\ -0.013}^{\ +0.014}\\[1ex]
	                -0.0057_{\ -0.0047}^{\ +0.0042} &  0.363_{\ -0.029}^{\ +0.032} \end{pmatrix}$ &
     $ \begin{pmatrix}  1.138_{\ -0.011}^{\ +0.011} & 0.654_{\ -0.006}^{\ +0.006}\\[1ex]
             -0.0066_{\ -0.0032}^{\ +0.0032} &  0.305_{\ -0.006}^{\ +0.006} \end{pmatrix}$ &
     $\begin{pmatrix}   1.212_{\ -0.009}^{\ +0.009} &  0.283_{\ -0.009}^{\ +0.009}\\[1ex]
             -0.018_{\ -0.007}^{\ +0.007} & 0.2292_{\ -0.0037}^{\ +0.0038}  \end{pmatrix}$\\
   				 \cmidrule(r){2-5} 
				 	  & - &
     $\begin{pmatrix}  1.121_{\ -0.019}^{\ +0.019} & -0.650_{\ -0.014}^{\ +0.013}\\[1ex]
             0.0057_{\ -0.0042}^{\ +0.0047} & 0.363_{\ -0.029}^{\ +0.032}  \end{pmatrix}$ &
     $\begin{pmatrix}  1.138_{\ -0.011}^{\ +0.011} &  -0.654_{\ -0.006}^{\ +0.006}\\[1ex]
             0.0066_{\ -0.0032}^{\ +0.0032} &  0.305_{\ -0.006}^{\ +0.006} \end{pmatrix}$ &
     $\begin{pmatrix}  1.2137_{\ -0.0002}^{\ +0.0002} & -0.7338_{\ -0.0002}^{\ +0.0002}\\[1ex]
             0.0148_{\ -0.0025}^{\ +0.0025} &  0.2202_{\ -0.0010}^{\ +0.0010}  \end{pmatrix}$\\
	\midrule					  
       \multirow{2}{*}{$45$} & + &
       $\begin{pmatrix} 0.539_{\ -0.026}^{\ +0.027} &  0.0165_{\ -0.0009}^{\ +0.0009}\\[1ex]
             0.243_{\ -0.135}^{\ +0.136} & 2.283_{\ -0.028}^{\ +0.029} \end{pmatrix}$ &
       $\begin{pmatrix}  0.488_{\ -0.008}^{\ +0.008} &  0.01414_{\ -0.00023}^{\ +0.00024}\\[1ex]
             -0.303_{\ -0.065}^{\ +0.061} &  2.187_{\ -0.056}^{\ +0.057} \end{pmatrix}$ &
       $\begin{pmatrix}  0.3623_{\ -0.0012}^{\ +0.0012} &  0.02601_{\ -0.00003}^{\ +0.00003}\\[1ex]
             -0.754_{\ -0.028}^{\ +0.028} & 3.005_{\ -0.010}^{\ +0.010} \end{pmatrix}$\\
         				 \cmidrule(r){2-5} 
				 	  & - &
       $\begin{pmatrix}  0.296_{\ -0.036}^{\ +0.040} &  -0.0237_{\ -0.0005}^{\ +0.0013}\\[1ex]
             -1.008_{\ -0.265}^{\ +0.220} &  0.778_{\ -0.112}^{\ +0.128} \end{pmatrix}$ &
       $\begin{pmatrix} 0.223_{\ -0.010}^{\ +0.010} &  -0.02644_{\ -0.00024}^{\ +0.00026}\\[1ex]
             -0.404_{\ -0.056}^{\ +0.053} &  0.855_{\ -0.025}^{\ +0.025}  \end{pmatrix}$ & 
       $\begin{pmatrix} 0.1717_{\ -0.0034}^{\ +0.0034} & -0.0296_{\ -0.0017}^{\ +0.0019}\\[1ex]
             -0.771_{\ -0.096}^{\ +0.092} &  0.807_{\ -0.061}^{\ +0.063}  \end{pmatrix}$\\
      \bottomrule
   \end{tabular}
   \caption{Values for the RG running coefficients at $\mu=3\,\GeV$ for the four doublets of operators in three different schemes ($\MSbar$, RI and a chosen SF scheme). We quote here, as our best result, the case $n_\gamma/n_\beta$=2/3 obtained by numerical integration. The systematic errors have been estimated by computing the maximal deviation between the central value and the values of the 2/2, +3/3 and -3/3 numerical solutions.}
   \label{tab:runnum}
\end{scriptsize}
\end{table}

\vspace{0.5cm} 

\begin{table}[h!]
\begin{scriptsize}
   \centering
   \begin{tabular}{c|cccc} 
      \toprule
      $\widetilde U^{2/2, \mbox{O}(g^2)}_{\mathcal{Q}_{1}\mathcal{Q}_2}(3~\rm GeV)$ & Fierz & RI & $\MSbar$ & SF \\  
      \midrule
      \multirow{2}{*}{$23$} & + & 
      $\begin{pmatrix}  1.120_{\ -0.019}^{\ +0.020} & 0.624_{\ -0.000}^{\ +0.040}\\[1ex]
             -0.0051_{\ -0.0053}^{\ +0.0036} & 0.352_{\ -0.018}^{\ +0.043} \end{pmatrix}$ &
     $ \begin{pmatrix} 1.139_{\ -0.012}^{\ +0.009} &  0.648_{\ -0.000}^{\ +0.012}\\[1ex]
             -0.0066_{\ -0.0032}^{\ +0.0032} &  0.301_{\ -0.003}^{\ +0.010} \end{pmatrix}$ &
     $\begin{pmatrix} 1.184_{\ -0.000}^{\ +0.038} & 0.290_{\ -0.015}^{\ +0.002}\\[1ex]
             -0.018_{\ -0.007}^{\ +0.007} &  0.2292_{\ -0.0037}^{\ +0.0038}\end{pmatrix}$\\
   				 \cmidrule(r){2-5} 
				 	  & - &
     $\begin{pmatrix}  1.120_{\ -0.019}^{\ +0.020} & -0.624_{\ -0.040}^{\ +0.000}\\[1ex]
             0.0051_{\ -0.0036}^{\ +0.0053} &  0.352_{\ -0.018}^{\ +0.043} \end{pmatrix}$ &
     $\begin{pmatrix}  1.139_{\ -0.012}^{\ +0.009} & -0.648_{\ -0.012}^{\ +0.000}\\[1ex]
             0.0066_{\ -0.0032}^{\ +0.0032} &  0.305_{\ -0.003}^{\ +0.010} \end{pmatrix}$ &
     $\begin{pmatrix} 1.219_{\ -0.005}^{\ +0.000} & -0.7346_{\ -0.0000}^{\ +0.0009}\\[1ex]
             0.0154_{\ -0.0030}^{\ +0.0019} & 0.2203_{\ -0.0011}^{\ +0.0009} \end{pmatrix}$\\
	\midrule					  
      \multirow{2}{*}{$45$} & + &
       $\begin{pmatrix} 0.528_{\ -0.015}^{\ +0.038} &  0.0164_{\ -0.0008}^{\ +0.0010}\\[1ex]
             -0.358_{\ -0.000}^{\ +0.737} & 2.276_{\ -0.021}^{\ +0.036} \end{pmatrix}$ &
       $\begin{pmatrix} 0.484_{\ -0.003}^{\ +0.013} &  0.01413_{\ -0.00023}^{\ +0.00025}\\[1ex]
             -0.533_{\ -0.000}^{\ +0.291} &  2.179_{\ -0.048}^{\ +0.065} \end{pmatrix}$ &
       $\begin{pmatrix} 0.3629_{\ -0.0018}^{\ +0.0006} &  0.02605_{\ -0.00008}^{\ +0.00000}\\[1ex]
             -0.382_{\ -0.400}^{\ +0.000} & 2.945_{\ -0.000}^{\ +0.070} \end{pmatrix}$\\
         				 \cmidrule(r){2-5} 
				 	  & - &
       $\begin{pmatrix} 0.266_{\ -0.006}^{\ +0.070} &  -0.0233_{\ -0.0010}^{\ +0.0008}\\[1ex]
             -0.596_{\ -0.678}^{\ +0.000} &  0.764_{\ -0.098}^{\ +0.143} \end{pmatrix}$ &
       $\begin{pmatrix}  0.215_{\ -0.001}^{\ +0.018} &  -0.02644_{\ -0.00025}^{\ +0.00026}\\[1ex]
             -0.294_{\ -0.166}^{\ +0.000} &  0.851_{\ -0.022}^{\ +0.029} \end{pmatrix}$ & 
       $\begin{pmatrix}  0.168_{\ -0.000}^{\ +0.007} & -0.0285_{\ -0.0028}^{\ +0.0008}\\[1ex]
             -0.670_{\ -0.197}^{\ +0.000} &  0.777_{\ -0.030}^{\ +0.094}\end{pmatrix}$\\
      \bottomrule
   \end{tabular}
   \caption{Values for the RG running coefficients at $\mu=3\,\GeV$ for the four doublets of operators in three different schemes ($\MSbar$, RI and a chosen SF scheme). We quote here the 2/2 result from the perturbative expansion at $\mbox{O}(g^2)$, which is the case usually considered in literature, both for phenomenological application and in lattice computations.   
 The systematic errors have been estimated by computing the maximal deviation between the central value and the values of the 2/2, 2/3, +3/3 and -3/3 numerical solutions. It is worth noticing the large asymmetric errors which occour in particular in the 45 Fierz $+$ and $-$ matrices (especially in the RI scheme).}
\label{tab:run}
\end{scriptsize}
\end{table}

\end{landscape}

\clearpage
\newpage

\clearpage
\newpage

\clearpage
\newpage

\clearpage
\newpage

\clearpage
\newpage

\newpage
\begin{appendix}
\section{Constraints on anomalous dimensions from chiral symmetry}
\label{app:symm}

In section 5.3 of~\cite{Donini:1999sf} the authors derive an identity
between the the renormalisation matrices for $(\cQ_2^+,\cQ_3^+)$ and
$(\cQ_2^-,\cQ_3^-)$, valid in the RI-MOM scheme considered in that paper.
Here we discuss how such an identity can be derived from generic considerations
based on chiral symmetry, and how (or, rather, under which conditions)
it can be generalised to other renormalisation schemes.

Let us consider a renormalised matrix element of the form
$\langle f|\bar Q_k^\pm|i\rangle$,
where $Q_k^\pm$ is a parity-even operator and $|i,f\rangle$ are stable
hadron states with the same, well-defined parity.
Simple examples would be the matrix elements of $\Delta F=2$ operators
providing the hadronic contribution to $K^0$--$\bar K^0$ or $B^0$--$\bar B^0$
oscillation amplitudes (cf.~\res{sec:renorm}).
Bare matrix elements can be extracted from suitable three-point Euclidean
correlation functions
\begin{gather}
\langle\cO_f(x)\,Q_k^\pm(0)\,\cO_i(y)\rangle =
\frac{1}{\cZ}\int D[\psi]D[\bar\psi]D[A]\,e^{-S}\,\cO_f(x)\,Q_k^\pm(0)\,\cO_i(y)
\end{gather}
where $\cO_{i,f}$ are interpolating operators for the external states $|i,f\rangle$.
If we perform a change of fermion variables of the form
\begin{gather}
\label{eq:twist}
\psi \to \psi'=e^{i\gamma_5 T}\psi\,,~~~~~
\bar\psi \to \bar\psi'=\bar\psi e^{i\gamma_5 T}\,,
\end{gather}
where $\psi$ is a fermion field with $\NF$ flavour components
and $T$ is a traceless matrix acting on flavour space, this will
induce a corresponding transformation $Q_k^\pm \to {Q_k'}^\pm$, $\cO_{i,f}\to\cO'_{i,f}$
of the involved composite operators. If the regularised theory employed
to define the path integral preserves exactly the ${\rm SU}(\NF)_{\rm A}$
axial chiral symmetry of the formal continuum theory, the equality
$\langle\cO_f(x)\,Q_k^\pm(0)\,\cO_i(y)\rangle=\langle\cO'_f(x)\,{Q_k'}^\pm(0)\,\cO'_i(y)\rangle$ will hold exactly; otherwise, it will only hold upon
renormalisation and removal of the cutoff. At the level of matrix
elements, one will then have
\begin{gather}
\label{eq:twist_me}
\langle f|\bar Q_k^\pm|i\rangle_{(\psi,\bar\psi)} =
\langle f|\bar {Q_k'}^\pm|i\rangle_{(\psi',\bar\psi')} \,,
\end{gather}
where the subscript remarks that the interpretation of the operator
depends on the fermion variables used on each side of the equation.
If the flavour matrix $T$ is not traceless, the argument will
still hold if the fermion fields entering composite operators
are part of a valence sector, employed only for the purpose of
defining suitable correlation functions.

The result in~\req{eq:twist}
is at the basis e.g. of the definition of twisted-mass QCD lattice regularisations,
and is discussed in more detail in~\cite{Frezzotti:2000nk,Pena:2004gb,Frezzotti:2004wz}.
Indeed, the rotation in \req{eq:twist} will in general transform
the mass term of the action. One crucial remark at this point is that,
if a mass-independent renormalisation scheme is used, renormalisation
constants for any given composite operator will be independent of
which fermion variables are employed in the computation of the matrix element.

Let us now consider a particular case of \req{eq:twist} given by
\begin{gather}
\label{eq:rtwist}
T = \frac{\pi}{4}\left(\ba{rrrr}
1 & 0 & 0 & 0 \\
0 & 1 & 0 & 0 \\
0 & 0 & 1 & 0 \\
0 & 0 & 0 & -1
\ea\right)\,,
\end{gather}
where $\psi=(\psi_1,\psi_2,\psi_3,\psi_4)^T$ comprises the four, formally
distinct flavours that enter $Q_k^\pm,\cQ_k^\pm$. Under this rotation,
the ten operators of the basis in \req{eq:rel_ops} transform as
\begin{gather}
\begin{split}
Q_1^\pm &\to i\cQ_1^\pm\,,\\
Q_2^\pm &\to -i\cQ_2^\mp\,,\\
Q_3^\pm &\to i\cQ_3^\mp\,,\\
Q_4^\pm &\to i\cQ_4^\pm\,,\\
Q_5^\pm &\to i\cQ_5^\pm\,.
\end{split}
\end{gather}
In the case of operators 1,4,5 the rotation is essentially trivial, in that
it preserves Fierz ($2 \leftrightarrow 4$ exchange) eigenstates. However,
in the rotation of operators 2,3 the Fierz eigenvalue is exchanged. One
thus has, at the level of renormalised matrix elements,
\begin{gather}
\label{eq:rotop}
\langle f|\bar {\mathbf{Q}}^+(\mu)|i\rangle_{(\psi,\bar\psi)} =
R\langle f|\bar {\boldsymbol{{\cQ}}}^-(\mu)|i\rangle_{(\psi',\bar\psi')} \,,
\end{gather}
where $\mathbf{Q}^+=(Q_2^+,Q_3^+)^T$, ${\boldsymbol{{\cQ}}}^-=(\cQ_2^-,\cQ_3^-)^T$,
and $R=-i\tau^3$. In this latter expression we have written explicitly the
renormalisation scale $\mu$. If we now use the RG evolution operators
discussed in~\res{sec:renorm} to run \req{eq:rotop} to another scale $\mu'$,
one then has (recall that the continuum anomalous dimensions of $Q_k^+$
and $\cQ_k^+$ --- respectively, $Q_k^-$
and $\cQ_k^-$ --- are the same)
\begin{gather}
\begin{split}
\langle f|\bar {\mathbf{Q}}^+(\mu')|i\rangle_{(\psi,\bar\psi)} &=
U^+(\mu',\mu)\langle f|\bar {\mathbf{Q}}^+(\mu)|i\rangle_{(\psi,\bar\psi)} \\&=
U^+(\mu',\mu)R\langle f|\bar {\boldsymbol{{\cQ}}}^-(\mu)|i\rangle_{(\psi',\bar\psi')} \\&=
U^+(\mu',\mu)R [U^-(\mu',\mu)]^{-1}\langle f|\bar {\boldsymbol{{\cQ}}}^-(\mu')|i\rangle_{(\psi',\bar\psi')} \\&=
U^+(\mu',\mu)R [U^-(\mu',\mu)]^{-1}R^{-1}\langle f|\bar {\mathbf{Q}}^+(\mu')|i\rangle_{(\psi,\bar\psi)}\,,
\end{split}
\end{gather}
which implies
\begin{gather}
U^+(\mu',\mu) = R U^-(\mu',\mu) R^{-1}~~~~\forall\mu,\mu'\,.
\end{gather}
It is then immediate that the anomalous dimension matrices entering $U^\pm$
are related as
\begin{gather}
\label{eq:id23}
\gamma^+ =
\left(\ba{cc}
\gamma_{22}^+ & \gamma_{23}^+ \\
\gamma_{32}^+ & \gamma_{33}^+
\ea\right) =
\tau^3 \gamma^- \tau^3 =
\left(\ba{cc}
\gamma_{22}^- & -\gamma_{23}^- \\
-\gamma_{32}^- & \gamma_{33}^-
\ea\right) \,.
\end{gather}

The correct interpretation of this identity is that, given an anomalous
dimension matrix for, say, $Q_{2,3}^+$ and $\cQ_{2,3}^+$,
one can use \req{eq:id23} to construct a correct anomalous dimension
matrix for $Q_{2,3}^-$ and $\cQ_{2,3}^-$, and vice versa.
However, it does {\em not} guarantee that, given two different renormalisation
conditions for each fierzing, the resulting matrices of
anomalous dimensions will satisfy \req{eq:id23}. This will only be
the case if the renormalisation conditions can be related to each
other by the rotation in \req{eq:rtwist}; otherwise, the result
of applying \req{eq:id23} to the $\gamma^-$ that follows from the condition
imposed on Fierz - operators will lead to value of $\gamma^+$ in a different
renormalisation scheme than the one defined by the renormalisation condition
imposed directly on Fierz + operators.

The RI-MOM conditions of~\cite{Donini:1999sf}, as well as typical $\MSbar$ 
renormalisation conditions, result in schemes that satisfy the identity
directly, since the quantities involved respect the underlying chiral symmetry ---
e.g. the amputated correlation functions used in RI-MOM rotate in a
similar way to the three-point functions discussed above. Indeed, the known
NLO anomalous dimensions in RI-MOM and $\MSbar$ given in~\reapp{app:gammacont},
as well as (within uncertainties) the non-perturbative values of RI-MOM
renormalisation constants, fulfill \req{eq:id23}. Our SF renormalisation
conditions, on the other hand, are not related among them via rotations
with $R$, due to the chiral symmetry-breaking effects induced by the non-trivial boundary
conditions imposed on the fields. As a consequence, the finite parts
of the matrices of SF renormalisation constants, and hence $\gamma_2^{\rm SF}$,
do not satisfy the identity.
It has to be stressed that, as a consequence of the existence of schemes
where \req{eq:id23} is respected, the identity is satisfied by the
universal matrices $\gamma_0^\pm$,
as can be readily checked in \req{eq:load}; therefore, the violation of
the identity in e.g. SF schemes appears only at $\cO(g_0^4)$ in perturbation theory.

\section{NLO anomalous dimensions in continuum schemes}
\label{app:gammacont}

The two-loop anomalous dimension matrices in the
RI-MOM scheme (in Landau gauge)~\cite{Ciuchini:1997bw,Buras:2000if} and $\MSbar$ 
scheme~\cite{Buras:2000if} are given by (the factor $(4\pi)^{-4}$ has been omitted below to simplify the notation): 
\begin{gather}
\begin{split}
\gamma^{+,(1);{\rm RI}}_{22} &= \frac{(297+16 \log (2)) N^2+45}{6 N^2}-N_f\frac{2 (15+4\log (2))}{3 N} \nonumber\,,\\
\gamma^{+,(1);{\rm RI}}_{23} &= \frac{2 \left(4 N^2 (45+2\log (2))-9\right)}{3 N}-N_f\frac{4}{3} (15+4\log (2)) \nonumber\,,\\
\gamma^{+,(1);{\rm RI}}_{32} &= \frac{(53+160 \log (2)) N^2+108}{12 N}-N_f\frac{2}{3} (1+2\log (2)) \nonumber\,,\\
\gamma^{+,(1);{\rm RI}}_{33} &= \frac{-379 N^4+5 (99+32 \log (2)) N^2+45}{6 N^2}+N_f\frac{2 \left(13 N^2-4 \log (2)-15\right)}{3 N} \,,\\
\gamma^{+,(1);{\rm RI}}_{44} &= \frac{-379 N^4+2 (261-88 \log (2)) N^3+140 (3+2\log (2)) N^2-4 (-6+60 \log (2)) N-81}{6 N^2}+\nonumber\\
&+\,N_f\frac{2 \left(13 N^2+(-15+8\log (2)) N-4\log (2)-15\right)}{3 N} \nonumber\,,\\
\gamma^{+,(1);{\rm RI}}_{45} &= \frac{(157-368 \log (2)) N^3+(-494+556 \log (2)) N^2-4 (-39+30 \log (2)) N-72}{36 N^2}+\nonumber\\
&+\,N_f\frac{((-11+16 \log (2)) N-20 \log (2)+28)}{18 N} \nonumber\,,\\
\gamma^{+,(1);{\rm RI}}_{54} &= \frac{4 \left((-165+16 \log (2)) N^3+(-230+76 \log (2)) N^2-4 (-39+30 \log (2)) N-72\right)}{3 N^2}+\nonumber\\
&+\,N_f\frac{8 ((15+16 \log (2)) N-20 \log (2)+28)}{3 N} \nonumber\,,\\
\gamma^{+,(1);{\rm RI}}_{55} &= \frac{343 N^4-2 (-343+616 \log (2)) N^3+4 (-95+142 \log (2)) N^2+(504+720 \log (2)) N-531}{18 N^2}+\nonumber\\
&-\,N_f\frac{2 \left(13 N^2+(41-56 \log (2)) N+52 \log (2)-11\right)}{9 N} \nonumber\,.
\end{split}
\end{gather}

\begin{gather}
\begin{split}
\gamma^{-,(1);{\rm RI}}_{22} &= \frac{15}{2 N^2}+\frac{8 \log (2)}{3}+\frac{99}{2}-N_f\frac{2 (15+4\log (2))}{3 N} \nonumber\,,\\
\gamma^{-,(1);{\rm RI}}_{23} &= -\frac{8}{3} (45+2\log (2)) N+\frac{6}{N}+N_f\frac{4}{3} (15+4\log (2)) \nonumber\,,\\
\gamma^{-,(1);{\rm RI}}_{32} &= -\frac{1}{12} (53+160 \log (2)) N-\frac{9}{N}+N_f\frac{2}{3} (1+2\log (2)) \nonumber\,,\\
\gamma^{-,(1);{\rm RI}}_{33} &= \frac{-379 N^4+5 (99+32 \log (2)) N^2+45}{6 N^2}+N_f\frac{2 \left(13 N^2-4 \log (2)-15\right)}{3 N} \,,\\
\gamma^{-,(1);{\rm RI}}_{44} &= \frac{-379 N^4+2 (-261+88 \log (2)) N^3+140 (3+2\log (2)) N^2+24 (-1+10\log (2)) N-81}{6 N^2}+\nonumber\\
&+\,N_f\frac{2 \left(13 N^2-(-15+8\log (2)) N-4\log (2)-15\right)}{3 N} \nonumber\,,\\
\gamma^{-,(1);{\rm RI}}_{45} &= \frac{(-157+368 \log (2)) N^3+(-494+556 \log (2)) N^2+12 (-13+10\log (2)) N-72}{36 N^2}+\nonumber\\
&-\,N_f\frac{((-11+16\log (2)) N+20\log (2)-28)}{18 N} \nonumber\,,\\
\gamma^{-,(1);{\rm RI}}_{54} &= -\frac{4 \left((-165+16 \log (2)) N^3+(230-76 \log (2)) N^2+(156-120 \log (2)) N+72\right)}{3 N^2}+\nonumber\\
&-\,N_f\frac{8 ((15+16 \log (2)) N+4 (-7+5\log (2)))}{3 N} \nonumber\,,\\
\gamma^{-,(1);{\rm RI}}_{55} &= \frac{343 N^4+14 (-49+88 \log (2)) N^3+4 (-95+142 \log (2)) N^2-72 (7+10\log (2)) N-531}{18 N^2}+\nonumber\\
&-\,N_f\frac{2 \left(13 N^2+(-41+56 \log (2)) N+52 \log (2)-11\right)}{9 N} \nonumber\,.
\end{split}
\end{gather}

\begin{gather}
\begin{split}
\gamma^{+,(1);{\MSbar}}_{22} &= \frac{15}{2 N^2}+\frac{137}{6}-N_f\frac{22}{3 N} \nonumber\,,\\
\gamma^{+,(1);{\MSbar}}_{23} &= \frac{200 N}{3}-\frac{6}{N}-N_f\frac{44}{3} \nonumber\,,\\
\gamma^{+,(1);{\MSbar}}_{32} &= \frac{71 N}{4}+\frac{9}{N}-N_f 2 \nonumber\,,\\
\gamma^{+,(1);{\MSbar}}_{33} &= -\frac{203 N^2}{6}+\frac{479}{6}+\frac{15}{2 N^2}+N_f\left(\frac{10 N}{3}-\frac{22}{3 N}\right) \,,\\
\gamma^{+,(1);{\MSbar}}_{44} &= -\frac{203 N^2}{6}+\frac{107 N}{3}+\frac{136}{3}-\frac{12}{N}-\frac{107}{2 N^2}+N_f\left(\frac{10 N}{3}-\frac{2}{3}-\frac{10}{3 N}\right) \nonumber\,,\\
\gamma^{+,(1);{\MSbar}}_{45} &= -\frac{N}{36}-\frac{31}{9}+\frac{9}{N}-\frac{4}{N^2}+N_f\left(\frac{1}{9 N}-\frac{1}{18}\right) \nonumber\,,\\
\gamma^{+,(1);{\MSbar}}_{54} &= -\frac{364 N}{3}-\frac{704}{3}-\frac{208}{N}-\frac{320}{N^2}+N_f\left(\frac{136}{3}+\frac{176}{3 N}\right) \nonumber\,,\\
\gamma^{+,(1);{\MSbar}}_{55} &= \frac{343 N^2}{18}+21 N-\frac{188}{9}+\frac{44}{N}+\frac{21}{2 N^2}+N_f\left(-\frac{26 N}{9}-6+\frac{2}{9 N}\right) \nonumber\,.
\end{split}
\end{gather}

\begin{gather}
\begin{split}
\gamma^{-,(1);{\MSbar}}_{22} &= \frac{15}{2 N^2}+\frac{137}{6}-N_f\frac{22}{3 N} \nonumber\,,\\
\gamma^{-,(1);{\MSbar}}_{23} &= -\frac{200 N}{3}+\frac{6}{N}+N_f\frac{44}{3} \nonumber\,,\\
\gamma^{-,(1);{\MSbar}}_{32} &= -\frac{71 N}{4}-\frac{9}{N}+N_f 2 \nonumber\,,\\
\gamma^{-,(1);{\MSbar}}_{33} &= -\frac{203 N^2}{6}+\frac{479}{6}+\frac{15}{2 N^2}+N_f\left(\frac{10 N}{3}-\frac{22}{3 N}\right) \,,\\
\gamma^{-,(1);{\MSbar}}_{44} &= -\frac{203 N^2}{6}-\frac{107 N}{3}+\frac{136}{3}+\frac{12}{N}-\frac{107}{2 N^2}+N_f\left(\frac{10 N}{3}+\frac{2}{3}-\frac{10}{3 N}\right) \nonumber\,,\\
\gamma^{-,(1);{\MSbar}}_{45} &= \frac{N}{36}-\frac{31}{9}-\frac{9}{N}-\frac{4}{N^2}+N_f\left(\frac{1}{18}+\frac{1}{9 N}\right)  \nonumber\,,\\
\gamma^{-,(1);{\MSbar}}_{54} &= \frac{364 N}{3}-\frac{704}{3}+\frac{208}{N}-\frac{320}{N^2}+N_f\left(\frac{176}{3 N}-\frac{136}{3}\right) \nonumber\,,\\
\gamma^{-,(1);{\MSbar}}_{55} &= \frac{343 N^2}{18}-21 N-\frac{188}{9}-\frac{44}{N}+\frac{21}{2 N^2}+N_f\left(-\frac{26 N}{9}+6+\frac{2}{9 N}\right) \nonumber\,.
\end{split}
\end{gather}

\section{Perturbative expansion of RG evolution for $\NF=3$}
\label{app:nf3}

It is well-known that the condition in \req{eq:J1} that determines the leading
non-trivial coefficient in the NLO perturbative expansion of the RG evolution
operator, \req{eq:Wpert}, is ill-behaved for the operators $Q_{2,3}^\pm$ for
$\NF=30$ and, more relevantly, for $\NF=3$ \cite{Ciuchini:1993vr,Ciuchini:1992tj}.
The reason is that, when \req{eq:J1} is written as a linear system, the $4 \times 4$
matrix that multiplies the vector of elements of $J_1$ has zero determinant,
rendering the system indeterminate.

A simple way to understand the anatomy of this problem in greater detail proceeds by
writing the explicit solution to \req{eq:J1} as a function of the parameter $\epsilon=3-\NF$;
if the NLO anomalous dimension matrix in the scheme under consideration is written as
\begin{gather}
\gamma_1^\pm = \frac{1}{(4\pi)^4}\left(
\ba{cc}
g_{22}^\pm & g_{23}^\pm \\
g_{32}^\pm & g_{33}^\pm \\
\ea
\right)
\end{gather}
then one finds
\begin{gather}
J_1^\pm = \frac{1}{\epsilon}\,J_{1,-1}^\pm + J_{1,0}^\pm + \epsilon J_{1,1}^\pm + \cO(\epsilon^2)\,,
\end{gather}
with
\begin{align}
J_{1,-1}^\pm &= \frac{1}{(4\pi)^2}\left(\ba{cc}
0 & \pm\frac{1}{2}(g_{22}^\pm - g_{33}^\pm) - \frac{3}{4}g_{23}^\pm + \frac{1}{3}g_{32}^\pm \\[1.0ex]
0 & 0
\ea\right)\,,\\
\label{eq:J10}
J_{1,0}^\pm &= \frac{1}{(4\pi)^2}\left(\ba{cc}
\frac{1}{162}(128-9g_{22}^\pm \mp 3g_{32}^\pm) & \frac{1}{27}(\pm 128 \mp g_{22}^\pm-g_{32}^\pm \pm g_{33}^\pm)\\[1.0ex]
-\frac{1}{36}g_{32}^\pm & \frac{1}{162}(-1024 \pm 3g_{32}^\pm-9g_{33}^\pm) \\
\ea\right)\,,\\
J_{1,1}^\pm &= \frac{1}{(4\pi)^2}\left(\ba{cc}
\frac{1}{4374}(172+18g_{22}^\pm \pm 9g_{32}^\pm) & \pm \frac{1}{2187}( 516+6g_{22}^\pm \pm 7g_{32}^\pm-6g_{33}^\pm) \\[1.0ex]
\frac{1}{972}g_{32}^\pm & \frac{1}{4374}(-1376 \mp 9g_{32}^\pm+18g_{33}^\pm)\\
\ea\right)\,.
\end{align}
In the limit $\epsilon\to 0$ the element $23$ of $J_1^\pm$ diverges; it is easy to check that the aforementioned
$4 \times 4$ matrix, consistently, has determinant $\propto\epsilon$.
A similar expansion of the 
matrices $\tilde U_{\rm\scriptscriptstyle LO}^\pm(\mu) \equiv [\alphas(\mu)]^{-\frac{\gamma_0^\pm}{2b_0}}$ yields
\begin{gather}
\tilde U_{\rm\scriptscriptstyle LO}^\pm(\mu) = \tilde U_{\rm\scriptscriptstyle LO,0}^\pm(\mu) + \epsilon\tilde U_{\rm\scriptscriptstyle LO,1}^\pm(\mu) + \cO(\epsilon^2)\,,
\end{gather}
with
\begin{align}
\tilde U_{\rm\scriptscriptstyle LO,0}^\pm(\mu) &= \left(\ba{cc}
\alphas^{-1/9}(\mu) & \mp\frac{2}{3}[\alphas^{8/9}(\mu)-\alphas^{-1/9}(\mu)] \\[1.0ex]
0 & \alphas^{8/9}(\mu)\\
\ea\right)\,, \\
\tilde U_{\rm\scriptscriptstyle LO,1}^\pm(\mu) &= \left(\ba{cc}
\frac{2}{243}\alphas^{-1/9}(\mu) & \pm\frac{4}{729}[8\alphas^{8/9}(\mu)+\alphas^{-1/9}(\mu)] \\[1.0ex]
0 & -\frac{16}{243}\alphas^{8/9}(\mu)\\
\ea\right)\log[\alphas(\mu)]\,.
\end{align}
When these expressions are plugged in \req{eq:Utilde}, and the perturbative expansion \req{eq:Wpert}
is used, one gets
\begin{gather}
\begin{split}
\tilde U^\pm(\mu) =&~ \tilde U_{\rm\scriptscriptstyle LO,0}^\pm(\mu) + \gbar^2(\mu)\left[
\frac{1}{\epsilon} \tilde U_{\rm\scriptscriptstyle LO,0}^\pm(\mu) J_{1,-1}^\pm +
\tilde U_{\rm\scriptscriptstyle LO,0}^\pm(\mu) J_{1,0}^\pm +
\tilde U_{\rm\scriptscriptstyle LO,1}^\pm(\mu) J_{1,-1}^\pm + \cO(\epsilon)
\right] \\ &~+ \cO(\gbar^4(\mu))\,,
\end{split}
\end{gather}
which is still divergent as $\epsilon \to 0$. This implies, in particular, that RGI
operators cannot be defined consistently using the above form of the perturbative expansion
for $W$. The RG evolution operator $U(\mu_2,\mu_1)=[\tilde U(\mu_2)]^{-1} \tilde U(\mu_1)$,
on the other hand, is finite: the divergent part has the form
\begin{gather}
\frac{1}{\epsilon}\left\{
\gbar^2(\mu_1)U_{\rm\scriptscriptstyle LO}^\pm(\mu_2,\mu_1) J_{1,-1}^\pm -
\gbar^2(\mu_2) J_{1,-1}^\pm U_{\rm\scriptscriptstyle LO}^\pm(\mu_2,\mu_1)
\right\}
= \frac{\pm\frac{1}{2}(g_{22}^\pm - g_{33}^\pm) - \frac{3}{4}g_{23}^\pm + \frac{1}{3}g_{32}^\pm}{4\pi\epsilon}
\,M\,,
\end{gather}
with
\begin{gather}
M=
U_{\rm\scriptscriptstyle LO}^\pm(\mu_2,\mu_1)
\left(\ba{cc}
0 & \alphas(\mu_2) \\
0 & 0
\ea\right) -
\left(\ba{cc}
0 & \alphas(\mu_1) \\
0 & 0
\ea\right)
U_{\rm\scriptscriptstyle LO}^\pm(\mu_2,\mu_1)
\,,
\end{gather}
and it is easy to check, using the explicit expression for $\tilde U_{\rm\scriptscriptstyle LO,0}^\pm(\mu)$
and the identity
$U_{\rm\scriptscriptstyle LO}^\pm(\mu_2,\mu_1)=[U_{\rm\scriptscriptstyle LO,0}^\pm(\mu_2)]^{-1}U_{\rm\scriptscriptstyle LO,0}^\pm(\mu_1)$,
that $M=0$.\footnote{This is completely analogous e.g. to the discussion leading to Eq.~(53) in~\cite{Kitahara:2016nld}.}
The full expression for $U(\mu_2,\mu_1)$ in the $\epsilon \to 0$ limit still receives contributions from $J_{1,-1}$, via
the products with the $\cO(\epsilon)$ terms in the expansion of $\tilde U_{\rm\scriptscriptstyle LO}$,
which actually give rise to the only dependence of the expanded $U(\mu_2,\mu_1)$ on $\gamma_1^\pm$.

A number of solutions to this problem have been proposed in the literature
\cite{Ciuchini:1993vr,Ciuchini:1992tj,Huber:2005ig,Kitahara:2016nld}, consisting of various
regularisation schemes to treat the singular terms in $3-\NF$. Here we note that the problem
can be entirely bypassed by using the numerical integration of the RG equation in \req{eq:rg_W},
as done in this paper to explore the case $\NF=2$ in detail. Indeed, applying exactly the same
procedure for $\NF=3$ --- i.e., solving \req{eq:rg_W} after having substituted the perturbative
expressions for $\gamma$ and $\beta$ to any prescribed order --- is well-behaved numerically,
which in turn allows to construct both the RG evolution matrix and RGI operators without trouble.
The only point in the procedure where the expansion coefficient $J_1$ may enter explicitly is
the initial condition in \req{eq:in_cond}, where for the $\NF=2$ case we have employed
$W(\mu_0)=\mathbf{1}+\gbar^2(\mu_0)J_1$ at some very high energy scale $\mu_0$. However,
this can be replaced by the initial condition $W(\mu_0)=\mathbf{1}$ at an even higher
scale, thus again avoiding the appearance of any singularity; it turns out that the required
value of $\gbar^2(\mu)$ has to be extremely small, such that the
systematics associated to the choice of coupling for the initial condition is negligible
at the level of the result run down to $\gbar^2(\mu) \sim 2$. This in turn requires using an
expensive numerical integrator to work across several orders of magnitude, which is easy e.g.
using standard Mathematica functions, provided proper care is taken to chose a stable
integrator.

\begin{figure}[t!]
\begin{center}
\includegraphics[width=150mm]{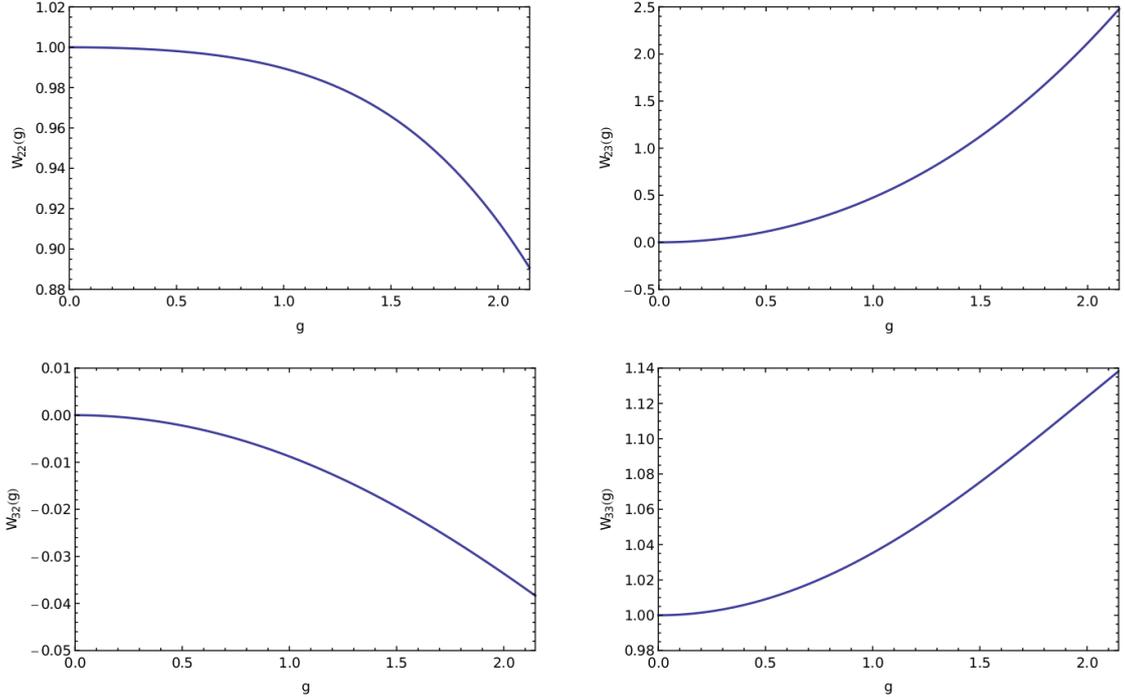}
\end{center}
\vspace{-5mm}
\caption{$W$ as a function of the coupling constant, with $\NF=3$ for operators $O_2$, $O_3$ fierz "$+$", in the $\MSbar$ scheme.}
\label{fig:Wnf3MSbar}
\end{figure}

As a crosscheck of the robustness of our numerical approach we have computed explicitly the function $W(\mu)$ for $\NF=3$,
using our numerical integration and $W=\mathbf{1}$ as an initial condition, set at an extremely small value of the coupling. 
Our result for $W$, displayed in \refig{fig:Wnf3MSbar}, can then be fitted to an ansatz where $J$ is taken to have a polynomial
dependence in $\gbar^2$, to check whether the first coefficient $J_1$ 
is compatible within systematic fit errors (obtained by trying different polynomial orders up to $\mathcal{O}(\gbar^8)$ and coupling 
values for the initial condition) with the one quoted in Eq.(2.30) of \cite{Kitahara:2016nld}. Note that in order to have a direct comparison 
it has to be taken into account that we are using a different relative normalization between operators $O_2$, $O_3$ than the one in \cite{Kitahara:2016nld} 
which translates into a factor $-2$ and $-1/2$ respectively for $[J_1^{\rm fit}]_{23}$, $[J_1^{\rm fit}]_{32}$ and that, 
since we are working with the renormalization constants instead of the Wilson coefficients, the convention used for the $J$ is this work corresponds to $J^T$ in \cite{Kitahara:2016nld} .
 What we obtain is
\begin{gather}
J_1^{\rm fit} = \frac{1}{(4\pi)^2}\left ( \ba{cc}
-1.0470(8) & 70.13(38)\\
-1.39583(1) & 5.78550(8) 
\ea \right )
\end{gather}
which is indeed well-compatible with the above-mentioned result. Note that the coefficient $23$ contains a precise numerical
value of the parameter $t$ employed in \cite{Kitahara:2016nld} to regularise the divergence of $J$ in $3-\NF$.

As a further crosscheck, we have also compared the result of computing the $\NF=2$ evolution with the two possible
initial conditions. The outcome is that, if the value of the coupling at which $W=\mathbf{1}$ is sufficiently small, the
two results are equal up to several significant figures down to values of the coupling $\gbar^2 \gtrsim 2$, where
the hadronic regimes is entered.

\clearpage

\section{Finite parts of RI-MOM renormalisation constants in Landau gauge}
\label{app:finite_RI}

In this appendix we gather the results for the finite part of the one-loop matching coefficients 
$[\cX_O^{(1)}]_{\rm RI;lat}$ between the lattice and the RI-MOM scheme in Landau gauge. They can 
be extracted from~\cite{Constantinou:2010zs} and are given by
\begin{eqnarray}
\lbrack\cX_{22}^{+,(1)}\rbrack_{\rm RI;lat} &=& 0.0272369 \,\icsw^2+0.0485167 \,\icsw-0.294894 \nonumber\,,\\
\lbrack\cX_{23}^{+,(1)}\rbrack_{\rm RI;lat} &=& 0.0218485 \,\icsw^2+0.0632421 \,\icsw+0.0753979 \nonumber\,,\\
\lbrack\cX_{32}^{+,(1)}\rbrack_{\rm RI;lat} &=& 0.00755569 \nonumber\,,\\
\lbrack\cX_{33}^{+,(1)}\rbrack_{\rm RI;lat} &=& -0.00553581 \,\icsw^2-0.0463464 \,\icsw-0.362656 \,,\\
\lbrack\cX_{44}^{+,(1)}\rbrack_{\rm RI;lat} &=& 0.00538842 \,\icsw^2-0.0147254 \,\icsw-0.351294 \nonumber\,,\\
\lbrack\cX_{45}^{+,(1)}\rbrack_{\rm RI;lat} &=& 0.000303451 \,\icsw^2+0.000878362 \,\icsw-0.00178318 \nonumber\,,\\
\lbrack\cX_{54}^{+,(1)}\rbrack_{\rm RI;lat} &=& -0.0728282 \,\icsw^2-0.210807 \,\icsw-0.266293 \nonumber\,,\\
\lbrack\cX_{55}^{+,(1)}\rbrack_{\rm RI;lat} &=& 0.0442301 \,\icsw^2+0.0977049 \,\icsw-0.290267 \nonumber\,.\\
&&\nonumber\\
&&\nonumber\\
\lbrack\cX_{22}^{-,(1)}\rbrack_{\rm RI;lat} &=& 0.0272369 \,\icsw^2+0.0485167 \,\icsw-0.294894 \nonumber\,,\\
\lbrack\cX_{23}^{-,(1)}\rbrack_{\rm RI;lat} &=& -0.0218485 \,\icsw^2-0.0632421 \,\icsw-0.0753979\nonumber\,,\\
\lbrack\cX_{32}^{-,(1)}\rbrack_{\rm RI;lat} &=& -0.00755569\nonumber\,,\\
\lbrack\cX_{33}^{-,(1)}\rbrack_{\rm RI;lat} &=& -0.00553581 \,\icsw^2-0.0463464 \,\icsw-0.362656\,,\\
\lbrack\cX_{44}^{-,(1)}\rbrack_{\rm RI;lat} &=& -0.01646 \,\icsw^2-0.0779674 \,\icsw-0.374019\nonumber\,,\\
\lbrack\cX_{45}^{-,(1)}\rbrack_{\rm RI;lat} &=& -0.00151725 \,\icsw^2-0.00439181 \,\icsw+0.0013602\nonumber\,,\\
\lbrack\cX_{54}^{-,(1)}\rbrack_{\rm RI;lat} &=& 0.0145656 \,\icsw^2+0.0421614 \,\icsw+0.24599\nonumber\,,\\
\lbrack\cX_{55}^{-,(1)}\rbrack_{\rm RI;lat} &=& 0.0223817 \,\icsw^2+0.0344629 \,\icsw-0.257729\nonumber\,.
\end{eqnarray}

\clearpage

\section{Finite parts of SF renormalisation constants}
\label{app:finite}

In this appendix we discuss how to determine the dependence on $a/L$ of the one-loop renormalization constants $Z^{(1)}$ defined in Section~\ref{sec:sf}. The approach is essentially an application of the present context to the techniques discussed in Appendix D of~\cite{Bode:1999sm}. \\
Defining $\ell=L/a$ we will hence consider $F(\ell)=Z^{(1)}$ as a pure function of $\ell=\{\ell_1,\dots,\ell_N\}$. We will also assume that all divergences have been removed from $F(\ell)$, which in general means linear divergences related to the additive renormalisation of quark masses and proportional to the one-loop value of the critical mass $m_{cr}^{(1)}$, and logarithmic divergences proportional to a LO anomalous dimension.
To ensure the robustness of our method we performed separate fits, and we checked, for each ansatz, the fitted value of $\gamma^{(0)}$
was the correct one within the available precision.
First of all a roundoff error has to be assigned to $F(\ell)$, which takes into account the uncertainties coming from the numerical computation itself. Following~\cite{Bode:1999sm}, we choose as an estimate for this error, in the case that the computation has been carried out in double precision, 
\begin{gather}
 \delta F(\ell) \equiv \epsilon(\ell) |F(\ell)| \, ,
 \end{gather}
\begin{gather}
\epsilon(\ell)=\left ( \frac{\ell}{2} \right )^3 \times 10^{-14}.
\end{gather} 
As showed in Section~\ref{sec:sf} the expected behaviour of $F(\ell)$ leads to the consideration of an asymptotic expansion of the form 
\begin{gather}
\label{eq:fit_ansatz}
F(\ell)=\alpha_0+\sum_{k=1}^{n}\frac{1}{\ell^k}(\alpha_k+\beta_k\log \ell)+R_n(\ell) \, ,
\end{gather}  
where the residue $R_n(\ell)$ is expected to decrease faster as $\ell \to \infty$ than any of the terms in the sum. To determine the coefficients $(\alpha_k,\beta_k)$ we minimise a quadratic form in the residues 
\begin{gather}
\label{eq:chi2}
\chi^2=(F-f\xi)^T(F-f\xi)\, ,
\end{gather}
where $F$ and $\xi$ are the $N-$ and $(2n+1)-$column vectors $(F(\ell_1),\dots,F(\ell))^T$ and \\$(\alpha_0,\alpha_1,\dots,\alpha_n,\beta_1,\dots,\beta_n)^T$, respectively, and $f$ is the $N\times(2n+1)$ matrix 
\begin{gather}
f=\begin{pmatrix} 
1 &  \ell_1^{-1} & \cdots & \ell_1^{-n} & \ell_1^{-1}\log \ell_1 & \cdots & \ell_1^{-n}\log \ell_1\\   
1 &  \ell_2^{-1} & \cdots & \ell_2^{-n} & \ell_2^{-1}\log \ell_2 & \cdots & \ell_2^{-n}\log \ell_2\\ 
\vdots & \vdots & \vdots & \vdots & \vdots & \vdots & \vdots \\ 
1 &  \ell_N^{-1} & \cdots & \ell_N^{-n} & \ell_N^{-1}\log \ell_N & \cdots & \ell_N^{-n}\log \ell_N
\end{pmatrix}
\end{gather}
Again following~\cite{Bode:1999sm}, we have not introduced a matrix of weights in the definition of $\chi^2$. A necessary condition to minimise $\chi^2$ is 
\begin{gather}
\label{eq:chi_min}
f\xi=PF
\end{gather}
where we have assumed that the columns of $f$ are linearly independent vectors (assuming $2n+1 \ll N$), and $P$ is the projector onto the subspace of $\mathbb{R}^N$ generated by them. Eq.~(\ref{eq:chi_min}) can be solved using the singular value decomposition of $f$, which has the form of 
\begin{gather}
f=USV^T
\end{gather}
where $U$ is an $N \times (2n+1)$ matrix such that
\begin{gather}
U^TU=\mathbf{1}\quad ,\quad UU^T=P
\end{gather}
$S$ is diagonal, and $(2n+1)\times(2n+1)$ matrix $V$ is orthonormal. With this decomposition one has
\begin{gather}
\xi=VS^{-1}U^TF \, .
\end{gather}
Finally, the uncertainty in the result for $\xi_k$ can be modelled using error propagation as
\begin{gather}
\delta \xi_k^2=\sum_{l=1}^N(VS^{-1}U^T)_{kl}^2(\delta F)_l^2 \, ,
\end{gather}
where $(\delta F)_k \equiv \delta F(\ell_k)$.\\
As a remark on the above method regarding practical applications, it has to be pointed out that the choice of Eq.~(\ref{eq:chi2}) for the quadratic form $\chi^2$ implies, in particular, that small values of $\ell$ might be given excessive weight. This problem has been dealt with by considering a range $[\ell_{min},\ell_{max}]$ with changing $\ell_{min}$. For this work the better convergence in results for $(\alpha_k,\beta_k)$ was given by $\ell_{min}=16$ and $\ell_{max}=46$.
The estimation of systematic uncertainty of the fitting procedure has be performed using the proposal by the authors of~\cite{Bode:1999sm}. We considered two independent fits at order $n$ and $n+1$, i.e. extending the Ansatz in Eq.~(\ref{eq:fit_ansatz}) by terms $1/\ell^{n+1}$ and $\log \ell / \ell^{n+1}$ with coefficients $\alpha_{n+1}$ and $\beta_{n+1}$ respectively. The systematic uncertainty of the finite part $r_0=\alpha_0$ is defined as the difference of the value of the parameter $\alpha_0$ extracted by the two different fits. In the present work we have used $n=2$ in the fit Ansatz for the $\mathcal{O}$(a)-improved data, and $n=3$ for unimproved ones.

\begin{center}
\begin{table}[h!]
\noindent\begin{center}\begin{tabular}{|c|c|cc|}\hline
 \hline
$\alpha$ & $s$ & $(r_0)_{23}^{+}(c_{sw}=0)$ & $(r_0)_{23}^{-}(c_{sw}=0)$ \\ 
 \hline
\multirow{12}{*}{0} & $1$ & $\begin{pmatrix}-0.2973(1) & 0.12889(6) \\0.02613(1) & -0.20350(10)\end{pmatrix}$ & $\begin{pmatrix}-0.3055(1) & 0.008223(4) \\-0.02778(1) & -0.19359(9)\end{pmatrix}$ \\
& $2$ & $\begin{pmatrix}-0.3027(1) & 0.13105(6) \\0.02322(1) & -0.20234(9)\end{pmatrix}$ & $\begin{pmatrix}-0.3212(2) & 0.03063(1) \\-0.03590(2) & -0.18199(8)\end{pmatrix}$ \\
& $3$ & $\begin{pmatrix}-0.3172(1) & 0.13685(7) \\0.03615(2) & -0.20751(10)\end{pmatrix}$ & $\begin{pmatrix}-0.3252(2) & 0.03643(2) \\-0.02962(1) & -0.19096(9)\end{pmatrix}$ \\
& $4$ & $\begin{pmatrix}-0.2991(1) & 0.11812(6) \\0.03093(1) & -0.17471(8)\end{pmatrix}$ & $\begin{pmatrix}-0.3104(1) & 0.03794(2) \\-0.03310(2) & -0.16164(8)\end{pmatrix}$ \\
& $5$ & $\begin{pmatrix}-0.3045(1) & 0.12028(6) \\0.02802(1) & -0.17355(8)\end{pmatrix}$ & $\begin{pmatrix}-0.3261(2) & 0.06035(3) \\-0.04123(2) & -0.15004(7)\end{pmatrix}$ \\
& $6$ & $\begin{pmatrix}-0.3190(2) & 0.12608(6) \\0.04095(2) & -0.17872(8)\end{pmatrix}$ & $\begin{pmatrix}-0.3302(2) & 0.06615(3) \\-0.03494(2) & -0.15901(7)\end{pmatrix}$ \\
\hline
\multirow{12}{*}{3/2} &$1$ & $\begin{pmatrix}-0.3100(1) & 0.12889(6) \\0.02613(1) & -0.2161(1)\end{pmatrix}$ & $\begin{pmatrix}-0.3181(2) & 0.008223(4) \\-0.02778(1) & -0.20623(10)\end{pmatrix}$ \\
& $2$ & $\begin{pmatrix}-0.3154(1) & 0.13105(6) \\0.02322(1) & -0.2150(1)\end{pmatrix}$ & $\begin{pmatrix}-0.3338(2) & 0.03063(1) \\-0.03590(2) & -0.19462(9)\end{pmatrix}$ \\
& $3$ & $\begin{pmatrix}-0.3299(2) & 0.13685(7) \\0.03615(2) & -0.2201(1)\end{pmatrix}$ & $\begin{pmatrix}-0.3379(2) & 0.03643(2) \\-0.02962(1) & -0.20360(9)\end{pmatrix}$ \\
& $4$ & $\begin{pmatrix}-0.3118(1) & 0.11812(6) \\0.03093(1) & -0.18734(9)\end{pmatrix}$ & $\begin{pmatrix}-0.3231(2) & 0.03794(2) \\-0.03310(2) & -0.17428(8)\end{pmatrix}$ \\
& $5$ & $\begin{pmatrix}-0.3172(1) & 0.12028(6) \\0.02802(1) & -0.18618(9)\end{pmatrix}$ & $\begin{pmatrix}-0.3387(2) & 0.06035(3) \\-0.04123(2) & -0.16267(8)\end{pmatrix}$ \\
& $6$ & $\begin{pmatrix}-0.3317(2) & 0.12608(6) \\0.04095(2) & -0.19135(9)\end{pmatrix}$ & $\begin{pmatrix}-0.3428(2) & 0.06615(3) \\-0.03494(2) & -0.17165(8)\end{pmatrix}$ \\
\hline
\multirow{12}{*}{1} & $1$ & $\begin{pmatrix}-0.3057(1) & 0.12889(6) \\0.02613(1) & -0.21192(10)\end{pmatrix}$ & $\begin{pmatrix}-0.3139(1) & 0.008223(4) \\-0.02778(1) & -0.20202(9)\end{pmatrix}$ \\
& $2$ & $\begin{pmatrix}-0.3111(1) & 0.13105(6) \\0.02322(1) & -0.21076(10)\end{pmatrix}$ & $\begin{pmatrix}-0.3296(2) & 0.03063(1) \\-0.03590(2) & -0.19041(9)\end{pmatrix}$ \\
& $3$ & $\begin{pmatrix}-0.3257(2) & 0.13685(7) \\0.03615(2) & -0.2159(1)\end{pmatrix}$ & $\begin{pmatrix}-0.3336(2) & 0.03643(2) \\-0.02962(1) & -0.19938(9)\end{pmatrix}$ \\
& $4$ & $\begin{pmatrix}-0.3075(1) & 0.11812(6) \\0.03093(1) & -0.18313(9)\end{pmatrix}$ & $\begin{pmatrix}-0.3188(2) & 0.03794(2) \\-0.03310(2) & -0.17007(8)\end{pmatrix}$ \\
& $5$ & $\begin{pmatrix}-0.3129(1) & 0.12028(6) \\0.02802(1) & -0.18197(9)\end{pmatrix}$ & $\begin{pmatrix}-0.3345(2) & 0.06035(3) \\-0.04123(2) & -0.15846(7)\end{pmatrix}$ \\
& $6$ & $\begin{pmatrix}-0.3275(2) & 0.12608(6) \\0.04095(2) & -0.18714(9)\end{pmatrix}$ & $\begin{pmatrix}-0.3386(2) & 0.06615(3) \\-0.03494(2) & -0.16744(8)\end{pmatrix}$ \\
\hline\end{tabular}\\
\end{center}
\caption{Numerical results of the 1-loop finite parts for operators $\text{VA-AV,PS-SP}$ in the $18$ \SF~renormalisation schemes under investigation defined by the source $s$ and the parameter $\alpha$ as in Eq.~(\ref{eq:Z1loop_mix}). }
\label{tab:r_op23_csw0}
\end{table}
\end{center}

\begin{center}
\begin{table}[h!]
\noindent\begin{center}\begin{tabular}{|c|c|cc|}\hline
 \hline
$\alpha$ & $s$ & $(r_0)_{23}^{+}(c_{sw}=1)$ & $(r_0)_{23}^{-}(c_{sw}=1)$ \\ 
 \hline
\multirow{12}{*}{0} & $1$ & $\begin{pmatrix}-0.22165(6) & 0.21392(6) \\0.026133(6) & -0.25536(2)\end{pmatrix}$ & $\begin{pmatrix}-0.22981(8) & -0.0767(1) \\-0.027786(8) & -0.24544(3)\end{pmatrix}$ \\
& $2$ & $\begin{pmatrix}-0.22703(5) & 0.21608(6) \\0.02324(1) & -0.25420(2)\end{pmatrix}$ & $\begin{pmatrix}-0.24545(4) & -0.05439(7) \\-0.035896(8) & -0.233856(10)\end{pmatrix}$ \\
& $3$ & $\begin{pmatrix}-0.24151(2) & 0.22187(7) \\0.03613(2) & -0.25936(3)\end{pmatrix}$ & $\begin{pmatrix}-0.24950(3) & -0.04859(6) \\-0.029622(1) & -0.24282(2)\end{pmatrix}$ \\
& $4$ & $\begin{pmatrix}-0.22344(6) & 0.20317(4) \\0.030919(6) & -0.22664(5)\end{pmatrix}$ & $\begin{pmatrix}-0.23475(7) & -0.04710(4) \\-0.033097(6) & -0.21357(5)\end{pmatrix}$ \\
& $5$ & $\begin{pmatrix}-0.22882(4) & 0.20532(5) \\0.0280232(1) & -0.22548(5)\end{pmatrix}$ & $\begin{pmatrix}-0.25039(3) & -0.02475(2) \\-0.04121(2) & -0.20199(7)\end{pmatrix}$ \\
& $6$ & $\begin{pmatrix}-0.24330(1) & 0.21111(6) \\0.04092(3) & -0.23064(4)\end{pmatrix}$ & $\begin{pmatrix}-0.25444(2) & -0.01896(3) \\-0.03493(1) & -0.21095(6)\end{pmatrix}$ \\
\hline
\multirow{12}{*}{3/2} & $1$ & $\begin{pmatrix}-0.23423(2) & 0.21392(6) \\0.026133(6) & -0.26795(7)\end{pmatrix}$ & $\begin{pmatrix}-0.24239(3) & -0.0767(1) \\-0.027786(8) & -0.25803(8)\end{pmatrix}$ \\
& $2$ & $\begin{pmatrix}-0.239614(2) & 0.21608(6) \\0.02324(1) & -0.26679(7)\end{pmatrix}$ & $\begin{pmatrix}-0.258036(7) & -0.05439(7) \\-0.035896(8) & -0.24644(6)\end{pmatrix}$ \\
& $3$ & $\begin{pmatrix}-0.25409(3) & 0.22187(7) \\0.03613(2) & -0.27195(8)\end{pmatrix}$ & $\begin{pmatrix}-0.26209(1) & -0.04859(6) \\-0.029622(1) & -0.25541(7)\end{pmatrix}$ \\
& $4$ & $\begin{pmatrix}-0.23602(1) & 0.20317(4) \\0.030919(6) & -0.239229(3)\end{pmatrix}$ & $\begin{pmatrix}-0.24733(2) & -0.04710(4) \\-0.033097(6) & -0.226161(3)\end{pmatrix}$ \\
& $5$ & $\begin{pmatrix}-0.241406(4) & 0.20532(5) \\0.0280232(1) & -0.238070(6)\end{pmatrix}$ & $\begin{pmatrix}-0.26298(2) & -0.02475(2) \\-0.04121(2) & -0.21458(3)\end{pmatrix}$ \\
& $6$ & $\begin{pmatrix}-0.25589(3) & 0.21111(6) \\0.04092(3) & -0.243229(6)\end{pmatrix}$ & $\begin{pmatrix}-0.26703(3) & -0.01896(3) \\-0.03493(1) & -0.22354(1)\end{pmatrix}$ \\
\hline
\multirow{12}{*}{1} & $1$ & $\begin{pmatrix}-0.23004(3) & 0.21392(6) \\0.026133(6) & -0.26375(5)\end{pmatrix}$ & $\begin{pmatrix}-0.23820(5) & -0.0767(1) \\-0.027786(8) & -0.25383(6)\end{pmatrix}$ \\
& $2$ & $\begin{pmatrix}-0.23542(1) & 0.21608(6) \\0.02324(1) & -0.26259(5)\end{pmatrix}$ & $\begin{pmatrix}-0.253840(9) & -0.05439(7) \\-0.035896(8) & -0.24225(4)\end{pmatrix}$ \\
& $3$ & $\begin{pmatrix}-0.24990(2) & 0.22187(7) \\0.03613(2) & -0.26775(6)\end{pmatrix}$ & $\begin{pmatrix}-0.2578952(9) & -0.04859(6) \\-0.029622(1) & -0.25121(6)\end{pmatrix}$ \\
& $4$ & $\begin{pmatrix}-0.23183(3) & 0.20317(4) \\0.030919(6) & -0.23503(2)\end{pmatrix}$ & $\begin{pmatrix}-0.24314(4) & -0.04710(4) \\-0.033097(6) & -0.22197(2)\end{pmatrix}$ \\
& $5$ & $\begin{pmatrix}-0.23721(1) & 0.20532(5) \\0.0280232(1) & -0.23387(2)\end{pmatrix}$ & $\begin{pmatrix}-0.258780(6) & -0.02475(2) \\-0.04121(2) & -0.21038(4)\end{pmatrix}$ \\
& $6$ & $\begin{pmatrix}-0.25169(2) & 0.21111(6) \\0.04092(3) & -0.23903(1)\end{pmatrix}$ & $\begin{pmatrix}-0.26283(1) & -0.01896(3) \\-0.03493(1) & -0.21934(3)\end{pmatrix}$ \\
\hline\end{tabular}\\
\end{center}
\caption{Numerical results of the 1-loop finite parts for operators $\text{VA-AV,PS-SP}$ in the $18$ \SF~renormalisation schemes under investigation defined by the source $s$ and the parameter $\alpha$ as in Eq.~(\ref{eq:Z1loop_mix}). These results have been computed including the clover term in the fermonic action. }
\label{tab:r_op23_csw1}
\end{table}
\end{center}

\begin{center}
\begin{table}[h!]
\noindent\begin{center}\begin{tabular}{|c|c|cc|}\hline
 \hline
$\alpha$ & $s$ & $(r_0)_{45}^{+}(c_{sw}=0)$ & $(r_0)_{45}^{-}(c_{sw}=0)$ \\ 
 \hline
\multirow{12}{*}{0} & $1$ & $\begin{pmatrix}-0.20786(10) & -0.008176(4) \\-0.16835(8) & -0.3844(2)\end{pmatrix}$ & $\begin{pmatrix}-0.18729(9) & 0.012475(6) \\0.3886(2) & -0.2278(1)\end{pmatrix}$ \\
&$2$ & $\begin{pmatrix}-0.20612(10) & -0.008902(4) \\-0.15539(8) & -0.3898(2)\end{pmatrix}$ & $\begin{pmatrix}-0.19077(9) & 0.010444(5) \\0.3826(2) & -0.2313(1)\end{pmatrix}$ \\
&$3$ & $\begin{pmatrix}-0.20882(10) & -0.007780(4) \\-0.15920(8) & -0.3882(2)\end{pmatrix}$ & $\begin{pmatrix}-0.18302(9) & 0.014967(7) \\0.3788(2) & -0.2335(1)\end{pmatrix}$ \\
&$4$ & $\begin{pmatrix}-0.18240(8) & -0.009086(4) \\-0.11374(6) & -0.3863(2)\end{pmatrix}$ & $\begin{pmatrix}-0.15486(7) & 0.014097(7) \\0.3496(2) & -0.2298(1)\end{pmatrix}$ \\
&$5$ & $\begin{pmatrix}-0.18066(8) & -0.009811(5) \\-0.10078(5) & -0.3917(2)\end{pmatrix}$ & $\begin{pmatrix}-0.15835(7) & 0.012065(6) \\0.3436(2) & -0.2333(1)\end{pmatrix}$ \\
&$6$ & $\begin{pmatrix}-0.18335(9) & -0.008689(4) \\-0.10459(5) & -0.3901(2)\end{pmatrix}$ & $\begin{pmatrix}-0.15059(7) & 0.016589(8) \\0.3398(2) & -0.2355(1)\end{pmatrix}$ \\

\hline
\multirow{12}{*}{3/2} & $1$ & $\begin{pmatrix}-0.2205(1) & -0.008176(4) \\-0.16835(8) & -0.3970(2)\end{pmatrix}$ & $\begin{pmatrix}-0.19992(9) & 0.012475(6) \\0.3886(2) & -0.2404(1)\end{pmatrix}$ \\
&$2$ & $\begin{pmatrix}-0.2188(1) & -0.008902(4) \\-0.15539(8) & -0.4024(2)\end{pmatrix}$ & $\begin{pmatrix}-0.20341(10) & 0.010444(5) \\0.3826(2) & -0.2439(1)\end{pmatrix}$ \\
&$3$ & $\begin{pmatrix}-0.2214(1) & -0.007780(4) \\-0.15920(8) & -0.4008(2)\end{pmatrix}$ & $\begin{pmatrix}-0.19565(9) & 0.014967(7) \\0.3788(2) & -0.2462(1)\end{pmatrix}$ \\
&$4$ & $\begin{pmatrix}-0.19503(9) & -0.009086(4) \\-0.11374(6) & -0.3989(2)\end{pmatrix}$ & $\begin{pmatrix}-0.16750(8) & 0.014097(7) \\0.3496(2) & -0.2424(1)\end{pmatrix}$ \\
&$5$ & $\begin{pmatrix}-0.19329(9) & -0.009811(5) \\-0.10078(5) & -0.4043(2)\end{pmatrix}$ & $\begin{pmatrix}-0.17098(8) & 0.012065(6) \\0.3436(2) & -0.2459(1)\end{pmatrix}$ \\
&$6$ & $\begin{pmatrix}-0.19599(9) & -0.008689(4) \\-0.10459(5) & -0.4028(2)\end{pmatrix}$ & $\begin{pmatrix}-0.16322(8) & 0.016589(8) \\0.3398(2) & -0.2481(1)\end{pmatrix}$ \\

\hline
\multirow{12}{*}{1} & $1$ & $\begin{pmatrix}-0.2163(1) & -0.008176(4) \\-0.16835(8) & -0.3928(2)\end{pmatrix}$ & $\begin{pmatrix}-0.19571(9) & 0.012475(6) \\0.3886(2) & -0.2362(1)\end{pmatrix}$ \\
&$2$ & $\begin{pmatrix}-0.2145(1) & -0.008902(4) \\-0.15539(8) & -0.3982(2)\end{pmatrix}$ & $\begin{pmatrix}-0.19920(9) & 0.010444(5) \\0.3826(2) & -0.2397(1)\end{pmatrix}$ \\
&$3$ & $\begin{pmatrix}-0.2172(1) & -0.007780(4) \\-0.15920(8) & -0.3966(2)\end{pmatrix}$ & $\begin{pmatrix}-0.19144(9) & 0.014967(7) \\0.3788(2) & -0.2419(1)\end{pmatrix}$ \\
&$4$ & $\begin{pmatrix}-0.19082(9) & -0.009086(4) \\-0.11374(6) & -0.3947(2)\end{pmatrix}$ & $\begin{pmatrix}-0.16329(8) & 0.014097(7) \\0.3496(2) & -0.2382(1)\end{pmatrix}$ \\
&$5$ & $\begin{pmatrix}-0.18908(9) & -0.009811(5) \\-0.10078(5) & -0.4001(2)\end{pmatrix}$ & $\begin{pmatrix}-0.16677(8) & 0.012065(6) \\0.3436(2) & -0.2417(1)\end{pmatrix}$ \\
&$6$ & $\begin{pmatrix}-0.19177(9) & -0.008689(4) \\-0.10459(5) & -0.3985(2)\end{pmatrix}$ & $\begin{pmatrix}-0.15901(7) & 0.016589(8) \\0.3398(2) & -0.2439(1)\end{pmatrix}$ \\
\hline\end{tabular}\\
\end{center}
\caption{Numerical results of the 1-loop finite parts for operators  $\text{PS+SP,T$\tilde{\text{T}}$}$  in the $18$ \SF~renormalisation schemes under investigation defined by the source $s$ and the parameter $\alpha$ as in Eq.~(\ref{eq:Z1loop_mix}). }
\label{tab:r_op45_csw0}
\end{table}
\end{center}

\begin{center}
\begin{table}[h!]
\noindent\begin{center}\begin{tabular}{|c|c|cc|}\hline
 \hline
$\alpha$ & $s$ & $(r_0)_{45}^{+}(c_{sw}=1)$ & $(r_0)_{45}^{-}(c_{sw}=1)$ \\ 
 \hline
\multirow{12}{*}{0} & $1$ & $\begin{pmatrix}-0.21719(1) & -0.0069948(4) \\-0.4517(3) & -0.24249(6)\end{pmatrix}$ & $\begin{pmatrix}-0.28168(3) & 0.006568(2) \\0.4451(2) & -0.17105(8)\end{pmatrix}$ \\
& $2$ & $\begin{pmatrix}-0.215453(7) & -0.007719(1) \\-0.4388(2) & -0.24787(4)\end{pmatrix}$ & $\begin{pmatrix}-0.28516(4) & 0.004541(6) \\0.4392(2) & -0.17453(6)\end{pmatrix}$ \\
& $3$ & $\begin{pmatrix}-0.21814(1) & -0.0065985(6) \\-0.4426(2) & -0.24629(4)\end{pmatrix}$ & $\begin{pmatrix}-0.27742(2) & 0.009054(4) \\0.4354(2) & -0.17674(6)\end{pmatrix}$ \\
& $4$ & $\begin{pmatrix}-0.19179(6) & -0.007902(2) \\-0.3972(1) & -0.24444(5)\end{pmatrix}$ & $\begin{pmatrix}-0.24934(5) & 0.008185(2) \\0.4062(1) & -0.17299(7)\end{pmatrix}$ \\
& $5$ & $\begin{pmatrix}-0.19006(6) & -0.008626(4) \\-0.38433(7) & -0.24982(3)\end{pmatrix}$ & $\begin{pmatrix}-0.25281(4) & 0.006158(2) \\0.40026(9) & -0.17647(6)\end{pmatrix}$ \\
& $6$ & $\begin{pmatrix}-0.19275(6) & -0.007506(2) \\-0.38813(7) & -0.24823(3)\end{pmatrix}$ & $\begin{pmatrix}-0.24507(6) & 0.010672(8) \\0.39646(9) & -0.17869(6)\end{pmatrix}$ \\
\hline
\multirow{12}{*}{3/2} & $1$ & $\begin{pmatrix}-0.22978(6) & -0.0069948(4) \\-0.4517(3) & -0.255079(8)\end{pmatrix}$ & $\begin{pmatrix}-0.29427(8) & 0.006568(2) \\0.4451(2) & -0.18363(3)\end{pmatrix}$ \\
& $2$ & $\begin{pmatrix}-0.22804(6) & -0.007719(1) \\-0.4388(2) & -0.260460(9)\end{pmatrix}$ & $\begin{pmatrix}-0.29775(9) & 0.004541(6) \\0.4392(2) & -0.18711(1)\end{pmatrix}$ \\
& $3$ & $\begin{pmatrix}-0.23073(6) & -0.0065985(6) \\-0.4426(2) & -0.258875(8)\end{pmatrix}$ & $\begin{pmatrix}-0.29001(7) & 0.009054(4) \\0.4354(2) & -0.18933(1)\end{pmatrix}$ \\
& $4$ & $\begin{pmatrix}-0.204380(9) & -0.007902(2) \\-0.3972(1) & -0.257024(3)\end{pmatrix}$ & $\begin{pmatrix}-0.2619224(2) & 0.008185(2) \\0.4062(1) & -0.18558(3)\end{pmatrix}$ \\
& $5$ & $\begin{pmatrix}-0.20264(1) & -0.008626(4) \\-0.38433(7) & -0.26241(1)\end{pmatrix}$ & $\begin{pmatrix}-0.265398(7) & 0.006158(2) \\0.40026(9) & -0.189057(9)\end{pmatrix}$ \\
& $6$ & $\begin{pmatrix}-0.205331(9) & -0.007506(2) \\-0.38813(7) & -0.26082(1)\end{pmatrix}$ & $\begin{pmatrix}-0.257660(10) & 0.010672(8) \\0.39646(9) & -0.191276(8)\end{pmatrix}$ \\
\hline
\multirow{12}{*}{1} & $1$ & $\begin{pmatrix}-0.22558(4) & -0.0069948(4) \\-0.4517(3) & -0.25088(2)\end{pmatrix}$ & $\begin{pmatrix}-0.29007(7) & 0.006568(2) \\0.4451(2) & -0.17944(5)\end{pmatrix}$ \\
& $2$ & $\begin{pmatrix}-0.22384(4) & -0.007719(1) \\-0.4388(2) & -0.256264(6)\end{pmatrix}$ & $\begin{pmatrix}-0.29355(7) & 0.004541(6) \\0.4392(2) & -0.18292(3)\end{pmatrix}$ \\
& $3$ & $\begin{pmatrix}-0.22653(4) & -0.0065985(6) \\-0.4426(2) & -0.254679(7)\end{pmatrix}$ & $\begin{pmatrix}-0.28581(6) & 0.009054(4) \\0.4354(2) & -0.18514(3)\end{pmatrix}$ \\
& $4$ & $\begin{pmatrix}-0.20018(2) & -0.007902(2) \\-0.3972(1) & -0.25283(2)\end{pmatrix}$ & $\begin{pmatrix}-0.25773(2) & 0.008185(2) \\0.4062(1) & -0.18138(4)\end{pmatrix}$ \\
& $5$ & $\begin{pmatrix}-0.19845(3) & -0.008626(4) \\-0.38433(7) & -0.258210(1)\end{pmatrix}$ & $\begin{pmatrix}-0.261202(9) & 0.006158(2) \\0.40026(9) & -0.18486(2)\end{pmatrix}$ \\
& $6$ & $\begin{pmatrix}-0.20114(2) & -0.007506(2) \\-0.38813(7) & -0.256624(2)\end{pmatrix}$ & $\begin{pmatrix}-0.25346(3) & 0.010672(8) \\0.39646(9) & -0.18708(2)\end{pmatrix}$ \\
\hline\end{tabular}\\
\end{center}
\caption{Numerical results of the 1-loop finite parts for operators $\text{PS+SP,T$\tilde{\text{T}}$}$ in the $18$ \SF~renormalisation schemes under investigation defined by the source $s$ and the parameter $\alpha$ as in Eq.~(\ref{eq:Z1loop_mix}). These results have been computed including the clover term in the fermonic action. }
\label{tab:r_op45_csw1}
\end{table}
\end{center}

\cleardoublepage
\section{NLO anomalous dimension matrices in SF schemes}
\label{app:nload}

\begin{center}
\noindent\begin{tabular}{|c|c|c|}
\hline
$\alpha$ & $s$ & ${\gamma^{(1)}}_{23}^{+}$ \\ 
 \hline
\multirow{12}{*}{0} & $1$ & $\begin{pmatrix}0.001519(10) + N_f[-0.000057850(2)] & 0.00983(2) + N_f[-0.00034710(1)] \\0.006188(1) + N_f[-0.000080203] & -0.006776(8) + N_f[-0.00001842(2)]\end{pmatrix}$ \\
& $2$ & $\begin{pmatrix}0.001080(8) + N_f[-0.000057850(2)] & 0.00936(2) + N_f[-0.00034710(1)] \\0.005504(3) + N_f[-0.000080203] & -0.006855(7) + N_f[-0.00001842(2)]\end{pmatrix}$ \\
& $3$ & $\begin{pmatrix}-0.001673(4) + N_f[-0.000057850(2)] & 0.00870(2) + N_f[-0.00034710(1)] \\0.008552(4) + N_f[-0.000080203] & -0.00651(1) + N_f[-0.00001842(2)]\end{pmatrix}$ \\
& $4$ & $\begin{pmatrix}0.000936(10) + N_f[-0.000057850(2)] & 0.00743(2) + N_f[-0.00034710(1)] \\0.007320(1) + N_f[-0.000080203] & -0.00290(2) + N_f[-0.00001842(2)]\end{pmatrix}$ \\
& $5$ & $\begin{pmatrix}0.000497(6) + N_f[-0.000057850(2)] & 0.00695(2) + N_f[-0.00034710(1)] \\0.0066351(1) + N_f[-0.000080203] & -0.00297(2) + N_f[-0.00001842(2)]\end{pmatrix}$ \\
& $6$ & $\begin{pmatrix}-0.002256(4) + N_f[-0.000057850(2)] & 0.00629(2) + N_f[-0.00034710(1)] \\0.009684(7) + N_f[-0.000080203] & -0.00263(2) + N_f[-0.00001842(2)]\end{pmatrix}$ \\
\hline
\multirow{12}{*}{3/2} & $1$ & $\begin{pmatrix}-0.000022(3) + N_f[-0.000057850(2)] & 0.00983(2) + N_f[-0.00034710(1)] \\0.006188(1) + N_f[-0.000080203] & -0.00832(2) + N_f[-0.00001842(2)]\end{pmatrix}$ \\
& $2$ & $\begin{pmatrix}-0.000461(1) + N_f[-0.000057850(2)] & 0.00936(2) + N_f[-0.00034710(1)] \\0.005504(3) + N_f[-0.000080203] & -0.00840(2) + N_f[-0.00001842(2)]\end{pmatrix}$ \\
& $3$ & $\begin{pmatrix}-0.003214(6) + N_f[-0.000057850(2)] & 0.00870(3) + N_f[-0.00034710(1)] \\0.008552(4) + N_f[-0.000080203] & -0.00805(3) + N_f[-0.00001842(2)]\end{pmatrix}$ \\
& $4$ & $\begin{pmatrix}-0.000605(3) + N_f[-0.000057850(2)] & 0.00743(1) + N_f[-0.00034710(1)] \\0.007320(1) + N_f[-0.000080203] & -0.004438(2) + N_f[-0.00001842(2)]\end{pmatrix}$ \\
& $5$ & $\begin{pmatrix}-0.0010440(6) + N_f[-0.000057850(2)] & 0.00695(1) + N_f[-0.00034710(1)] \\0.0066351(1) + N_f[-0.000080203] & -0.004516(2) + N_f[-0.00001842(2)]\end{pmatrix}$ \\
& $6$ & $\begin{pmatrix}-0.003797(7) + N_f[-0.000057850(2)] & 0.00629(2) + N_f[-0.00034710(1)] \\0.009684(7) + N_f[-0.000080203] & -0.004167(4) + N_f[-0.00001842(2)]\end{pmatrix}$ \\
\hline
\multirow{12}{*}{1} & $1$ & $\begin{pmatrix}0.000492(5) + N_f[-0.000057850(2)] & 0.00983(2) + N_f[-0.00034710(1)] \\0.006188(1) + N_f[-0.000080203] & -0.00780(2) + N_f[-0.00001842(2)]\end{pmatrix}$ \\
& $2$ & $\begin{pmatrix}0.000053(3) + N_f[-0.000057850(2)] & 0.00936(2) + N_f[-0.00034710(1)] \\0.005504(3) + N_f[-0.000080203] & -0.00788(2) + N_f[-0.00001842(2)]\end{pmatrix}$ \\
& $3$ & $\begin{pmatrix}-0.002700(4) + N_f[-0.000057850(2)] & 0.00870(2) + N_f[-0.00034710(1)] \\0.008552(4) + N_f[-0.000080203] & -0.00753(2) + N_f[-0.00001842(2)]\end{pmatrix}$ \\
& $4$ & $\begin{pmatrix}-0.000092(5) + N_f[-0.000057850(2)] & 0.00743(1) + N_f[-0.00034710(1)] \\0.007320(1) + N_f[-0.000080203] & -0.003924(7) + N_f[-0.00001842(2)]\end{pmatrix}$ \\
& $5$ & $\begin{pmatrix}-0.000530(2) + N_f[-0.000057850(2)] & 0.00695(1) + N_f[-0.00034710(1)] \\0.0066351(1) + N_f[-0.000080203] & -0.004002(7) + N_f[-0.00001842(2)]\end{pmatrix}$ \\
& $6$ & $\begin{pmatrix}-0.003283(5) + N_f[-0.000057850(2)] & 0.00629(2) + N_f[-0.00034710(1)] \\0.009684(7) + N_f[-0.000080203] & -0.003654(6) + N_f[-0.00001842(2)]\end{pmatrix}$ \\
\hline\end{tabular}\\
\end{center}


\begin{center}
\noindent\begin{tabular}{|c|c|c|}
\hline
$\alpha$ & $s$ & ${\gamma^{(1)}}_{23}^{-}$ \\ 
 \hline
\multirow{12}{*}{0} & $1$ & $\begin{pmatrix}0.00051(1) + N_f[-0.000118203(2)] & -0.00803(4) + N_f[0.00070922(1)] \\-0.006546(2) + N_f[0.000063796] & -0.00660(1) + N_f[0.00056285(2)]\end{pmatrix}$ \\
& $2$ & $\begin{pmatrix}-0.002017(7) + N_f[-0.000118203(2)] & -0.00577(2) + N_f[0.00070922(1)] \\-0.008464(2) + N_f[0.000063796] & -0.004564(4) + N_f[0.00056285(2)]\end{pmatrix}$ \\
& $3$ & $\begin{pmatrix}-0.002036(5) + N_f[-0.000118203(2)] & -0.00609(2) + N_f[0.00070922(1)] \\-0.0069803(3) + N_f[0.000063796] & -0.006138(8) + N_f[0.00056285(2)]\end{pmatrix}$ \\
& $4$ & $\begin{pmatrix}-0.00049(1) + N_f[-0.000118203(2)] & -0.00498(2) + N_f[0.00070922(1)] \\-0.007802(1) + N_f[0.000063796] & -0.00229(2) + N_f[0.00056285(2)]\end{pmatrix}$ \\
& $5$ & $\begin{pmatrix}-0.003025(6) + N_f[-0.000118203(2)] & -0.00272(1) + N_f[0.00070922(1)] \\-0.009719(5) + N_f[0.000063796] & -0.00026(3) + N_f[0.00056285(2)]\end{pmatrix}$ \\
& $6$ & $\begin{pmatrix}-0.003045(4) + N_f[-0.000118203(2)] & -0.00305(1) + N_f[0.00070922(1)] \\-0.008236(3) + N_f[0.000063796] & -0.00183(2) + N_f[0.00056285(2)]\end{pmatrix}$ \\
\hline
\multirow{12}{*}{3/2} & $1$ & $\begin{pmatrix}-0.001026(6) + N_f[-0.000118203(2)] & -0.00803(4) + N_f[0.00070922(1)] \\-0.006546(2) + N_f[0.000063796] & -0.00814(3) + N_f[0.00056285(2)]\end{pmatrix}$ \\
& $2$ & $\begin{pmatrix}-0.003558(2) + N_f[-0.000118203(2)] & -0.00577(2) + N_f[0.00070922(1)] \\-0.008464(2) + N_f[0.000063796] & -0.00611(2) + N_f[0.00056285(2)]\end{pmatrix}$ \\
& $3$ & $\begin{pmatrix}-0.003577(2) + N_f[-0.000118203(2)] & -0.00609(2) + N_f[0.00070922(1)] \\-0.0069803(3) + N_f[0.000063796] & -0.00768(2) + N_f[0.00056285(2)]\end{pmatrix}$ \\
& $4$ & $\begin{pmatrix}-0.002034(3) + N_f[-0.000118203(2)] & -0.00498(1) + N_f[0.00070922(1)] \\-0.007802(1) + N_f[0.000063796] & -0.003835(2) + N_f[0.00056285(2)]\end{pmatrix}$ \\
& $5$ & $\begin{pmatrix}-0.004566(5) + N_f[-0.000118203(2)] & -0.002722(8) + N_f[0.00070922(1)] \\-0.009719(5) + N_f[0.000063796] & -0.00180(1) + N_f[0.00056285(2)]\end{pmatrix}$ \\
& $6$ & $\begin{pmatrix}-0.004586(5) + N_f[-0.000118203(2)] & -0.00305(1) + N_f[0.00070922(1)] \\-0.008236(3) + N_f[0.000063796] & -0.003374(6) + N_f[0.00056285(2)]\end{pmatrix}$ \\
\hline
\multirow{12}{*}{1} & $1$ & $\begin{pmatrix}-0.000512(8) + N_f[-0.000118203(2)] & -0.00803(4) + N_f[0.00070922(1)] \\-0.006546(2) + N_f[0.000063796] & -0.00763(2) + N_f[0.00056285(2)]\end{pmatrix}$ \\
& $2$ & $\begin{pmatrix}-0.003044(2) + N_f[-0.000118203(2)] & -0.00577(2) + N_f[0.00070922(1)] \\-0.008464(2) + N_f[0.000063796] & -0.00559(1) + N_f[0.00056285(2)]\end{pmatrix}$ \\
& $3$ & $\begin{pmatrix}-0.0030635(3) + N_f[-0.000118203(2)] & -0.00609(2) + N_f[0.00070922(1)] \\-0.0069803(3) + N_f[0.000063796] & -0.00717(2) + N_f[0.00056285(2)]\end{pmatrix}$ \\
& $4$ & $\begin{pmatrix}-0.001521(6) + N_f[-0.000118203(2)] & -0.00498(1) + N_f[0.00070922(1)] \\-0.007802(1) + N_f[0.000063796] & -0.003321(7) + N_f[0.00056285(2)]\end{pmatrix}$ \\
& $5$ & $\begin{pmatrix}-0.004052(3) + N_f[-0.000118203(2)] & -0.002722(8) + N_f[0.00070922(1)] \\-0.009719(5) + N_f[0.000063796] & -0.00129(2) + N_f[0.00056285(2)]\end{pmatrix}$ \\
& $6$ & $\begin{pmatrix}-0.004072(3) + N_f[-0.000118203(2)] & -0.00305(1) + N_f[0.00070922(1)] \\-0.008236(3) + N_f[0.000063796] & -0.00286(1) + N_f[0.00056285(2)]\end{pmatrix}$ \\
\hline\end{tabular}\\
\end{center}


\begin{center}
\noindent\begin{tabular}{|c|c|c|}
 \hline
$\alpha$ & $s$ & ${\gamma^{(1)}}_{45}^{+}$ \\ 
 \hline
\multirow{12}{*}{0} & $1$ & $\begin{pmatrix}0.002303(3) + N_f[0.00012884(1)] & -0.00169802(7) + N_f[0.0000026054(2)] \\0.00172(8) + N_f[0.00179861(4)] & 0.00081(2) + N_f[-0.00035752(1)]\end{pmatrix}$ \\
& $2$ & $\begin{pmatrix}0.002685(2) + N_f[0.00012884(1)] & -0.0018770(3) + N_f[0.0000026054(2)] \\0.00336(7) + N_f[0.00179861(4)] & -0.00001(1) + N_f[-0.00035752(1)]\end{pmatrix}$ \\
& $3$ & $\begin{pmatrix}0.002077(4) + N_f[0.00012884(1)] & -0.0015930(3) + N_f[0.0000026054(2)] \\0.00232(7) + N_f[0.00179861(4)] & 0.00046(1) + N_f[-0.00035752(1)]\end{pmatrix}$ \\
& $4$ & $\begin{pmatrix}0.00558(2) + N_f[0.00012884(1)] & -0.0019027(6) + N_f[0.0000026054(2)] \\0.00795(6) + N_f[0.00179861(4)] & 0.00040(1) + N_f[-0.00035752(1)]\end{pmatrix}$ \\
& $5$ & $\begin{pmatrix}0.00597(2) + N_f[0.00012884(1)] & -0.002082(1) + N_f[0.0000026054(2)] \\0.00959(4) + N_f[0.00179861(4)] & -0.00043(1) + N_f[-0.00035752(1)]\end{pmatrix}$ \\
& $6$ & $\begin{pmatrix}0.00536(1) + N_f[0.00012884(1)] & -0.0017978(6) + N_f[0.0000026054(2)] \\0.00856(4) + N_f[0.00179861(4)] & 0.000048(10) + N_f[-0.00035752(1)]\end{pmatrix}$ \\
\hline
\multirow{12}{*}{3/2} & $1$ & $\begin{pmatrix}0.00076(2) + N_f[0.00012884(1)] & -0.00169802(7) + N_f[0.0000026054(2)] \\0.00172(8) + N_f[0.00179861(4)] & -0.000727(3) + N_f[-0.00035752(1)]\end{pmatrix}$ \\
& $2$ & $\begin{pmatrix}0.00114(1) + N_f[0.00012884(1)] & -0.0018770(3) + N_f[0.0000026054(2)] \\0.00336(7) + N_f[0.00179861(4)] & -0.001556(3) + N_f[-0.00035752(1)]\end{pmatrix}$ \\
& $3$ & $\begin{pmatrix}0.00054(2) + N_f[0.00012884(1)] & -0.0015930(3) + N_f[0.0000026054(2)] \\0.00232(7) + N_f[0.00179861(4)] & -0.001082(3) + N_f[-0.00035752(1)]\end{pmatrix}$ \\
& $4$ & $\begin{pmatrix}0.004044(3) + N_f[0.00012884(1)] & -0.0019027(5) + N_f[0.0000026054(2)] \\0.00795(3) + N_f[0.00179861(4)] & -0.001138(2) + N_f[-0.00035752(1)]\end{pmatrix}$ \\
& $5$ & $\begin{pmatrix}0.004426(5) + N_f[0.00012884(1)] & -0.002082(1) + N_f[0.0000026054(2)] \\0.00959(2) + N_f[0.00179861(4)] & -0.001966(5) + N_f[-0.00035752(1)]\end{pmatrix}$ \\
& $6$ & $\begin{pmatrix}0.003817(3) + N_f[0.00012884(1)] & -0.0017978(5) + N_f[0.0000026054(2)] \\0.00856(2) + N_f[0.00179861(4)] & -0.001492(5) + N_f[-0.00035752(1)]\end{pmatrix}$ \\
\hline
\multirow{12}{*}{1} & $1$ & $\begin{pmatrix}0.00128(1) + N_f[0.00012884(1)] & -0.00169802(7) + N_f[0.0000026054(2)] \\0.00172(8) + N_f[0.00179861(4)] & -0.000214(7) + N_f[-0.00035752(1)]\end{pmatrix}$ \\
& $2$ & $\begin{pmatrix}0.00166(1) + N_f[0.00012884(1)] & -0.0018770(3) + N_f[0.0000026054(2)] \\0.00336(7) + N_f[0.00179861(4)] & -0.001042(3) + N_f[-0.00035752(1)]\end{pmatrix}$ \\
& $3$ & $\begin{pmatrix}0.00105(1) + N_f[0.00012884(1)] & -0.0015930(3) + N_f[0.0000026054(2)] \\0.00232(7) + N_f[0.00179861(4)] & -0.000568(3) + N_f[-0.00035752(1)]\end{pmatrix}$ \\
& $4$ & $\begin{pmatrix}0.004557(7) + N_f[0.00012884(1)] & -0.0019027(6) + N_f[0.0000026054(2)] \\0.00795(4) + N_f[0.00179861(4)] & -0.000624(6) + N_f[-0.00035752(1)]\end{pmatrix}$ \\
& $5$ & $\begin{pmatrix}0.004940(9) + N_f[0.00012884(1)] & -0.002082(1) + N_f[0.0000026054(2)] \\0.00959(3) + N_f[0.00179861(4)] & -0.001453(2) + N_f[-0.00035752(1)]\end{pmatrix}$ \\
& $6$ & $\begin{pmatrix}0.004331(7) + N_f[0.00012884(1)] & -0.0017978(5) + N_f[0.0000026054(2)] \\0.00856(2) + N_f[0.00179861(4)] & -0.000979(1) + N_f[-0.00035752(1)]\end{pmatrix}$ \\
\hline\end{tabular}\\
\end{center}


\begin{center}
\noindent\begin{tabular}{|c|c|c|}
 \hline
$\alpha$ & $s$ & ${\gamma^{(1)}}_{45}^{-}$ \\ 
 \hline
\multirow{12}{*}{0} & $1$ & $\begin{pmatrix}-0.01620(2) + N_f[0.00069509(2)] & 0.001678(1) + N_f[-0.0000064430(8)] \\0.00879(7) + N_f[-0.002125401(8)] & 0.00907(1) + N_f[-0.0004158697(7)]\end{pmatrix}$ \\
& $2$ & $\begin{pmatrix}-0.01676(2) + N_f[0.00069509(2)] & 0.001156(2) + N_f[-0.0000064430(8)] \\0.00887(6) + N_f[-0.002125401(8)] & 0.008779(9) + N_f[-0.0004158697(7)]\end{pmatrix}$ \\
& $3$ & $\begin{pmatrix}-0.01560(1) + N_f[0.00069509(2)] & 0.002266(2) + N_f[-0.0000064430(8)] \\0.00841(6) + N_f[-0.002125401(8)] & 0.008299(9) + N_f[-0.0004158697(7)]\end{pmatrix}$ \\
& $4$ & $\begin{pmatrix}-0.01236(2) + N_f[0.00069509(2)] & 0.001914(1) + N_f[-0.0000064430(8)] \\0.00755(4) + N_f[-0.002125401(8)] & 0.00896(1) + N_f[-0.0004158697(7)]\end{pmatrix}$ \\
& $5$ & $\begin{pmatrix}-0.01292(2) + N_f[0.00069509(2)] & 0.0013917(10) + N_f[-0.0000064430(8)] \\0.00762(3) + N_f[-0.002125401(8)] & 0.008664(8) + N_f[-0.0004158697(7)]\end{pmatrix}$ \\
& $6$ & $\begin{pmatrix}-0.01176(2) + N_f[0.00069509(2)] & 0.002502(3) + N_f[-0.0000064430(8)] \\0.00716(3) + N_f[-0.002125401(8)] & 0.008184(8) + N_f[-0.0004158697(7)]\end{pmatrix}$ \\
\hline
\multirow{12}{*}{3/2} & $1$ & $\begin{pmatrix}-0.01774(3) + N_f[0.00069509(2)] & 0.001678(1) + N_f[-0.0000064430(8)] \\0.00879(7) + N_f[-0.002125401(8)] & 0.007530(5) + N_f[-0.0004158697(7)]\end{pmatrix}$ \\
& $2$ & $\begin{pmatrix}-0.01830(4) + N_f[0.00069509(2)] & 0.001156(2) + N_f[-0.0000064430(8)] \\0.00887(6) + N_f[-0.002125401(8)] & 0.007238(3) + N_f[-0.0004158697(7)]\end{pmatrix}$ \\
& $3$ & $\begin{pmatrix}-0.01714(3) + N_f[0.00069509(2)] & 0.002266(2) + N_f[-0.0000064430(8)] \\0.00841(6) + N_f[-0.002125401(8)] & 0.006758(3) + N_f[-0.0004158697(7)]\end{pmatrix}$ \\
& $4$ & $\begin{pmatrix}-0.013901(2) + N_f[0.00069509(2)] & 0.0019137(8) + N_f[-0.0000064430(8)] \\0.00755(3) + N_f[-0.002125401(8)] & 0.007415(4) + N_f[-0.0004158697(7)]\end{pmatrix}$ \\
& $5$ & $\begin{pmatrix}-0.014461(4) + N_f[0.00069509(2)] & 0.0013917(6) + N_f[-0.0000064430(8)] \\0.00762(2) + N_f[-0.002125401(8)] & 0.007123(2) + N_f[-0.0004158697(7)]\end{pmatrix}$ \\
& $6$ & $\begin{pmatrix}-0.013305(5) + N_f[0.00069509(2)] & 0.002502(2) + N_f[-0.0000064430(8)] \\0.00716(2) + N_f[-0.002125401(8)] & 0.006643(2) + N_f[-0.0004158697(7)]\end{pmatrix}$ \\
\hline
\multirow{12}{*}{1} & $1$ & $\begin{pmatrix}-0.01722(3) + N_f[0.00069509(2)] & 0.001678(1) + N_f[-0.0000064430(8)] \\0.00879(7) + N_f[-0.002125401(8)] & 0.008043(7) + N_f[-0.0004158697(7)]\end{pmatrix}$ \\
& $2$ & $\begin{pmatrix}-0.01778(3) + N_f[0.00069509(2)] & 0.001156(2) + N_f[-0.0000064430(8)] \\0.00887(6) + N_f[-0.002125401(8)] & 0.007751(5) + N_f[-0.0004158697(7)]\end{pmatrix}$ \\
& $3$ & $\begin{pmatrix}-0.01663(2) + N_f[0.00069509(2)] & 0.002266(2) + N_f[-0.0000064430(8)] \\0.00841(6) + N_f[-0.002125401(8)] & 0.007271(5) + N_f[-0.0004158697(7)]\end{pmatrix}$ \\
& $4$ & $\begin{pmatrix}-0.013388(7) + N_f[0.00069509(2)] & 0.0019137(10) + N_f[-0.0000064430(8)] \\0.00755(3) + N_f[-0.002125401(8)] & 0.007928(6) + N_f[-0.0004158697(7)]\end{pmatrix}$ \\
& $5$ & $\begin{pmatrix}-0.013947(4) + N_f[0.00069509(2)] & 0.0013917(6) + N_f[-0.0000064430(8)] \\0.00762(3) + N_f[-0.002125401(8)] & 0.007637(4) + N_f[-0.0004158697(7)]\end{pmatrix}$ \\
& $6$ & $\begin{pmatrix}-0.01279(1) + N_f[0.00069509(2)] & 0.002502(2) + N_f[-0.0000064430(8)] \\0.00716(3) + N_f[-0.002125401(8)] & 0.007156(4) + N_f[-0.0004158697(7)]\end{pmatrix}$ \\
\hline\end{tabular}\\
\end{center}

\end{appendix}
\bibliographystyle{JHEPjus}
\bibliography{biblio}
\end{document}